\documentclass[aps,prb,a4paper, amsfonts, amssymb, amsmath, reprint, showkeys, nofootinbib, twoside]{revtex4-2}

\usepackage[english]{babel}
\usepackage[utf8]{inputenc}

\usepackage{amsthm}
\usepackage{mathtools}
\usepackage{physics}
\usepackage{xcolor}
\usepackage{graphicx}
\usepackage[left=23mm,right=13mm,top=35mm,columnsep=15pt]{geometry} 
\usepackage{adjustbox}
\usepackage{placeins}
\usepackage[T1]{fontenc}
\usepackage{lipsum}
\usepackage{csquotes}

\begin{document}

\title{ Spectral features of magnetic domain walls on surface of 3D
  topological insulators.}

\author{I. P. Rusinov} \email[Correspondence email address:
]{rusinovip@gmail.com}
\affiliation{Tomsk State University,
Tomsk, 634050 Russia} \affiliation{St. Petersburg State University,
    199034 St. Petersburg, Russia}

\author{V. N. Men'shov} \affiliation{NRC Kurchatov Institute, Kurchatov
  Sqr. 1, 123182 Moscow, Russia} \affiliation{Tomsk State University,
  Tomsk, 634050 Russia} \affiliation{St. Petersburg State University,
  199034 St. Petersburg, Russia}

\author{E. V. Chulkov} \affiliation{St. Petersburg State University,
  199034 St. Petersburg, Russia} \affiliation{Donostia International
  Physics Center (DIPC), 20018 San Sebasti\'{a}n/Donostia, Spain}
\affiliation{ Departamento de Pol\'{i}meros y Materiales Avanzados:
  F\'{i}sica, Qu\'{i}mica y Tecnolog\'{i}a, Facultad de Ciencias
  Qu\'{i}micas, Universidad del Pa\'{i}s Vasco UPV/EHU, 20080 San
  Sebasti\'{a}n/Donostia, Basque Country, Spain }

\date{\today} 

\begin{abstract}
We present a theoretical investigation of electron states hosted by
magnetic domain walls on the 3D topological insulator surface. The
consideration includes the domain walls with distinct vectorial and
spatial textures.  The study is carried out on the basis of the
Hamiltonian for quasi-relativistic fermions by using a continual
approach and tight-binding calculations. We derive the spectral
characteristics and spatial localization of the the one-dimensional
low-energy states appearing at the domain walls. The antiphase domain
walls are shown to generate the topologically protected chiral states
with linear dispersion, the group velocity and spin-polarization
direction of which depend on an easy axis orientation. In the case of
an easy plane anisotropy, we predict a realization of a dispersionless
state, flat band in the energy spectrum, that is spin-polarized along
the surface normal. Modification of the surface states in the
multi-domain case, which is approximated by a periodic set of domain
walls, is described as well. We find that the magnetic domain walls
with complex internal texture, such as N\'{e}el-like or Bloch-like
walls, also host the topological states, although their spectrum and
spin structure can be changed compared with the sharp wall case.
\end{abstract}


\maketitle

\section{INTRODUCTION}

The effect of quantized conductivity without an external magnetic
field was revealed and studied in topological insulators (TIs) that
are time-reversal invariant semiconductors with strong spin-orbit
coupling~\cite{hasan2010colloquium,
  Qi2011colloquium,Ando2013colloquium}.  However, namely the breaking
of time-reversal symmetry (TRS) through the introduction of a magnetic
order in TIs provides conducive ground for an emergence of fascinating
phenomena such as quantum anomalous Hall effect
(QAHE)~\cite{Weng2015_adv, Kou_2015,Chang_2016, LiuZhangQi_2016,
  Tokura2019}, axion insulator state~\cite{Tokura2019}, and Majorana
fermions~\cite{He294}, which would allow extending potential of
spintronic applications. At present, there are several viable
approaches for creating a magnetic order in TIs on the basis of
tetradymite-like semiconductors~\cite{MS_JL_2019}. Accordingly, five
alternative platforms suitable for realization of phenomena associated
with quantized transverse conductivity can be outlined as follows.  1)
QAHE was first detected in thin films composed of a few
quintuple-layers (QLs) of TIs Cr$_x$(Bi,Sb)$_{2 -
  x}$Te$_3$~\cite{Chang167} and V$_x$(Bi,Sb)$_{2 -
  x}$Te$_3$~\cite{Bestwick_2015}, where randomly dissolved moments of
transition metal atoms form ferromagnetic (FM) long-range order.  2)
By using magnetic modulation doping, when the rich-Cr/V-doped thin
layers are inserted near both the surfaces of (Bi,Sb)$_2$Te$_3$ films,
experimentalists have succeeded in observing QAHE~\cite{Mogi_2015,
  Okada2016, Mogi1669} and axion insulator
state~\cite{Mogi2017,Xiao_2018,Mogi1669}. 3) The
magnetic-proximity-effect-induced QAHE has been implemented in the
(Zn,Cr)Te/(Bi,Sb)$_2$Te$_3$/(Zn,Cr)Te heterostructure due to a fine
tuning of the composition of FM insulator interfaced with
TI~\cite{Watanabe2019}. 4) The giant exchange gap at the Dirac point
of the surface state could be achieved due to a magnetic extension
effect, when a thin layer of FM insulator is deposited on the surface
of a nonmagnetic TI, which are both structurally and compositionally
compatible with each other~\cite{Otrokov_HO2017,
  Otrokov.jetpl2017}. Perhaps this effect is responsible for anomalous
Hall regime at temperatures of several Kelvin observed in the magnetic
topological bulk crystals in which Mn ions self-organize into a
periodic MnBi$_2$Te$_4$/Bi$_2$Te$_3$ superlattice~\cite{Deng2021}. 5)
Recently, the existence of quantized Hall conductivity, accompanied by
zero longitudinal resistance, was experimentally demonstrated in the
thin flakes of an intrinsic antiferromagnetic (AFM) TI MnBi$_2$Te$_4$
with odd numbers of septuple-layers (SLs), concretely 5 SLs, under
zero extrinsic magnetic field~\cite{Deng895}.  This is a typical
behavior of QAHE, theoretically predicted for intrinsic AFM TI in
Ref.~\citenum{Otrokov.prl2019}. In turn, the flakes of MnBi$_2$Te$_4$
with even number of SLs, namely 6 SLs, have exhibited axion insulator
state at zero magnetic field~\cite{Liu2020}.

Despite rapid progress in this field, the exploration of magnetic TIs
with in-plane easy axis~\cite{Shikin2018, Rakhmilevich_2018,
  Chen_2015_adv} or with magnetic moment canted towards the
surface~\cite{Assaf_2015,LeeZhu_2019} remains scarce. The recent
study~\cite{Petrov_2021} has predicted the in-plane sublattice
magnetization in vanadium-based family of AFM TIs
V(Bi,Sb)$_2$(Se,Te)$_4$, which could significantly broaden the base
for the search of exotic fermion states at the TI surface.

It should be stressed that, in the magnetic TI samples prepared for
spin-dependent transport measurements, the physical boundaries such as
surfaces, interfaces and side faces play a particular
role~\cite{Tokura2019,MS_JL_2019}. Conceptually, given that the
surface/interface states are fully gapped at the Dirac point due to
the spontaneous magnetization normal to the surface, the topologically
protected edge chiral channels running along the side faces are
responsible for the quantized Hall conductivity~\cite{Weng_Rui_2015,
  Kou_2015, Chang_2016, LiuZhangQi_2016, Tokura2019, MS_JL_2019}. In
reality, the magnetization distribution at the surface can be highly
distinct from the bulk magnetic order. It is thought that this
distinction can be caused by various factors. One of them are thermal
magnetization fluctuations which, as a rule, are enhanced in the
surface region of a magnetic~\cite{Getzlaff_2008}. At the same time,
the magnetic fluctuations are frozen by static symmetry-reducing
imperfections~\cite{Bertotti_1998}. The structural defects and
compositional disorder inevitably occur during the exfoliation or the
epitaxial growth of the TI samples~\cite{Gong_2019, Shikin2020,
  Saas_2020}. For instance, the surface of an exfoliated
MnBi$_2$Te$_4$ flake can even be subjected to strong chemical
reconstruction, so that the topmost SL becomes rather as a Mn-doped
Bi$_2$Te$_3$ and a Mn$_x$Bi$_y$Te double layer with a clear van der
Waals gap in between~\cite{Hou2020}. These and other symmetry-reducing
imperfections can dominate locally the exchange and dipolar
interactions between moments as well as magnetic anisotropy at the
surface. In turn, an interplay between the exchange coupling and
anisotropy drives nucleation and energetic stability of the magnetic
landscape at the terminating surface including complex space-varying
textures of magnetization. It is also important to emphasize that, in
tetradymite TI materials, the topologically protected surface states
are mostly localized inside a few terminating QLs/SLs, i.e., roughly
speaking, inside the same region where the magnetic order is sensitive
to the perturbations and therefore can be remarkably modified compared
with the bulk magnetic order.

The existence of magnetic DWs on the surface of magnetic TIs and the
1D electron states associated with them is experimentally
confirmed~\cite{Saas_2020, LiuMinhao2016, Lachmane1500740}. Using
magnetic force microscopy, Sass and coauthors presented microscopic
evidence of the DWs on the as-grown (0001)-surface of MnBi$_2$Te$_4$,
which is consistent with opposite surface magnetizations of antiphase
domains or terraces separated by SL steps~\cite{Saas_2020}. The
emergence of multidomain states at topological phase transitions under
external field sweep was found in various magnetic systems based on
TIs~\cite{LiuMinhao2016, WangLian2014, Xiao_2018, Lachmane1500740,
  Mogi1669, Allen14511, WuXiao2020}.  The inconsistency of
spectroscopy (ARPES) results~\cite{Stability1, Stability2, Stability3,
  Stability4} with the earlier ones~\cite{OtrokovN2019,
  Zeugner_2019_AFM, LeeZhu_2019, Vidal_2019_MBT} indicates the
possibility of a spatially inhomogeneous structure of the surface
magnetization in MnBi$_2$Te$_4$. In addition, the magnetic DWs on the
AFM TI surface could also be induced intentionally using a magnetic
force microscope tip~\cite{Yasuda2017_MTI} or by spatially modulated
external magnetic field due to Meissner repulsion from a bulk
superconductor, as it has been realized in Cr-doped TI
(Bi,Sb)$_2$Te$_3$~\cite{Rosen2017}. On the theoretical side, the issue
on particular quasiparticle states on the magnetic DWs of TIs was
raised previously in Refs.~\citenum{Araki_2016, Zhang_MDW_2019,
  VarnavaVander_MDW, Dugaev_2020, Tavazza_2021, Petrov_2021}.

In this work, we study how the magnetic inhomogeneities such as DWs
can modify a picture of the topological surface states. The DW spatial
textures are modelled as static one-dimensional boundaries between
magnetic domains of different orientations on the TI surface. We
discuss the properties of the DW induced fermion states in the context
of the real space localization and the spin-resolved spectral
function. We demonstrate that such DWs host topologically protected
chiral fermion states, the energy spectrum of which can evolve from a
linear dispersion to a flat band depending on the easy axis
orientation. The results are obtained by means of two complementary
methods, one based on the effective low-energy Dirac equation
description and the other based on a numerical tight-binding approach.

The rest of the paper is organized as follows. In Sec.~II we present
the microscopic model of DWs on the magnetic TI surface and outline
methodological details of the description of the electron states bound
to DWs. In Sec.~III, we successively investigate the characteristics
of the low-energy surface states generated by a single magnetic
antiphase DWs under different directions of easy axis, a pair of the
antiphase DWs, and the multi-domain periodic configurations of
magnetization. Here we also study how the noncollinear and noncoplanar
textures of the magnetization modify the spectral properties of the
surface states. Finally, in Sec.~IV, we summarize the obtained results
and discuss possible manifestations of DWs in spectroscopy and
transport properties of TIs.

\section{MODEL AND METHODS}

\begin{figure}
	\centering
	\includegraphics[width=0.8\columnwidth]{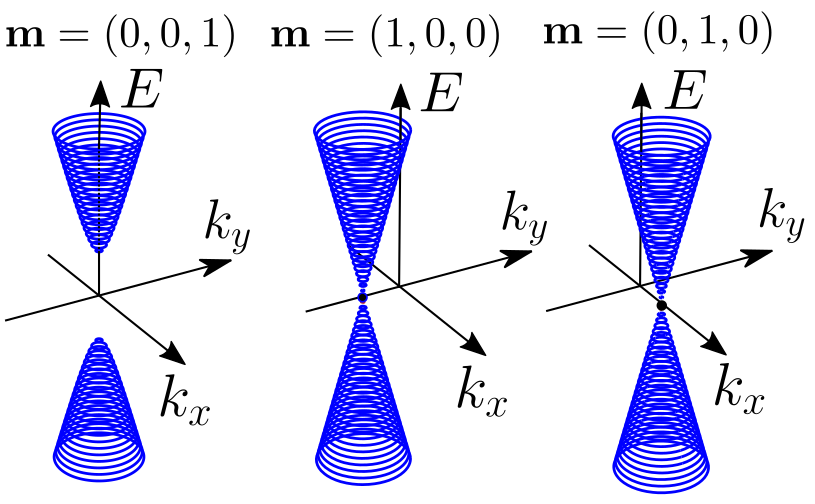}
	\caption{ Energy dispersion in the momentum space $(k_x,k_y)$
          for topological surface state subjected to the uniform
          surface magnetization oriented along axes the $z$-axis (left
          panel), the $x$-axis (central panel) and the $y$-axis (right
          panel).  }
	\label{fig:fig1}
\end{figure}

The essential low-energy physics on the 3D TI surface can be described
by massless two-component fermions with the spin-momentum
locking~\cite{hasan2010colloquium, Qi2011colloquium,
  Ando2013colloquium}. In the presence of the exchange field at the
surface, the motion of the fermions can be modelled with effective 2D
Hamiltonian:
\begin{equation}
  H(\mathbf{k}) = -\nu [\mathbf{k} \times \boldsymbol{\sigma}]_z +
  J (\mathbf{M} \cdot \boldsymbol{\sigma}) .
  \label{eq:main}
\end{equation}
Here, the first term is the lowest-order expansion of the Dirac
quasiparticle energy in the small in-plane momentum
$\mathbf{k}=(k_x,k_y)$ around the $\overline{\Gamma}$-point in the
surface Brillouin zone (BZ), $\nu$ is the Fermi velocity (one assumes
$\hbar=1$), $\boldsymbol{\sigma}$ is trio of Pauli matrices $\sigma_i$
($i=x,y,z$) for the spin degree of freedom. In the tetradymite TI
systems, both in the case of randomly distributed \emph{3d} transition
metal dopants and in the case of regularly aligned layers of magnetic
ions, the exchange coupling is mediated by \emph{p}-orbitals of the TI
host. It can lead to appearance of a magnetic ordering and hence a
surface magnetization $\mathbf{M}(x,y)$. The latter causes a spin
polarization of the the Dirac fermions owing to an effective
interaction $J$. We present the corresponding exchange energy by the
second term in Eq.~(\ref{eq:main}) that allows us to capture the DW
fingerprints in the spectral properties of the surface electrons in
any of the five platforms for magnetic ordering in TI listed above. It
should be noted that the interaction strength $J$ depends profoundly
on strategy of the formation magnetic order in the system. We also
suppose that the interaction $J$ is the same in the in-plane and
out-of-plane direction (for certainty $J>0$) and ignore the
particle–hole asymmetry in the surface bands. We suggest that the
surface states modelled with Eq.~(\ref{eq:main}) reside in the
projected bulk band gap. It should be noted, we do not include
possible hexagonal warping term~\cite{LFU_hex_warping_2009,
  TRauch_DualTop_2014}, which could be avoided at small momentum near
the BZ center.

If the surface magnetization is uniform, the energy spectrum of the
Hamiltonian~(\ref{eq:main}) is given by the relation $E^2 = (\nu k_x -
J M_y )^2 + (\nu k_y + J M_x)^2 + J^2 M^2_z$.  Thus, when the vector
$\mathbf{M}$ is oriented in the surface plane, the 2D electron states
show a gapless cone spectrum with the Dirac point shifted from the BZ
origin in the direction perpendicular to $\mathbf{M}$. In the case of
out-of-plane easy axis, the surface states are gapped.  The
corresponding 2D spectral relations for the three orientations of the
uniform surface magnetization are illustrated in
Fig.~\ref{fig:fig1}. 
However, for topological surface states, an interplay of the band
topology and magnetic ordering is not reduced to these spectral
features. Indeed, the topological indices of the surface states of a
magnetic TI might be closely linked with the sign of either the
exchange energy gap or exchange shift of the Dirac cone in 2D BZ,
depending on the magnetic easy axis direction. In the other words, the
fermions moving on the magnetically inhomogeneous surface can acquire
the spatially varying topological indices. Thus, DW, across which the
surface magnetization changes its sign, is also a border separating
domains with different topological indices. Thereby, such a border is
expected to host peculiar topologically protected 1D state.

As emphasized above, despite the presence of magnetic order in the TI
bulk, the terminating surface can display rather considerable local
variations of the magnetization $\mathbf{M}(x,y)$, including DWs with
various textures. The magnetic DWs of certain types can bind the Dirac
fermions. In order to elucidate this issue, we address a particular
class of magnetic configurations in which the $\mathbf{M}$, magnitude
is fixed, $|\mathbf{M}(x,y)|=M_0$, while the vector
$\mathbf{M}(x,y)/M_0 = \mathbf{m}(x,y)$ is a function of the position
at the surface plane. Furthermore, we restrict our study to a 1D model
in the sense that the magnetization undergoes large-scale modulation
in only one spatial direction, for definiteness $\mathbf{m}(x,y) \to
\mathbf{m}(x)$, while retaining lattice periodicity in the orthogonal
direction and, consequently, conserving momentum $k_y$.  We consider
several representative orientation configurations of the surface
magnetization. We do not analyze here the energy of these
configurations, but make use of them as a playground for the
exploration of substantive characteristics of the topological surface
states. At first, we will address isolated antiphase DWs of zero width
such as $\mathbf{m}(x) = \mathbf{m}_0 \operatorname{sgn}(x)$, which
may differ in the $\mathbf{m}_0$-direction with respect to the
crystallographic axes of the system. If the distance between the
neighboring DWs is not too large, the DW induced bound states overlap
and hybridize. This aspect is explored by using configurations of a
pair of DWs and a 1D periodic domain lattice. We also consider
non-collinear magnetic DWs at the TI surface and study the bound
states appearing at the isolated 180$^{\circ}$ N\'{e}el-type and
Bloch-type DWs, which profiles are simulated by piece wise unit
vectors $\mathbf{m}$.

Our $k$-linear model approach, Eq.~(\ref{eq:main}), has the advantage
that it allows us to readily explain main trends in modification of
the topological surface states caused by perturbations of
magnetization that are not easily accounted within \emph{ab initio}
simulations.  We employ both the continuum model analysis and the
tight-binding calculations.  The former is used to find explicit
low-energy solutions for the Dirac-like equation. The latter are based
on a lattice regularization of the $\mathbf{k} \cdot \mathbf{p}$
Hamiltonian (\ref{eq:main}) via the substitution $k_{x,y} \to (1/a)
\sin(k_{x,y} a)$ where $a$ is 2D square lattice constant. The
tight-binding simulation is featured by the dimensionless parameter
$JM_0a/\nu$. To avoid an appearance of false Dirac points at the
boundary of BZ, known as the double-fermion
problem~\cite{Resende_2017}, we include the Willson mass term
$\frac{2w\sigma_z}{a^2} [2 -\cos(k_x a) - \cos(k_y a)]$, which does
not significantly affect the surface state behavior at small
momenta. The tight-binding calculations are performed using recursive
technique for the Green functions~\cite{Sancho:85, Henk:93}.  Having
obtained the retarded Green function ($G^R(E,k_y)$) for a given
magnetization distribution, we can analyze further the corresponding
momentum-resolved one-particle spectral function $\rho(E,k_y)=(-1/\pi)
\Im\Tr G^R (E,k_y)$ and that for each spin polarization
$S_i(E,k_y)=(-1/\pi) \Im\Tr \sigma_i G^R (E,k_y)$. Furthermore, we can
obtain the total density of states (DOS) $\rho(E) = (2\pi/a)
\int_{-\pi/a}^{\pi/a} dk_y \rho(E,k_y)$, and the net spin polarization
$S_i(E) = (2\pi/a) \int_{-\pi/a}^{\pi/a} dk_y S_i(E,k_y) $ as function
of the state energy, $E$. The spatial distribution of the surface
electron state is evaluated by the local density of states
$\rho(E,x,k_y) = (-1/\pi) \Im [E + i\epsilon - H(x,k_y)]^{-1}$ at
energy $E$, where $i\epsilon$ is a small broadening width. The results
of the tight-binding numerical simulations are presented in Figs. 2-9.

\section{THE MAIN RESULTS}
\subsection{Single antiphase domain wall}

\begin{figure*}
	\centering
	\includegraphics[width=1.7\columnwidth]{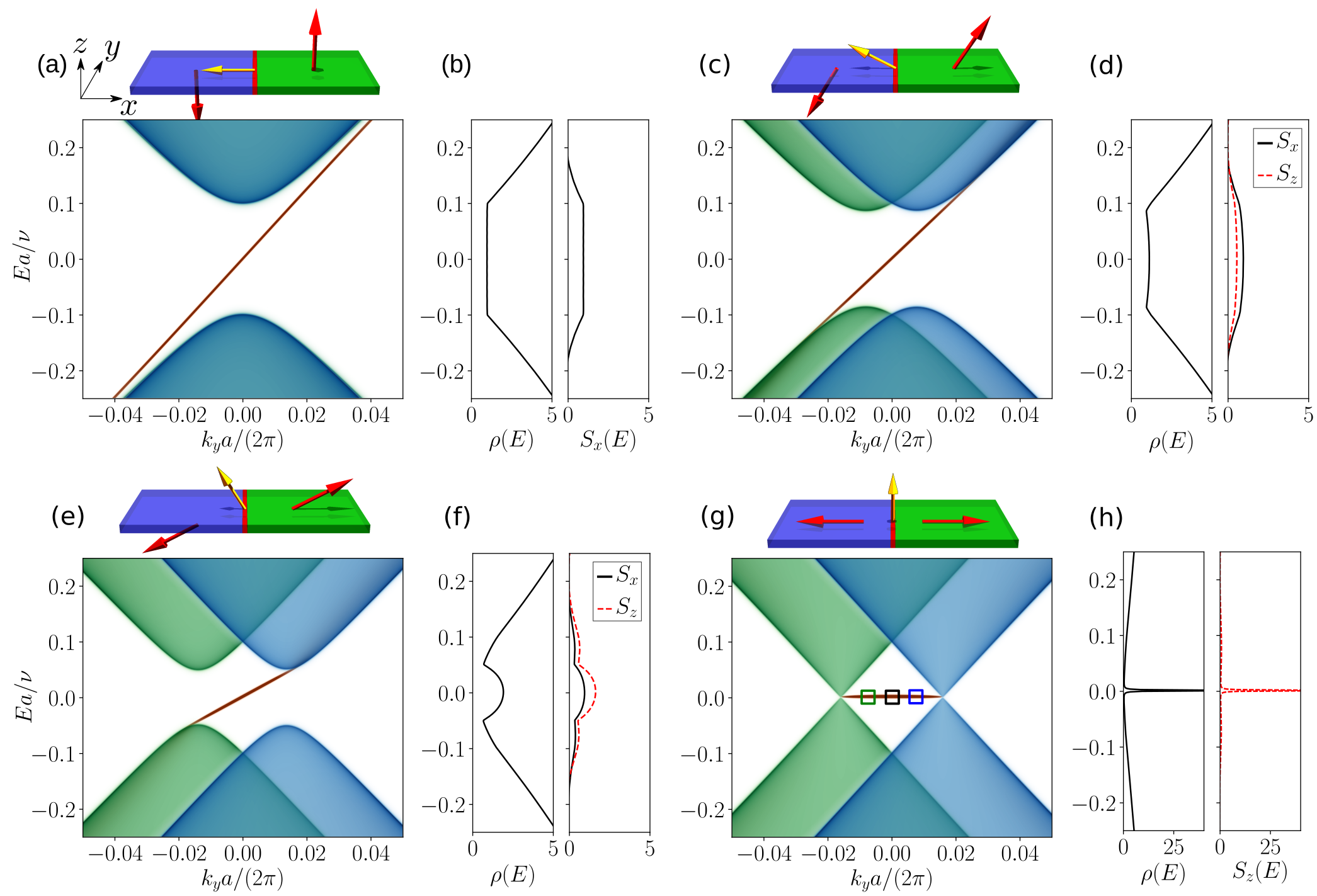}
	\caption{ Bound electron states induced by single sharp
          antiphase DW on the magnetic TI surface for increasing angle
          $\theta$. Panels (a,b), (c,d), (e,f), (g,h) correspond to
          $\theta=0$, $\pi/6$, $\pi/3$, $\pi/2$, respectively.
          Projected topological surface bands associated with left and
          right magnetic domains are in blue and green, respectively.
          In the panel (a) these bands coincide.  Dependence $E(k_y)$
          for the DW induced bound state obtained in terms of spectral
          function is in red. Total DOS, $\rho(E)$, and the spin
          polarization, $S_i(E)$, are shown as well. The top of the
          panels (a), (c), (e), (g) contains the schematic image of
          magnetic texture of antiphase DW indicated by arrows.}
	\label{fig:fig2}
\end{figure*}

\begin{figure}
	\centering
	\includegraphics[width=0.65\columnwidth]{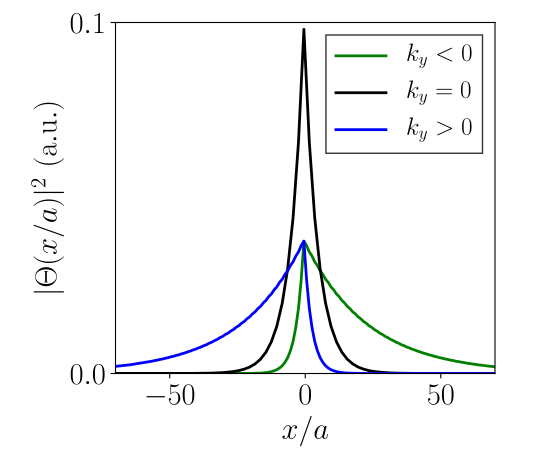}
	\caption{ Spatial profiles of the probability density of the
          planar DW bound state for distinct momenta $k_y$ indicated
          with small color squares on the dispersion curves in
          Fig.~\ref{fig:fig2} (g).}
	\label{fig:fig3}
\end{figure}

We begin by considering the electron state bound to a single collinear
DW, when the two its sides differ in sign of the magnetization. For
definiteness, DW is assumed to extend along the $y$-direction being
centered at $x=0$. The spatial configuration of such an antiphase DW
is merely given by a step function profile, $\mathbf{m}(x,y) =
\mathbf{m}(x) = \mathbf{m}_0 \operatorname{sgn}(x) $, i.e., the DW
magnetization changes sharply its direction to just the opposite one
under crossing over the boundary $(x=0,y)$, while the magnetization
within the domains stays uniform. We describe the unit vector
$\mathbf{m}_0=(\cos(\phi)\sin(\theta), \sin(\phi)\sin(\theta),
\cos(\theta))$ with spherical coordinates, where the polar angle
$\theta$ and azimuthal angle $\phi$ set the magnetization orientation
with respect to the $z$-axis and in the surface plane $(x,y)$,
respectively. The vector $\mathbf{m}_0$ is typically aligned with the
magnetic easy axis. The infinitely thin DW approximation is justified,
provided that the DW width is significantly shorter than the
localization scale of the electron state.

The system homogeneity along the $y$-direction implies that the Bloch
excitation with wave vector $k_y$ propagates along DW. Therefore, upon
replacement $k_x \to -i \partial_x$, the problem is reduced to the 1D
real-space Dirac equation $H(x,k_y) \Theta(x,k_y) = E(k_y) \Theta(x,
k_y)$ for the bispinor envelope function $\Theta(x,k_y) =
(\varphi(x,k_y), \chi(x,k_y))^\mathrm{T}$, which must decay into the
domain regions, i.e. $\Theta(|x| \to \infty ,k_y)=0$. On the other
hand, the function $\Theta(x,k_y)$ satisfies the boundary conditions
that ensure its continuity at $x=0$ and the derivative jump which
reads
  \begin{multline}
i \sigma_y [ \partial_x \Theta(x, k_y) \rvert_{0^{+}} - \partial_x
  \Theta(x,k_y) \rvert_{0^{-}} ] = \\ = 2 k_0 (\mathbf{m}_0 \cdot
\boldsymbol{\sigma}) \Theta(0,k_y),
\label{eq:Hamd}
\end{multline}
  where $k_0=JM_0/\nu$ is the characteristic momentum.

As a prime example of the eigen-problem we address the case when the
magnetization is tilted in the $(x,z)$ plane by angle $\theta$ and the
$m_y$ component is absent, i.e. $\sin(\phi)=0$. The <<tail-to-tail>>
DW, the two sides of which differ by angle $\pi$ (for definiteness $0
\le \theta \le \pi/2$) harbors the bound state featured by the
envelope function
\begin{equation}
  \Theta(x,k_y) = \Theta_0 \left(
  \begin{array}{c}
    \alpha \\
    1
  \end{array}
  \right)
  \sum_{\pm} h(\pm x) \exp(\mp p_\pm x)
  \label{eq:psi_t2t}
\end{equation}
and the energy dispersion
\begin{equation}
  E(k_y) = \nu k_y \cos(\theta).
  \label{eq:e_t2t}
\end{equation}
Here $h(x)$ is the Heviside function, the momenta $p_\pm = k_0 \pm k_y
\sin(\theta)$ determine the localization length of the bound state,
$\Theta_0$ is a normalization constant,
$\alpha=\tan((\pi/2-\theta)/2)$.  The spectral branch (\ref{eq:e_t2t})
exists within the momentum interval $|k_y \sin(\theta))| < k_0$, where
$p_{\pm} > 0$, in other words, the 1D linearly dispersive mode
(\ref{eq:e_t2t}) meets the 2D Dirac cones when $|k_y \sin(\theta))| =
k_0$.

Thus, we find that at the TI surface there is only one chiral bound
state per single magnetic DW, propagating along DW with group velocity
$\nu_{*}=\nu \cos(\theta)$.  The properties of the DW induced state
(\ref{eq:psi_t2t})--(\ref{eq:e_t2t}) alter as one changes the tilt
angle $\theta$. When the domain magnetization is parallel to the
surface normal, the emergent state is massless linearly dispersing and
completely spin-polarized along the $x$-axis fermion with energy
$E=\nu k_y$ and spinor construction $\alpha=1$. On the contrary, when
the vector $\mathbf{m}$ is parallel to the $x$-axis, a heavy fermion
with $E=0$ and spin polarization along the $z$-axis ($\alpha=0$)
appears. In any case, i.e. at arbitrary angle $\theta$, one can see
that the spin polarization of the bound state
(\ref{eq:psi_t2t})--(\ref{eq:e_t2t}) is exactly perpendicular to both
the DW propagation direction and the domain magnetization,
$(\mathbf{m}_0 \cdot \langle \boldsymbol{\sigma} \rangle)=0$.  In this
sense, besides spin-momentum coupling, the fermion state possesses an
peculiar sort of chirality when the fermion spin localized at DW is
tightly locked to the surface magnetization. The density of states
(DOS) for the solution (\ref{eq:psi_t2t})--(\ref{eq:e_t2t}) is given
by
\begin{equation}
  \rho(E)=\frac{a}{2\pi\nu\cos(\theta)} h(\nu k_0 \cot(\theta) - |E|).
  \label{eq:dos_t2t}
\end{equation}
The DOS (\ref{eq:dos_t2t}) is transformed from the constant value
$\rho(E)=a/(2 \pi \nu)$ for $\theta=0$ to the utterly narrow peak
$\rho(E)=(a k_0/\pi) \delta(E)$ for $\theta=\pi/2$, where $\delta(E)$ is
delta-function.

We have also numerically verified the existence and peculiarities of
the DW induced bound states using the lattice approximation. In the
tight-binding calculations, the model dimensionless parameter is
chosen as $J M_0 a / \nu=0.1$ only for convenience, in order to
facilitate the visualization of the results. One can easily verify
that the same qualitative results would be observed for any value of
this parameter. In Fig.~\ref{fig:fig2}, we plot representative
pictures of the spectral dependencies, $E(k_y)$, the total DOS,
$\rho(E)$, and the spin polarization, $S_i(E)$, for four selected
values of the angle $\theta$; besides, the corresponding DW
configurations are schematically depicted. What we are most interested
in is the low-lying states in the middle of the projected 1D BZ $|k_y|
< \pi/a$ with the momenta that are small compared to momentum of the
order of the reciprocal lattice length. First, the energy spectrum of
these states comprises pronounced projections of the two Dirac cones,
originated from the corresponding magnetic domains, separated by
momentum $2 k_0 \sin(\theta)$ and gapped by energy $2JM_0
\cos(\theta)$. Second, and most important, there is a 1D massless
linearly dispersing in-gap chiral fermion branch (at $\cos(\theta) \ne
0$) associated with the DW induced bound state.  The properties of the
emergent DW states controlled by the polar angle are drastically
different in the cases when the easy axis is normal to the surface or
lies in its plane. As seen in Fig.~\ref{fig:fig2} (a,b), at
$\cos(\theta)=1$, the fully spin-polarized along the $x$-axis
dispersionless mode with constant DOS spans the exchange gap of the
size $2 J M_0$ in the Dirac cone. Away from the angle $\theta=0$, the
state gradually changes its character: the group velocity,
$\nu_{*}=\nu \cos(\theta)$, of the bound state reduces linearly with
the $m_z$-component, the spin-polarization deviates from the $x$-axis
in the $(x,z)$ plane, and DOS as a function of energy concentrates
around $E=0$. Particular cases are demonstrated in Fig.~\ref{fig:fig2}
 for $\theta=\pi/6$ (c,d panels) and for $\theta=\pi/3$ (e,f
panels). Eventually, at $\cos(\theta)=0$, one observes in
Fig.~\ref{fig:fig2}(g) a perfectly flat band with energy $E=0$
that bridges the two the Dirac cones with nodes at momenta $\pm
k_0$. The corresponding heavy fermion state, fully spin-polarized
along the $z$-axis, manifests in a dramatic enhancement of DOS
(Fig.~\ref{fig:fig2} (h). In Fig.~\ref{fig:fig3} we depict the local
density profile of the flat band state for three momenta $k_y$.

Thus, combining analytic model and tight-binding numerical simulations
in the long wave-length limit we consistently demonstrate that the
antiphase DW at any polar orientation $\theta$ supports the in-gap
chiral degenerate bound state. The chirality consists in locking of
three vectors that are mutually orthogonal to each other, that is to
say they constitute the orthogonal trio: the domain polarization
$\mathbf{m}_0$ (oriented as a rule along a favoring axis of a
magnetocrystalline anisotropy in the surface plane), the state spin
polarization $\langle \boldsymbol{\sigma} \rangle$, and the
propagation direction, i.e. the wave-vector $\mathbf{y}k_y$. The
inversion of the DW magnetization $\mathbf{m}(x)$, such as
<<tail-to-tail>>$\leftrightarrow$<<head-to-head>>, entails the
alteration of the vectors $\langle\boldsymbol{\sigma}\rangle$ and
$\mathbf{y} k_y$ to just the opposite ones. In other words, the
fermions moving on the <<tail-to-tail>> and <<head-to-head>> DWs have
the opposite chiralities.

The spectral and spin features of the DW induced state are highly
sensitive to the orientation of $\mathbf{m}_0$ with respect to the
surface normal. Notably also, the existence of the eigen localized
solution (\ref{eq:psi_t2t})--(\ref{eq:e_t2t}) is guaranteed by the
antiphase DW texture, which allows us to consider such a magnetic
defect as a topological boundary and, consequently, the emergent
in-gap quasiparticle state as topologically protected one. When the
magnetization points perpendicular to the TI surface, an exchange
interaction induced magnetization opens up the gap of the size
$2JM_0$. At the boundary between up and down out-of-plane magnetic
domains, according to the known arguments~\cite{Namura2011,
  Upadhyaya_2016}, a chiral linearly dispersive state appears
[Fig.~\ref{fig:fig2} (a)] because of the inversion of the gap from
$+2JM_0$ to $-2JM_0$.  To provide a compelling evidence that the DW
induced state (\ref{eq:psi_t2t})--(\ref{eq:e_t2t}) has a topological
origin at any polar angle we use the reasoning similar to that given
in Ref.~\cite{Petrov_2021, VIU_1995}. In the general case, which is
addressed here, the role of the <<effective gap>> could be played by
the momentum dependent energy parameter $\Delta(x,k_y) = 2\nu [p_{+}
  h(x) - p_{-} h(-x)]=2[JM_0 \operatorname{sgn}(x) + \nu k_y
  \sin(\theta)]$. Provided that $k_y$ is restricted to the realm of
the existence of the bound state (\ref{eq:psi_t2t})--(\ref{eq:e_t2t})
in the 1D BZ, $|k_y \sin(\theta)| < k_0$, the parameter
$\Delta(x,k_y)$ changes its sign just at the magnetic DW. On the
contrary, if $|k_y \sin(\theta)| > k_0$, the parameter $\Delta(x,k_y)$
has the same sign on both sides of DW. Thus, the 1D gapless surface
bound state (\ref{eq:psi_t2t})--(\ref{eq:e_t2t}) appears due to the
<<gap>> $\Delta(x,k_y)$ closing at DW, which is guaranteed by the
texture of the antiphase DW at any polar orientation $\theta$. In the
partial case of the easy plane, $\theta=\pi/2$, the linearly
dispersing branch of the surface spectrum, Fig.~\ref{fig:fig2}
(a,c,e), transforms into the flat-band branch with energy $E=0$
connecting the Dirac points at $k_y=\pm k_0$, Fig.~\ref{fig:fig2}
(g). It should be acknowledged that the existence of a 2D
dispersionless state, so-called <<heavy fermion>>, was predicted by
Volkov and Pankratov in the case of a supersymmetric DW in a
ferroelectric semiconductor A$^4$B$^6$~\cite{VP_1986}.

When the surface magnetization remains in the $(y,z)$ plane, the
antiphase DW featured by the vector $\mathbf{m}_0 = (0, \sin(\theta),
\cos(\theta))$ hosts massless linearly dispersing chiral fermion for
any angle $0 \le \theta \le \pi/2$. In according to the boundary
conditions of Eq.~(\ref{eq:Hamd}), it is described by the envelope
function $\Theta(x,k_y) \sim \exp(-k_0 e^{i\theta} |x|)$ and spectrum
$E(k_y) = \nu k_y$ and is fully-spin-polarized along the $x$-axis. In
the case of the magnetization is oriented along DW,
i.e. $\theta=\pi/2$ and $\varphi=\pi/2$, the bound state disappears.

\begin{figure}
	\centering
	\includegraphics[width=\columnwidth]{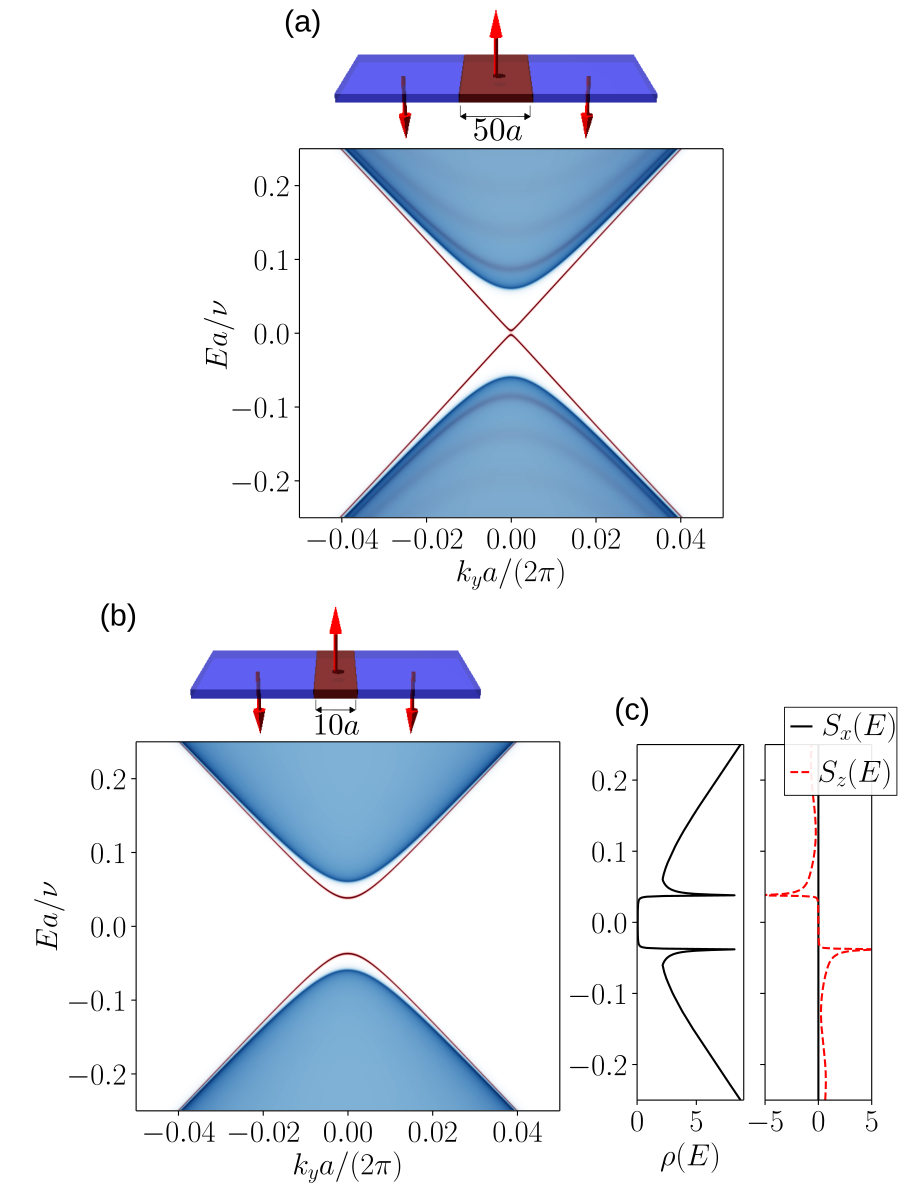}
	\caption{ Bound electron states induced by a pair of single
          sharp antiphase DWs on the magnetic TI surface for
          out-of-plane easy axis. Dependence $E(k_y)$ for the induced
          bound state obtained in terms of the spectral function is
          present in panels (a,b) in red for two different distances
          between DWs, $L=50a$ and $L=10a$. Total DOS, $\rho(E)$, and
          the spin polarization, $S_i(E)$, are shown in panel (c). The
          top of panels (a) and (b) contains the schematic image of
          out-of-plane antiphase magnetic texture, where the domain
          magnetizations are indicated by arrows.}
	\label{fig:fig4}
\end{figure}

\begin{figure}
	\centering
	\includegraphics[width=\columnwidth]{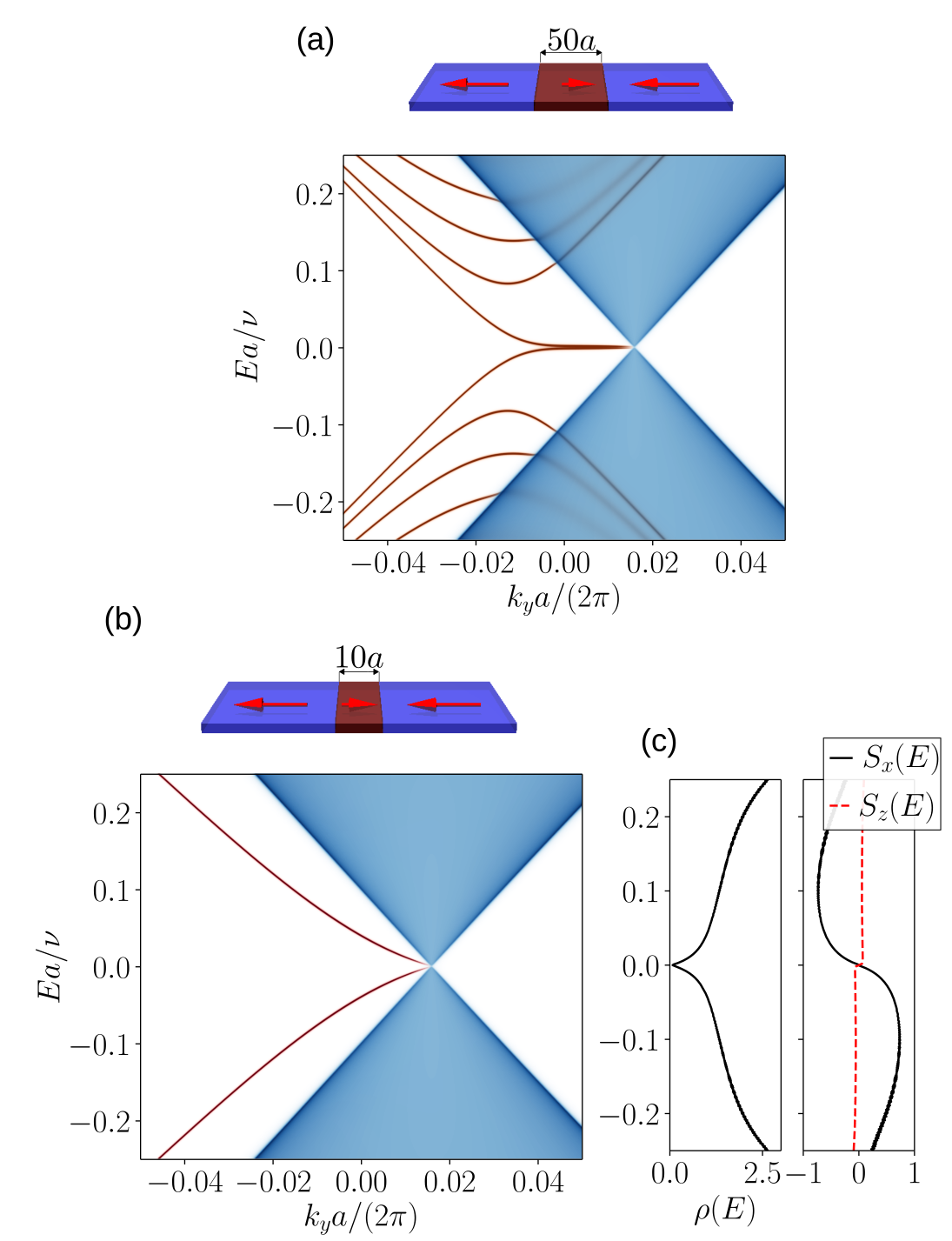}
	\caption{ The same as Fig.~\ref{fig:fig4}, but for the case of
          the in-plane easy axis.}
	\label{fig:fig5}
\end{figure}

\subsection{\label{sec:dw} The flat band state for a single domain wall with
  in-plane magnetization}

Now we turn to consideration of the magnetic TI surface where domains
are oriented in the basic plane due to preferable easy-plane
anisotropy. In order to understand what happens with the surface state
we set the single azimuthal DW that separates two regions with
differing magnetization orientations specified by angles $\phi_{+}$
and $\phi_{-}$. The magnetization spatial profile reads:
$\mathbf{m}(x)=(\cos(\phi_{+}),\sin(\phi_{+}),0) h(x) +
(\cos(\phi_{-}),\sin(\phi_{-}),0) h(-x)$. We find readily the
DW induced state pinned to energy $E=0$ and express its envelope
function as
\begin{align}
  \begin{split}
    \Theta(x) &=
    \left(
    \begin{array}{c}
      \varphi_0 \\
      0
    \end{array}
    \right)
    \sum_{\pm} h(\pm x) \exp(q_{\pm} x), \\
    & \quad -k_0 \cos(\phi_{-}) < k_y < -k_0 \cos(\phi_+); \\ 
    \Theta(x) &=
    \left(
    \begin{array}{c}
      0 \\
      \chi_0
    \end{array}
    \right)
    \sum_{\pm} h(\pm x) \exp(-q^{*}_{\pm} x), \\
    & \quad -k_0 \cos(\phi_{+}) < k_y < -k_0 \cos(\phi_-) \\
  \end{split}
  \label{eq:fb_envelope}
\end{align}
where the momenta $q_{\pm}=k_y + k_0 \exp(i\phi_{\pm})$ determine the
localization length of the bound state, $\varphi_0$ and $\chi_0$ are
normalization constants. From here we conclude that the azimuthal DW
generates the dispersionless bound state with perfect spin
polarization aligned perpendicular to the surface plane unless only
the $M_x$ components coincide. The realm of the flat band connecting
the projections of the two Dirac cones in the 1D BZ is determined by
angles $\phi_{+}$ and $\phi_{-}$. The condition $\phi_{+}=-\phi_{-}$
or $\phi_{+},\phi_{-}=\pm \pi/2$ means disappearing of the bound
state. On the contrary, in the case of the antiphase DW, when either
$\phi_{+}=0$, $\phi_{-}=\pi$ or $\phi_{+}=\pi$, $\phi_{-}=0$, the
state occupies the widest possible realm in the momentum space, $|k_y|
< k_0$, Fig.~\ref{fig:fig2} (g). The DOS for the flat band state of
Eq.~(\ref{eq:fb_envelope}) is given by $\rho(E) = [ak_0/(2\pi)] \left|
\cos(\phi_{+}) - \cos(\phi_{-}) \right| \delta(E) $.  Our
consideration demonstrates that the 1D flat band state is robust even
when the orientations of the magnetic domains are rotated with respect
to each other in the basic plane.

\subsection{Pair of antiphase domain walls}

If the distance between two DWs is not too large, the DW induced bound
states can overlap and hybridize, effectively leading to a deformation
of their spectra. To analyze this modification we describe the surface
magnetization distribution as composed of three different regions
divided by the sharp DWs fixed at $x=\pm L/2$. In the case of the
out-of-plane anisotropy, the up-magnetized stripe domain of a finite
width $L$ is placed between the two half-infinite down-magnetized
domains, which can be represented as $\mathbf{m}(x)=(0,0,-1)
h(|x|-L/2) + (0,0,1) h(L/2-|x|)$. When the overlap of the tails of the
states originating from the DWs is small, two 1D linear spectral
branches appear within the exchange gap $2JM_0$, which are slightly
gapped $\sim JM_0 \exp(-k_0 L)$ at $k_y=0$ due to the states
hybridization, Fig.~\ref{fig:fig4} (a). If the DWs are relatively
close to each other, $Lk_0 \ll 1$, the large overlap leads to
formation of a pair of weakly localized states with parabolic-like
bands (Fig.~\ref{fig:fig4} (b)), whose edges are near the the exchange
gap edges, $JM_0 - |E| \sim JM_0 (k_0 L)^2$ at $k_y=0$. The DOS and
spin polarization of these states are depicted in Fig.~\ref{fig:fig4}
(c).

In the case of the in-plane anisotropy, the magnetization
configuration includes the <<tail-to-tail>> and <<head-to-head>>
DWs. For the sake of definiteness, we assume that the side domain
regions magnetized along the $x$-axis are separated by the
intermediate region with the opposite magnetization direction, which
is described by $\mathbf{m}(x) = (-1,0,0) h(|x|-L/2) + (1,0,0) h(L/2 -
|x|)$. In the spectral picture of Fig.~\ref{fig:fig5} (a) one can see
that, at large enough width $L$, one of the cones, associated with the
middle domain, is weakly pronounced, while the heavy fermion spectral
branch acquires weak dispersion given by $E(k_y) \approx \sqrt{2JM_0
  \nu (k_0 - k_y)} \exp(-(k_0 - k_y) L)$. When $Lk_0 \ll 1$, both this
cone and the bound state disappear, Fig.~\ref{fig:fig5} (b). The
respective DOS and spin polarization are depicted in
Fig.~\ref{fig:fig5} (c).  Our analysis shows that the DW induced flat
band state is highly sensitive to the DW-to-DW distance.

\subsection{Periodical array of domain walls}

\begin{figure}
	\centering
	\includegraphics[width=\columnwidth]{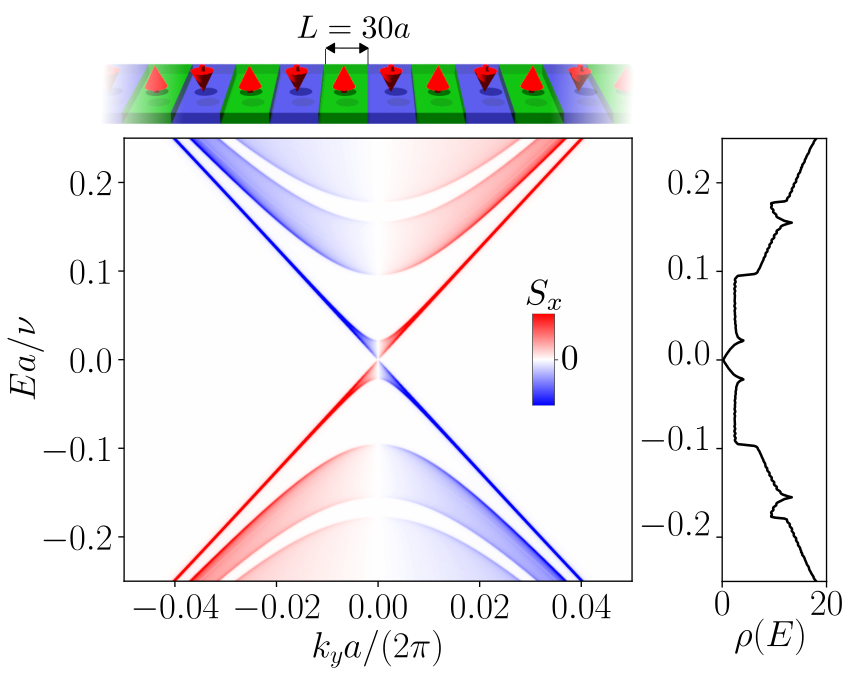}
	\caption{ Electron states induced by periodical array of sharp
          antiphase DWs on the magnetic TI surface with the
          out-of-plane easy axis. The spin-resolved spectral function
          is calculated for texture with the period $2L$, where
          $L=30a$.  The amplitude and sign of the spin polarization
          $S_x(E)$ of the energy states $E(k_y)$ are coded in red-blue
          color scheme.  The top of the panel contains the schematic
          image of the corresponding magnetic texture, where the
          domain magnetizations are indicated by arrows. Total DOS,
          $\rho(E)$, is shown as well.}
	\label{fig:fig6}
\end{figure}

\begin{figure*}
	\centering
	\includegraphics[width=1.8\columnwidth]{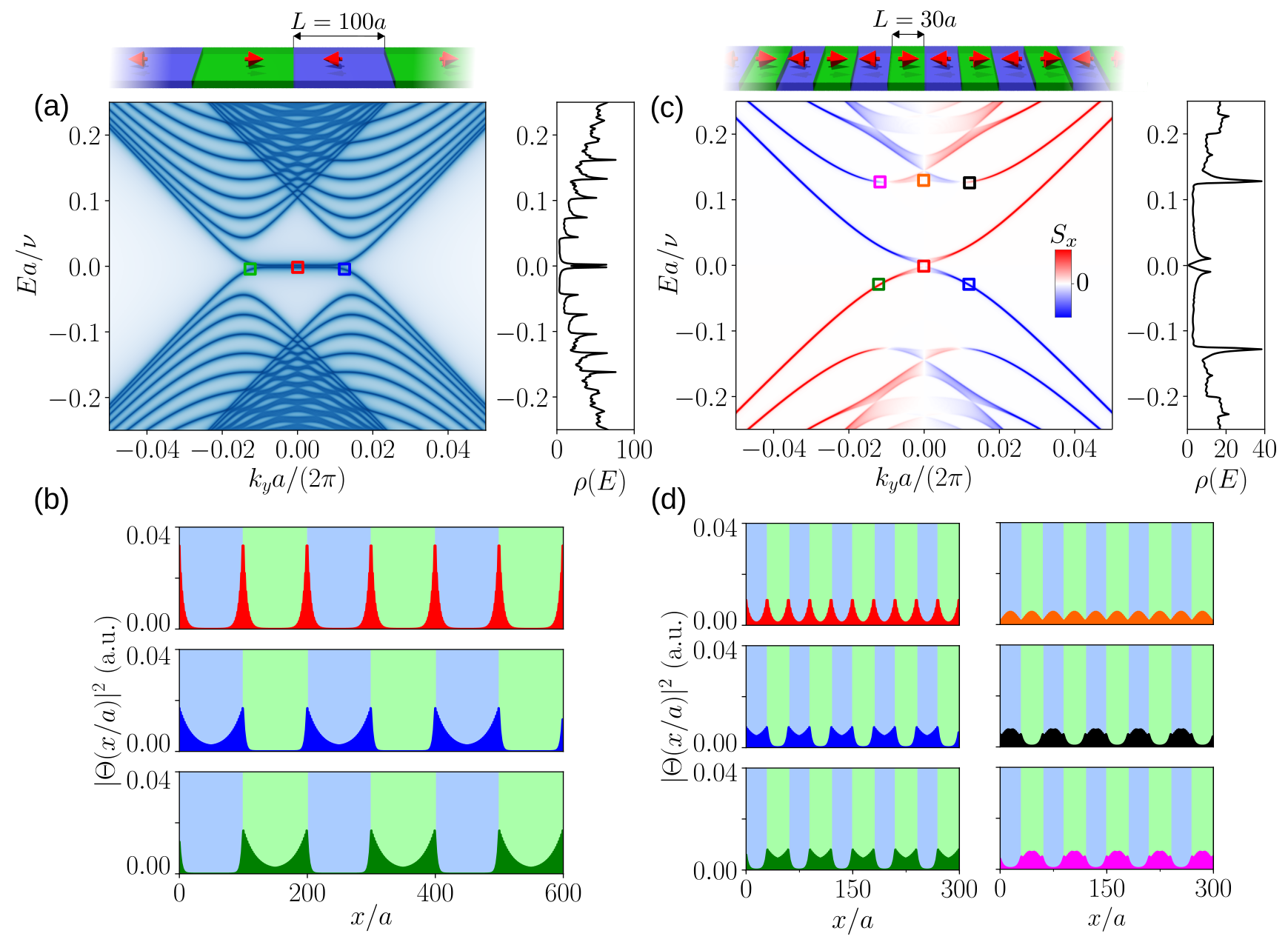}
	\caption{ Electron states induced by periodical array of sharp
          antiphase DWs on the magnetic TI surface for the in-plane
          easy axis. The calculated spin-resolved spectral function is
          shown for the texture with period $2L$, where $L=100a$, the
          panel (a), and $L=30a$ in the panel (c). In the latter case,
          the amplitude and sign of the spin polarization $S_x(E,k_y)$
          of the energy states $E(k_y)$ are coded in red-blue color
          scheme. The top of the panels (a) and (c) contains the
          schematic image of the corresponding magnetic texture, where
          the domain magnetizations are indicated by arrows. Total
          DOSs, $\rho(E)$, are shown as well. The panels (b) and (d)
          present spatial profiles of the probability density of the
          states with distinct momenta $k_y$ indicated with small
          squares of the corresponding color on the dispersion curves
          in panels (a) and (c), respectively.  }
	\label{fig:fig78}
\end{figure*}


In low-temperature phase, the surface of an as-grown magnetic material
sample should host a set of domains with different magnetization
directions~\cite{Bertotti_1998, Hubert_book, Nolting_book}.  Here the
question arises on how such a multi-domain magnetic texture, in
particular the domain length-scale, affects the electronic and spin
structure of the TI surface states. To figure that out, we address the
1D periodically modulated collinear textures of the surface
magnetization which consist of alternating magnetization stripes
aligned in the opposite directions: $\mathbf{m}(x) = \mathbf{m}_0
\operatorname{sgn} \left[ \sin(\pi x/L) \right]$, where $2L$ is the
texture period. The domain stripes have finite width along the
$x$-axis and infinite length along the $y$-axis and are separated from
one another by sharp antiphase DWs. The net surface magnetization is
supposed to be vanished. The domain polarization $\mathbf{m}_0$ is
fixed to an easy axis. To be more specific, we study the modification
of the topological surface states under the magnetic texture
$\mathbf{m}(x)$ for two cases: the out-of-plane anisotropy, when
$\mathbf{m}_0=(0,0,1)$, and the in-plane anisotropy, when
$\mathbf{m}_0=(1,0,0)$.

The tight-binding calculations are performed using an enlarged unit
cell, which is the stripe of width $2L = Na$ in the $x$-direction,
where $N$ is the number of sites per the unit cell. The latter
contains two anti-aligned domains. As we have already clarified, due
to the local variation of exchange field, an isolated antiphase DW
supports the bound state. As the magnetization forms a multi-domain
texture, the surface states stemming from the neighboring DWs
hybridize. Hence, it is natural to expect that the surface Dirac-like
spectrum rebuilds owing to the formation of the Bloch-like subbands
$E_n(k_x,k_y)$ in 2D BZ restricted to the realm $|k_y| < \pi/a$,
$|k_x| < \pi/(2L)$.

In the case of the $z$-easy axis, the 2D Dirac surface states are
gapped out by the domain magnetizations pointing outwards and inwards
from the sample surface, whereas the single DW produces the 1D in-gap
state. As the DWs are arranged regularly, the low-energy spectrum can
be expressed in the form $E^{\pm}_0(k_x,k_y) = \pm \sqrt{\nu^2 k_y^2 +
  \xi^2(k_x) + \Delta^2}$. In the regime where the typical
localization length $k^{-1}_0$ is smaller than the inter-DW distance,
the dispersion in $k_x$ appears due to the weak overlap of the
envelope function tails, $\xi(k_x) \sim \nu k_0 \exp(-k_0 L) \sin(k_x
L)$, the tiny hybridization gap $2\Delta$ between the branches
$E^{+}(k_x,k_y)$ and $E^{-}(k_x,k_y)$ is of the next order of
magnitude in the overlap.  The situation when the inter-DW distance is
comparable to the typical localization length $k^{-1}_0$ is
illustrated by the spectral picture in Fig.~\ref{fig:fig6} for
$L=30a$. Moreover, at the BZ boundary $|k_x|= \pi/(2L)$, the
remarkable gap opens up in the spectral density of the propagating
states about energy $|E|\approx 0.16 \nu/a $ at $k_y \approx 0$.  The
$x$-component of spin polarization $S_x(E,k_y)$ is depicted in
Fig.~\ref{fig:fig6} by means of color intensity of the spectral
branches. This image tells about the spin-momentum locking of the
surface states, excluding the narrow realm in the vicinity $k_y
\approx 0$.

More complicated situation arises in the case of the magnetization
lying in the surface plane. When the adjacent DWs lie at quite some
distance from one another, $L \gg k^{-1}_0$ (Fig.~\ref{fig:fig78} (a)
for $L=100a$), the flatness and spin polarization of the low-energy
states near $E=0$ changes very little as compared with the single-DW
case. There occur the dispersion $\xi(k_x) \sim \nu k_0 \exp(-k_0 L)
\sin(k_x L)$ along $k_x$ and the tiny gap $2\Delta$ between the
branches $E_0^{+}(k_x,|k_y| < k_0)$ and $E_0^{-}(k_x,|k_y| < k_0)$.
As follows from Fig.~\ref{fig:fig78} (b), the spatial profile of the
probability density of the low-energy state at $k_y \approx 0$ is
significantly different from that at $|k_y| \approx k_0$.  With regard
to the delocalized states, as seen in Fig.~\ref{fig:fig78} (a) for the
spectral density projected onto $k_y$, the continual cone spectrum is
split into a set of subbands, $E_n(|k_y|=k_0) \approx \pi \nu n/(2L)$,
$n=1,2,3...$. As the inter-DW distance is decreased to intermediate
value $L \sim k_0^{-1}$, we find that the low-energy spectrum loses
the flat behavior becoming parabolic along momentum $k_y$ which can be
seen in Fig.~\ref{fig:fig78} (c) for $L=30a$. Interestingly, while the
bands at zero energy disappear, two new flat-like bands at the finite
energy $|E| \approx \pm 0.13 \nu/a$ emerge due to a rebuilding of the
cone states in the exchange field of the periodic texture with
$L=30a$, Fig.~\ref{fig:fig78} (c). Correspondingly, two sharp peaks
are present in the full DOS. One also notes that the component
$S_x(E,k_y)$ reduces near $k_y \approx 0$, where the electron spin
density is polarized normal to the surface. The real space behavior of
some states with different eigen-values is shown in
Fig.~\ref{fig:fig78} (d).

Our analysis shows that the spectral characteristics of the low-energy
states in the case of the $z$-easy axis are generally stable in
relation to varying inter-DW distance. On the contrary, in the case of
the in-plane anisotropy, the dispersionless states bound to DW are
highly sensitive to change of the distance between neighboring
DWs. When $L \lesssim k_0^{-1}$, these states are interfering with
each other shifting the flat band level from zero energy. As a result,
the spectrum acquires the curvature and the gap. Moreover, the drastic
redistribution of DOS of the 2D cone states takes place under the
magnetization spatial modulation.

\subsection{\label{sec:nc} Noncollinear and noncoplanar domain walls}

\begin{figure}
	\centering \includegraphics[width=\columnwidth]{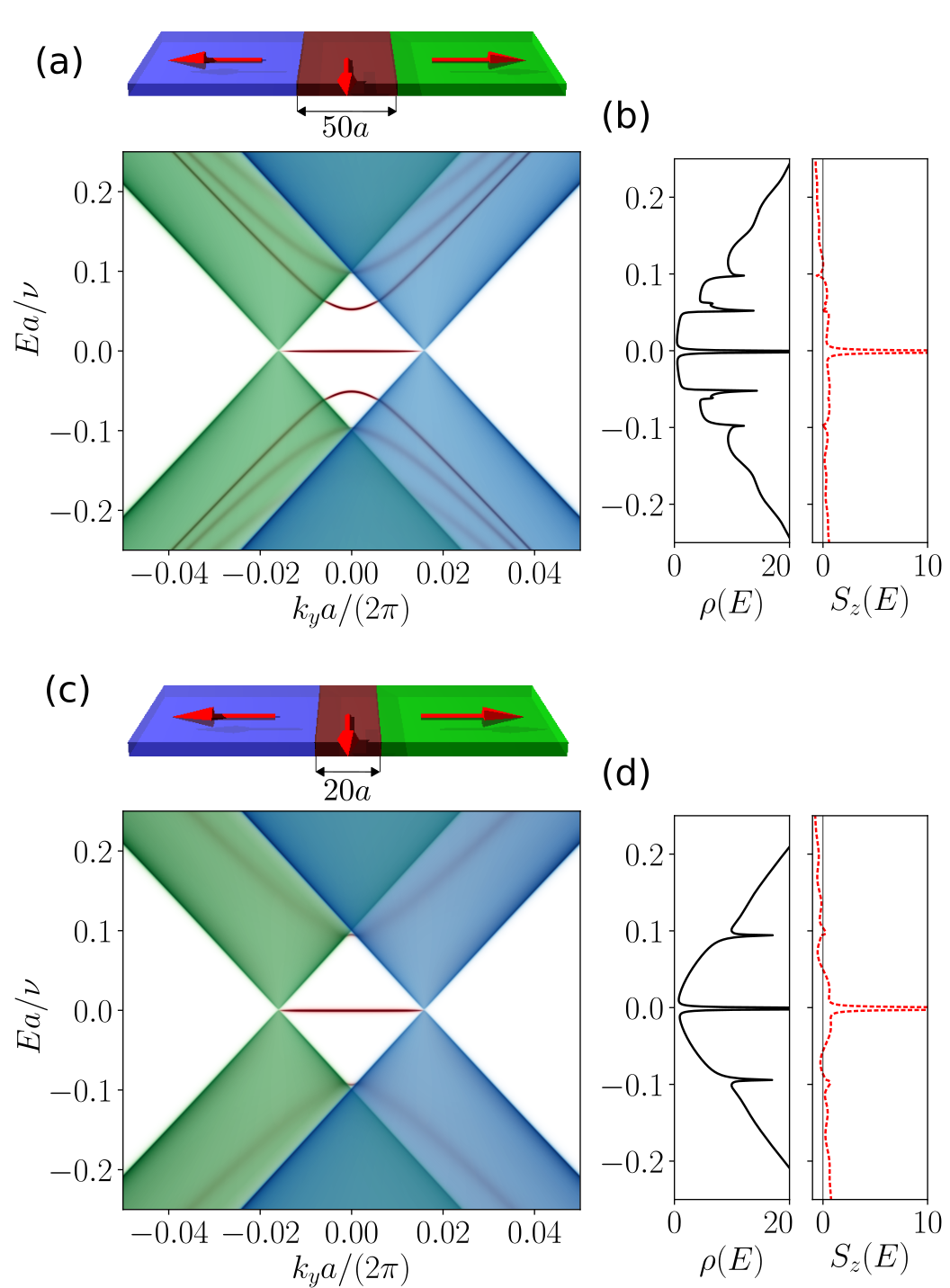}
	\caption{ Bound electron states induced by the
          N\'{e}el-like texture on the magnetic TI surface with the
          easy in-plane anisotropy. The spin-resolved spectral
          function is calculated for the texture with the middle
          region width $2l$, where $2l=50a$ in the panel (a) and
          $2l=20a$ in the panel (b). The top of the panels (a) and (c)
          contains the schematic image of the corresponding magnetic
          texture, where the right and left side regions are in green
          and blue, respectively, the middle region is in brown, the
          arrows indicate the moments directions. The projected
          surface band structure associated with the left and right
          sides of the texture is in blue and green, respectively. The
          dependencies $E(k_y)$ for the DW induced bound states are
          associated with red curves. The total DOS, $\rho(E)$, and
          the spin polarization, $S_z(E)$, are shown as well.}
	\label{fig:fig9}
\end{figure}

\begin{figure}
	\centering
	\includegraphics[width=\columnwidth]{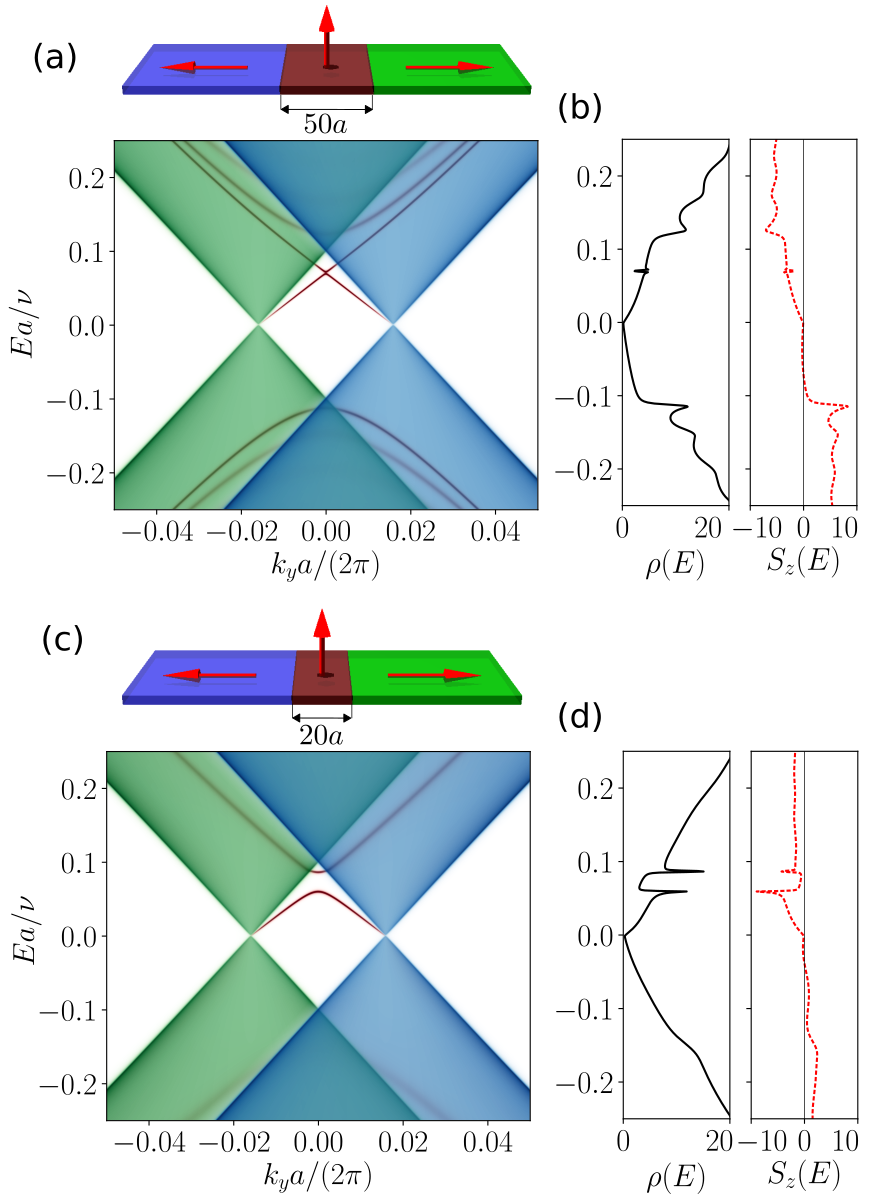}
	\caption{ The same as Fig.~\ref{fig:fig9}, but for the
          case of  polar  N\'{e}el-like texture.}
	\label{fig:fig10}
\end{figure}

Above, we have considered the antiphase DW described by the collinear
magnetization texture $\mathbf{m}(x)= \mathbf{m}_0
\operatorname{sgn}(x)$. Such a sharp profile is believed to be a
reasonable approximation if the DW size is much smaller than the
fermion localization length $\sim k_0^{-1}$, in this case, an internal
structure of the DW is not important.  Now we turn attention to more
complex DW textures displaying noncollinear and noncoplanar
arrangement of the surface magnetization varying over finite scale.
To explore the properties of the bound electron states induced by such
textures we address several typical examples of 1D 180$^{\circ}$
noncollinear and noncoplanar configurations, where the magnetization
orientation varies only along the $x$-axis, $\mathbf{m}(x) =
(\mathbf{m}_x(x), \mathbf{m}_y(x), \mathbf{m}_z(x))$, and the
magnitude is fixed, $|\mathbf{m}(x)|=1$. In order to avoid time
consuming calculations, we employ the approach, where the spatial
profile $\mathbf{m}(x)$ is composed of three differently oriented
regions: $\mathbf{m}(x) = \mathbf{m}_{\text{s}} h(|x|-l)
\operatorname{sgn}(x) + \mathbf{m}_{\text{m}} h(l - |x|)$,
$\mathbf{m}_{\text{s}}$ and $\mathbf{m}_{\text{m}}$ are independent of
the coordinate, the boundaries between the regions are perpendicular
to the $x$-axis.  The magnetizations of the side regions, $\pm
\mathbf{m}_{\text{s}}$, are antiparallel to each other and orthogonal
to that of the middle region, $\mathbf{m}_{\text{m}}$,
i.e. $(\mathbf{m}_{\text{s}} \cdot \mathbf{m}_{\text{m}})=0$. The
approximation simulates a helical rotation or cycloidal one of the
magnetization vector about the DW normal moving from one side to
another for the Bloch-like texture or the N\'{e}el-like one,
respectively. In the framework of our approach, the bound states
created by these textures are described by solution of the eigenvalue
problem $H(x,k_y) \Theta_n(x,k_y) = E_n(k_y) \Theta_n(x,k_y)$ with the
Hamiltonian (\ref{eq:main}) under boundary conditions at $x=l$ and
$x=-l$ presented in Eq.~(\ref{eq:Hamd}).

In the present approach, the coplanar vectors
$\mathbf{m}_{\text{s}}=(1,0,0)$ and $\mathbf{m}_{\text{m}}=(0,1,0)$
set the N\'{e}el-like DW on the TI surface.  The schematic of the
magnetization profiles along the DW can be seen at the top of
Fig~\ref{fig:fig9}. A hallmark of this N\'{e}el-like DW is an
existence of the dispersionless and strongly spin-polarized state with
energy $E_0(k_y)=0$ within the momentum interval $|k_y| < k_0$, at any
width $2l$ of the middle region. When compared with the case of the
sharp DW $\mathbf{m}=(1,0,0)\operatorname{sgn}(x)$ referred to
Eq.~\ref{eq:psi_t2t} at $\theta=\pi/2$, the presence of the middle
region is reflected in the phase shift of the envelope function away
from the DW core: $\Theta_0(x \to \pm \infty, k_y) \sim \exp( \mp |
k_0 + k_y| x \pm i k_0 l)$. The spectral characteristics of the
surface electrons are illustrated by the tight-binding simulations
displayed in Fig.~\ref{fig:fig9}. They confirm clearly the existence
of the flat band state and indicate a large sharp peak in both full
DOS $\rho(E)$ and the spin polarization $S_z(E)$ near zero-energy. It
should be noted that in the case of a broad middle region, $l >
k_0^{-1}$, the DW also supports the spin-degenerate gapped states with
quasi-parabolic bands $E_n(k_y) = E_n(-k_y)$ within the local gap in
the 2D projected bands (Fig.~\ref{fig:fig9}), here $n=\pm 1,\pm
2,... \pm N$. The eigenstates have opposite energies with respect to
the zero energy, $E_n(k_y) = -E_{-n}(k_y)$. With increasing the width,
$2l$, a number of the gapped states, $2N$, grows. The bound states
with $n \ne 0$ manifest in modification of DOS at the edges of the
quasi-parabolic bands. In the opposite case of the relatively sharp
DW, when $l < k_0^{-1}$, the additional states with $n \ne 0$ do not
occur. The specified above trends can be observed in
Fig.~\ref{fig:fig9} by comparing the spectral features for $2l=50a$
and $2l=20a$.

We performed a similar analysis of the bound states for the
noncoplanar surface magnetic texture defined by the vectors
$\mathbf{m}_{\text{s}}=(0,0,1)$ and $\mathbf{m}_{\text{m}}=(0,1,0)$,
which is here referred to as the Bloch-like DW.  We argue that, at any
width of the middle region, the Bloch-like DW hosts the strongly
spin-polarized state with linear momentum-energy relationship $E_0
(k_y) = \nu k_y$ spanning the exchange gap. The asymptotic behavior of
the envelope function of the state away from the DW core, given by the
relation $\Theta_0(x \to \pm \infty) \sim \exp(\mp k_0 (x-il))$ shows
that the existence of the middle region results in a phase shift of
the function. Furthermore, if the middle region is so wide that $l >
k_0^{-1}$, the additional bound gapped states with quasi-parabolic
dispersion, $E_n(k_y) = E_{n}(-k_y)$ ($n=\pm 1, \pm 2,... \pm N$),
appear in the gap. These states are doubly degenerate in spin, their
spectrum is mirrored with respect to zero energy, $E_n(k_y) =
-E_{-n}(k_y)$. With increasing $2l$, a number of the gapped states,
$2N$, increases.  In the case, when $l < k_0^{-1}$, the additional
states with $n \ne 0$ are absent.

Let us consider two the polar N\'{e}el-like textures, where the
surface magnetization is confined in the
$(x,z)$-plane. Fig.~\ref{fig:fig10} displays the spectral picture of
the surface electron states for magnetic configuration given by
$\mathbf{m}_{\text{s}}$=(1,0,0) and $\mathbf{m}_{\text{m}}$=(0,0,1).
As seen in Fig.~\ref{fig:fig10}, in the case of the relatively wide
middle region of $2l=50a$, a pair of quasi-linear bands appears within
the local gap starting from the Dirac points at $k_y=\pm k_0$ and
crossing at $k_y=0$. This band crossing is accompanied by an opening
of a tiny hybridization gap $\sim \exp(-2 k_0 l)$, which divides the
states with opposite spatial parity, $E_0(k_y)$ and $E_{-1}(k_y)$.  As
it can be inferred by comparing the spectra at different values of the
width $2l$ (see, for example, the plots for $2l=50a$ and $2l=20a$ in
Fig.~\ref{fig:fig10}), when the middle region is narrowed the gap
$|E_0(0) - E_{-1}(0)|$ enlarges. At the same time, the low-energy
dispersive state $E_0(k_y)$ sinks deeper into the local gap, and
eventually in the small $2l$ limit, it takes the form of the perfect
flat band close to zero-energy connecting the two Dirac points,
$E_0(k_y) \approx -2\nu k_0^2 l$ at $k_0 l \to 0$.  Note also, that
the state $E_0(k_y)$ is partially spin-polarized at the finite width
$2l$, however, it becomes fully spin-polarized state if the DW middle
region is completely shrinked. In turn, the state $E_{-1}(k_y)$ moves
towards the band continuum edge and leaves the local gap below a
certain value of the width $2l=k_0^{-1}$ ($E_{-1}(k_y) = -\nu k_0$ at
$2k_0 l =1$).

One can also show that the two in-gap bound states are hosted by the
N\'{e}el-like DW whose texture is defined by the vectors
$\mathbf{m}_{\text{s}}=(0,0,1)$ and
$\mathbf{m}_{\text{m}}=(1,0,0)$. Provided that the middle region is
narrow enough, $k_0 l \ll 1$, the one of these states leaves the
exchange gap for the band continuum, while the other remains within
the gap as the chiral fermion with the perfect linear dispersion
$E_0(k_y)=\nu k_y$. It is evident that a single 90$^{\circ}$ DW, which
contains the $(0,\pm1,0)$-oriented domain, does not support an in-gap
bound state. However, noncollinear and noncoplanar 180$^{\circ}$
textures, in which the middle region magnetization
$\mathbf{m}_{\text{m}}$ is aligned along $y$-axis, can generate from
one to several in-gap states. In any case, the 180$^{\circ}$ N\'{e}el-
or Bloch-type DWs investigated certainly provides the existence of the
topologically protected chiral state $E_0(k_y)$ inside the exchange gap.

\section{DISCUSSION AND SUMMARY}

We have shown that the real-space magnetic textures, such as surface
DWs, are capable of affecting substantially the momentum-space
behavior of low-energy quasiparticles in TIs. Thus, any particular
magnetization texture $\mathbf{M}(x,y)$ should provide its own
spectral portrait for the surface electron states in the surface
exchange gap. The averaged spectral function $A(E,\mathbf{k}) = -
(1/\pi) \Im \Tr \langle G^{R}(E,\mathbf{k} | \mathbf{M}(x,y) )
\rangle$, in which the contributions of all the textures are collected
accordingly to their statistic weight, is thought to be observed in
the spectroscopy measurements. Such an assumption may qualitatively
explain controversial ARPES results for the surface spectra of
intrinsic AFM TI MnBi$_2$Te$_4$ with out-of-plane easy axis obtained
by various experimental groups~\cite{OtrokovN2019, Shikin2020,
  Stability2, Stability1, Stability3}. In the seminal
work~\cite{OtrokovN2019} a clear exchange gap of about 70~meV at the
$\overline{\Gamma}$-point separating the upper and lower parts of the
surface cone was revealed, which is close to the theoretically
predicted value 88~meV. However, Shikin et al.~\cite{Shikin2020} have
found in the ARPES spectra both large (60–70~meV) and reduced
(<20~meV) surface gaps at the Dirac point. Some other ARPES
measurements~\cite{Stability2, Stability1, Stability3} have detected
the gapless Dirac states at the surface of MnBi$_2$Te$_4$ below the
N\'{e}el temperature instead of the expected exchange gapping of the
states. Within the developed theory, we speculate that in the
experimental findings cited here above the ARPES surface spectra may
contain a superposition of the DW-induced spectral features like the
ones presented in Figs.~\ref{fig:fig2}, \ref{fig:fig4}, \ref{fig:fig6}
averaged spatially over multiple DWs caught out in the light spot of a
micron scale. If the concentration of the DWs is relatively low, they
can hardly ensure a visible spectral signal inside the large exchange
gap, which is associated with almost homogeneous out-of-plane
magnetization of the broad domains. With the increasing concentration
of DWs, one can note the appearance of the Dirac-like gapless
dispersion crossing the exchange energy gap. Under the higher
concentration, the Dirac-like spectrum becomes gapped (i.e., with the
<<reduced>> gap in the terms of Ref.~\cite{Shikin2020} at the
$\overline{\Gamma}$-point owing to the hybridization of the states
bound to the neighboring DWs (see Fig.~\ref{fig:fig4} and the
corresponding comments). From that standpoint, the disagreement of the
ARPES findings for the surface spectra of intrinsic AFM TI
MnBi$_2$Te$_4$ can be attributed to the growth conditions and magnetic
treatment of tested samples.

The disagreement between some spectroscopic probes indicating a
relatively large exchange gap at the TI surface and the low
temperature range of the QAHE realization remains an important
matter~\cite{KimDenliner2020}. The quantized conductivity effects in
real systems are affected by a complex network of topologically
non-trivial and trivial conducting channels running along boundaries,
separating regions of distinct topological phases~\cite{MSC_2019}.
Our results demonstrate the presence of the DW-induced states
embodying the conducting channels at the magnetic surface of TI and
describe their characteristics. In magneto-transport measurements,
when the chemical potential is fixed within the energy gap, the
conductivity is realized by means of both a carrier propagation along
the edge channels of the sample and a percolation through 1D channels
caused by the network of the magnetic DWs~\cite{Kawamura_2018,
  WuXiao2020}. Therefore the measured conductivity can display
remarkable deviation from the expected quantized value.

We have revealed a unique possibility to tune the group velocity and
spin-polarization orientation of the DW-induced states by reorienting
the easy axis direction, which could be useful to design new devices
with controlled propagation velocity and chirality for the robust
excitations responsible for quantized dissipationless conductivity. We
have predicted that, when the moments lie within the easy-plane
parallel to the TI surface plane, both the head-to-head and
tail-to-tail planar textures generate the chiral zero-energy flat band
state in the single-particle spectrum. Due to the flatness of the
energy dispersion and respectively the large peak of DOS this state
could attract the particular attention from the perspective of
enhanced electron correlations. Thereby, the surface of the planar
magnetic TI carrying the DWs could provide unique platform for
realizing phase transitions driven by instabilities of the unusual
quasiparticles~--- the heavy chiral fermions with nontrivial
topology. In particular, in such a system, it would be interesting to
seek for the unconventional superconductivity with spin triplet
pairing~\cite{Mackanzie_2003} or triplet excitonic
insulator~\cite{Sethi_2021}.

To summarize: We have provided the physical picture to understand the
nature and properties of the DW-induced states at the TI magnetic
surface. Implying a rich domain texture of the surface we have shown
the existence of various topologically protected states stemming from
the DWs. The relevance of our results is based on the fact that the
properties of the DW-induced states are explained in terms of general
approach, and their description is material independent. It can be
applied to the surfaces of intrinsic magnetic TIs and the interfaces
in hetrostructures with induced magnetization under any direction of
anisotropy axis. Further investigations of the interplay between
domain texture and the surface fermion states in the magnetic TIs,
especially via different experimental techniques, are strongly desired
to open the way to realize novel quantum phenomena.

\section{ACKNOWLEDGMENTS}

This work was supported by the Saint Petersburg State University
(project ID No. 73028629) and Russian Science Foundation (Grant
No. 18-12-00169).

\bibliography{bibfile}

\providecommand{\noopsort}[1]{}\providecommand{\singleletter}[1]{#1}%
\begin{thebibliography}{72}%
\makeatletter
\providecommand \@ifxundefined [1]{%
 \@ifx{#1\undefined}
}%
\providecommand \@ifnum [1]{%
 \ifnum #1\expandafter \@firstoftwo
 \else \expandafter \@secondoftwo
 \fi
}%
\providecommand \@ifx [1]{%
 \ifx #1\expandafter \@firstoftwo
 \else \expandafter \@secondoftwo
 \fi
}%
\providecommand \natexlab [1]{#1}%
\providecommand \enquote  [1]{``#1''}%
\providecommand \bibnamefont  [1]{#1}%
\providecommand \bibfnamefont [1]{#1}%
\providecommand \citenamefont [1]{#1}%
\providecommand \href@noop [0]{\@secondoftwo}%
\providecommand \href [0]{\begingroup \@sanitize@url \@href}%
\providecommand \@href[1]{\@@startlink{#1}\@@href}%
\providecommand \@@href[1]{\endgroup#1\@@endlink}%
\providecommand \@sanitize@url [0]{\catcode `\\12\catcode `\$12\catcode
  `\&12\catcode `\#12\catcode `\^12\catcode `\_12\catcode `\%12\relax}%
\providecommand \@@startlink[1]{}%
\providecommand \@@endlink[0]{}%
\providecommand \url  [0]{\begingroup\@sanitize@url \@url }%
\providecommand \@url [1]{\endgroup\@href {#1}{\urlprefix }}%
\providecommand \urlprefix  [0]{URL }%
\providecommand \Eprint [0]{\href }%
\providecommand \doibase [0]{https://doi.org/}%
\providecommand \selectlanguage [0]{\@gobble}%
\providecommand \bibinfo  [0]{\@secondoftwo}%
\providecommand \bibfield  [0]{\@secondoftwo}%
\providecommand \translation [1]{[#1]}%
\providecommand \BibitemOpen [0]{}%
\providecommand \bibitemStop [0]{}%
\providecommand \bibitemNoStop [0]{.\EOS\space}%
\providecommand \EOS [0]{\spacefactor3000\relax}%
\providecommand \BibitemShut  [1]{\csname bibitem#1\endcsname}%
\let\auto@bib@innerbib\@empty
\bibitem [{\citenamefont {Hasan}\ and\ \citenamefont
  {Kane}(2010)}]{hasan2010colloquium}%
  \BibitemOpen
  \bibfield  {author} {\bibinfo {author} {\bibfnamefont {M.~Z.}\ \bibnamefont
  {Hasan}}\ and\ \bibinfo {author} {\bibfnamefont {C.~L.}\ \bibnamefont
  {Kane}},\ }\href@noop {} {\bibfield  {journal} {\bibinfo  {journal} {Reviews
  of Modern Physics}\ }\textbf {\bibinfo {volume} {82}},\ \bibinfo {pages}
  {3045} (\bibinfo {year} {2010})}\BibitemShut {NoStop}%
\bibitem [{\citenamefont {Qi}\ and\ \citenamefont
  {Zhang}(2011)}]{Qi2011colloquium}%
  \BibitemOpen
  \bibfield  {author} {\bibinfo {author} {\bibfnamefont {X.-L.}\ \bibnamefont
  {Qi}}\ and\ \bibinfo {author} {\bibfnamefont {S.-C.}\ \bibnamefont {Zhang}},\
  }\href {https://doi.org/10.1103/RevModPhys.83.1057} {\bibfield  {journal}
  {\bibinfo  {journal} {Rev. Mod. Phys.}\ }\textbf {\bibinfo {volume} {83}},\
  \bibinfo {pages} {1057} (\bibinfo {year} {2011})}\BibitemShut {NoStop}%
\bibitem [{\citenamefont {Ando}(2013)}]{Ando2013colloquium}%
  \BibitemOpen
  \bibfield  {author} {\bibinfo {author} {\bibfnamefont {Y.}~\bibnamefont
  {Ando}},\ }\href {https://doi.org/10.7566/JPSJ.82.102001} {\bibfield
  {journal} {\bibinfo  {journal} {Journal of the Physical Society of Japan}\
  }\textbf {\bibinfo {volume} {82}},\ \bibinfo {pages} {102001} (\bibinfo
  {year} {2013})}\BibitemShut {NoStop}%
\bibitem [{\citenamefont {Weng}\ \emph
  {et~al.}(2015{\natexlab{a}})\citenamefont {Weng}, \citenamefont {Yu},
  \citenamefont {Hu}, \citenamefont {Dai},\ and\ \citenamefont
  {Fang}}]{Weng2015_adv}%
  \BibitemOpen
  \bibfield  {author} {\bibinfo {author} {\bibfnamefont {H.}~\bibnamefont
  {Weng}}, \bibinfo {author} {\bibfnamefont {R.}~\bibnamefont {Yu}}, \bibinfo
  {author} {\bibfnamefont {X.}~\bibnamefont {Hu}}, \bibinfo {author}
  {\bibfnamefont {X.}~\bibnamefont {Dai}},\ and\ \bibinfo {author}
  {\bibfnamefont {Z.}~\bibnamefont {Fang}},\ }\href
  {https://doi.org/10.1080/00018732.2015.1068524} {\bibfield  {journal}
  {\bibinfo  {journal} {Advances in Physics}\ }\textbf {\bibinfo {volume}
  {64}},\ \bibinfo {pages} {227} (\bibinfo {year}
  {2015}{\natexlab{a}})}\BibitemShut {NoStop}%
\bibitem [{\citenamefont {Kou}\ \emph {et~al.}(2015)\citenamefont {Kou},
  \citenamefont {Fan}, \citenamefont {Lang}, \citenamefont {Upadhyaya},\ and\
  \citenamefont {Wang}}]{Kou_2015}%
  \BibitemOpen
  \bibfield  {author} {\bibinfo {author} {\bibfnamefont {X.}~\bibnamefont
  {Kou}}, \bibinfo {author} {\bibfnamefont {Y.}~\bibnamefont {Fan}}, \bibinfo
  {author} {\bibfnamefont {M.}~\bibnamefont {Lang}}, \bibinfo {author}
  {\bibfnamefont {P.}~\bibnamefont {Upadhyaya}},\ and\ \bibinfo {author}
  {\bibfnamefont {K.~L.}\ \bibnamefont {Wang}},\ }\href
  {https://doi.org/https://doi.org/10.1016/j.ssc.2014.10.022} {\bibfield
  {journal} {\bibinfo  {journal} {Solid State Communications}\ }\textbf
  {\bibinfo {volume} {215-216}},\ \bibinfo {pages} {34} (\bibinfo {year}
  {2015})}\BibitemShut {NoStop}%
\bibitem [{\citenamefont {Chang}\ and\ \citenamefont {Li}(2016)}]{Chang_2016}%
  \BibitemOpen
  \bibfield  {author} {\bibinfo {author} {\bibfnamefont {C.-Z.}\ \bibnamefont
  {Chang}}\ and\ \bibinfo {author} {\bibfnamefont {M.}~\bibnamefont {Li}},\
  }\href {https://doi.org/10.1088/0953-8984/28/12/123002} {\bibfield  {journal}
  {\bibinfo  {journal} {Journal of Physics: Condensed Matter}\ }\textbf
  {\bibinfo {volume} {28}},\ \bibinfo {pages} {123002} (\bibinfo {year}
  {2016})}\BibitemShut {NoStop}%
\bibitem [{\citenamefont {Liu}\ \emph {et~al.}(2016{\natexlab{a}})\citenamefont
  {Liu}, \citenamefont {Zhang},\ and\ \citenamefont {Qi}}]{LiuZhangQi_2016}%
  \BibitemOpen
  \bibfield  {author} {\bibinfo {author} {\bibfnamefont {C.-X.}\ \bibnamefont
  {Liu}}, \bibinfo {author} {\bibfnamefont {S.-C.}\ \bibnamefont {Zhang}},\
  and\ \bibinfo {author} {\bibfnamefont {X.-L.}\ \bibnamefont {Qi}},\ }\href
  {https://doi.org/10.1146/annurev-conmatphys-031115-011417} {\bibfield
  {journal} {\bibinfo  {journal} {Annual Review of Condensed Matter Physics}\
  }\textbf {\bibinfo {volume} {7}},\ \bibinfo {pages} {301} (\bibinfo {year}
  {2016}{\natexlab{a}})}\BibitemShut {NoStop}%
\bibitem [{\citenamefont {Tokura}\ \emph {et~al.}(2019)\citenamefont {Tokura},
  \citenamefont {Yasuda},\ and\ \citenamefont {Tsukazaki}}]{Tokura2019}%
  \BibitemOpen
  \bibfield  {author} {\bibinfo {author} {\bibfnamefont {Y.}~\bibnamefont
  {Tokura}}, \bibinfo {author} {\bibfnamefont {K.}~\bibnamefont {Yasuda}},\
  and\ \bibinfo {author} {\bibfnamefont {A.}~\bibnamefont {Tsukazaki}},\ }\href
  {https://doi.org/10.1038/s42254-018-0011-5} {\bibfield  {journal} {\bibinfo
  {journal} {Nature Reviews Physics}\ }\textbf {\bibinfo {volume} {1}},\
  \bibinfo {pages} {126} (\bibinfo {year} {2019})}\BibitemShut {NoStop}%
\bibitem [{\citenamefont {He}\ \emph {et~al.}(2017)\citenamefont {He},
  \citenamefont {Pan}, \citenamefont {Stern}, \citenamefont {Burks},
  \citenamefont {Che}, \citenamefont {Yin}, \citenamefont {Wang}, \citenamefont
  {Lian}, \citenamefont {Zhou}, \citenamefont {Choi}, \citenamefont {Murata},
  \citenamefont {Kou}, \citenamefont {Chen}, \citenamefont {Nie}, \citenamefont
  {Shao}, \citenamefont {Fan}, \citenamefont {Zhang}, \citenamefont {Liu},
  \citenamefont {Xia},\ and\ \citenamefont {Wang}}]{He294}%
  \BibitemOpen
  \bibfield  {author} {\bibinfo {author} {\bibfnamefont {Q.~L.}\ \bibnamefont
  {He}}, \bibinfo {author} {\bibfnamefont {L.}~\bibnamefont {Pan}}, \bibinfo
  {author} {\bibfnamefont {A.~L.}\ \bibnamefont {Stern}}, \bibinfo {author}
  {\bibfnamefont {E.~C.}\ \bibnamefont {Burks}}, \bibinfo {author}
  {\bibfnamefont {X.}~\bibnamefont {Che}}, \bibinfo {author} {\bibfnamefont
  {G.}~\bibnamefont {Yin}}, \bibinfo {author} {\bibfnamefont {J.}~\bibnamefont
  {Wang}}, \bibinfo {author} {\bibfnamefont {B.}~\bibnamefont {Lian}}, \bibinfo
  {author} {\bibfnamefont {Q.}~\bibnamefont {Zhou}}, \bibinfo {author}
  {\bibfnamefont {E.~S.}\ \bibnamefont {Choi}}, \bibinfo {author}
  {\bibfnamefont {K.}~\bibnamefont {Murata}}, \bibinfo {author} {\bibfnamefont
  {X.}~\bibnamefont {Kou}}, \bibinfo {author} {\bibfnamefont {Z.}~\bibnamefont
  {Chen}}, \bibinfo {author} {\bibfnamefont {T.}~\bibnamefont {Nie}}, \bibinfo
  {author} {\bibfnamefont {Q.}~\bibnamefont {Shao}}, \bibinfo {author}
  {\bibfnamefont {Y.}~\bibnamefont {Fan}}, \bibinfo {author} {\bibfnamefont
  {S.-C.}\ \bibnamefont {Zhang}}, \bibinfo {author} {\bibfnamefont
  {K.}~\bibnamefont {Liu}}, \bibinfo {author} {\bibfnamefont {J.}~\bibnamefont
  {Xia}},\ and\ \bibinfo {author} {\bibfnamefont {K.~L.}\ \bibnamefont
  {Wang}},\ }\href {https://doi.org/10.1126/science.aag2792} {\bibfield
  {journal} {\bibinfo  {journal} {Science}\ }\textbf {\bibinfo {volume}
  {357}},\ \bibinfo {pages} {294} (\bibinfo {year} {2017})}\BibitemShut
  {NoStop}%
\bibitem [{\citenamefont {Men'shov}\ \emph
  {et~al.}(2019{\natexlab{a}})\citenamefont {Men'shov}, \citenamefont
  {Shvets},\ and\ \citenamefont {Chulkov}}]{MS_JL_2019}%
  \BibitemOpen
  \bibfield  {author} {\bibinfo {author} {\bibfnamefont {V.~N.}\ \bibnamefont
  {Men'shov}}, \bibinfo {author} {\bibfnamefont {I.~A.}\ \bibnamefont
  {Shvets}},\ and\ \bibinfo {author} {\bibfnamefont {E.~V.}\ \bibnamefont
  {Chulkov}},\ }\href {https://doi.org/10.1134/S002136401924007X} {\bibfield
  {journal} {\bibinfo  {journal} {JETP Letters}\ }\textbf {\bibinfo {volume}
  {110}},\ \bibinfo {pages} {771} (\bibinfo {year}
  {2019}{\natexlab{a}})}\BibitemShut {NoStop}%
\bibitem [{\citenamefont {Chang}\ \emph {et~al.}(2013)\citenamefont {Chang},
  \citenamefont {Zhang}, \citenamefont {Feng}, \citenamefont {Shen},
  \citenamefont {Zhang}, \citenamefont {Guo}, \citenamefont {Li}, \citenamefont
  {Ou}, \citenamefont {Wei}, \citenamefont {Wang}, \citenamefont {Ji},
  \citenamefont {Feng}, \citenamefont {Ji}, \citenamefont {Chen}, \citenamefont
  {Jia}, \citenamefont {Dai}, \citenamefont {Fang}, \citenamefont {Zhang},
  \citenamefont {He}, \citenamefont {Wang}, \citenamefont {Lu}, \citenamefont
  {Ma},\ and\ \citenamefont {Xue}}]{Chang167}%
  \BibitemOpen
  \bibfield  {author} {\bibinfo {author} {\bibfnamefont {C.-Z.}\ \bibnamefont
  {Chang}}, \bibinfo {author} {\bibfnamefont {J.}~\bibnamefont {Zhang}},
  \bibinfo {author} {\bibfnamefont {X.}~\bibnamefont {Feng}}, \bibinfo {author}
  {\bibfnamefont {J.}~\bibnamefont {Shen}}, \bibinfo {author} {\bibfnamefont
  {Z.}~\bibnamefont {Zhang}}, \bibinfo {author} {\bibfnamefont
  {M.}~\bibnamefont {Guo}}, \bibinfo {author} {\bibfnamefont {K.}~\bibnamefont
  {Li}}, \bibinfo {author} {\bibfnamefont {Y.}~\bibnamefont {Ou}}, \bibinfo
  {author} {\bibfnamefont {P.}~\bibnamefont {Wei}}, \bibinfo {author}
  {\bibfnamefont {L.-L.}\ \bibnamefont {Wang}}, \bibinfo {author}
  {\bibfnamefont {Z.-Q.}\ \bibnamefont {Ji}}, \bibinfo {author} {\bibfnamefont
  {Y.}~\bibnamefont {Feng}}, \bibinfo {author} {\bibfnamefont {S.}~\bibnamefont
  {Ji}}, \bibinfo {author} {\bibfnamefont {X.}~\bibnamefont {Chen}}, \bibinfo
  {author} {\bibfnamefont {J.}~\bibnamefont {Jia}}, \bibinfo {author}
  {\bibfnamefont {X.}~\bibnamefont {Dai}}, \bibinfo {author} {\bibfnamefont
  {Z.}~\bibnamefont {Fang}}, \bibinfo {author} {\bibfnamefont {S.-C.}\
  \bibnamefont {Zhang}}, \bibinfo {author} {\bibfnamefont {K.}~\bibnamefont
  {He}}, \bibinfo {author} {\bibfnamefont {Y.}~\bibnamefont {Wang}}, \bibinfo
  {author} {\bibfnamefont {L.}~\bibnamefont {Lu}}, \bibinfo {author}
  {\bibfnamefont {X.-C.}\ \bibnamefont {Ma}},\ and\ \bibinfo {author}
  {\bibfnamefont {Q.-K.}\ \bibnamefont {Xue}},\ }\href
  {https://doi.org/10.1126/science.1234414} {\bibfield  {journal} {\bibinfo
  {journal} {Science}\ }\textbf {\bibinfo {volume} {340}},\ \bibinfo {pages}
  {167} (\bibinfo {year} {2013})}\BibitemShut {NoStop}%
\bibitem [{\citenamefont {Bestwick}\ \emph {et~al.}(2015)\citenamefont
  {Bestwick}, \citenamefont {Fox}, \citenamefont {Kou}, \citenamefont {Pan},
  \citenamefont {Wang},\ and\ \citenamefont
  {Goldhaber-Gordon}}]{Bestwick_2015}%
  \BibitemOpen
  \bibfield  {author} {\bibinfo {author} {\bibfnamefont {A.~J.}\ \bibnamefont
  {Bestwick}}, \bibinfo {author} {\bibfnamefont {E.~J.}\ \bibnamefont {Fox}},
  \bibinfo {author} {\bibfnamefont {X.}~\bibnamefont {Kou}}, \bibinfo {author}
  {\bibfnamefont {L.}~\bibnamefont {Pan}}, \bibinfo {author} {\bibfnamefont
  {K.~L.}\ \bibnamefont {Wang}},\ and\ \bibinfo {author} {\bibfnamefont
  {D.}~\bibnamefont {Goldhaber-Gordon}},\ }\href
  {https://doi.org/10.1103/PhysRevLett.114.187201} {\bibfield  {journal}
  {\bibinfo  {journal} {Phys. Rev. Lett.}\ }\textbf {\bibinfo {volume} {114}},\
  \bibinfo {pages} {187201} (\bibinfo {year} {2015})}\BibitemShut {NoStop}%
\bibitem [{\citenamefont {Mogi}\ \emph {et~al.}(2015)\citenamefont {Mogi},
  \citenamefont {Yoshimi}, \citenamefont {Tsukazaki}, \citenamefont {Yasuda},
  \citenamefont {Kozuka}, \citenamefont {Takahashi}, \citenamefont {Kawasaki},\
  and\ \citenamefont {Tokura}}]{Mogi_2015}%
  \BibitemOpen
  \bibfield  {author} {\bibinfo {author} {\bibfnamefont {M.}~\bibnamefont
  {Mogi}}, \bibinfo {author} {\bibfnamefont {R.}~\bibnamefont {Yoshimi}},
  \bibinfo {author} {\bibfnamefont {A.}~\bibnamefont {Tsukazaki}}, \bibinfo
  {author} {\bibfnamefont {K.}~\bibnamefont {Yasuda}}, \bibinfo {author}
  {\bibfnamefont {Y.}~\bibnamefont {Kozuka}}, \bibinfo {author} {\bibfnamefont
  {K.~S.}\ \bibnamefont {Takahashi}}, \bibinfo {author} {\bibfnamefont
  {M.}~\bibnamefont {Kawasaki}},\ and\ \bibinfo {author} {\bibfnamefont
  {Y.}~\bibnamefont {Tokura}},\ }\href {https://doi.org/10.1063/1.4935075}
  {\bibfield  {journal} {\bibinfo  {journal} {Applied Physics Letters}\
  }\textbf {\bibinfo {volume} {107}},\ \bibinfo {pages} {182401} (\bibinfo
  {year} {2015})}\BibitemShut {NoStop}%
\bibitem [{\citenamefont {Okada}\ \emph {et~al.}(2016)\citenamefont {Okada},
  \citenamefont {Takahashi}, \citenamefont {Mogi}, \citenamefont {Yoshimi},
  \citenamefont {Tsukazaki}, \citenamefont {Takahashi}, \citenamefont {Ogawa},
  \citenamefont {Kawasaki},\ and\ \citenamefont {Tokura}}]{Okada2016}%
  \BibitemOpen
  \bibfield  {author} {\bibinfo {author} {\bibfnamefont {K.~N.}\ \bibnamefont
  {Okada}}, \bibinfo {author} {\bibfnamefont {Y.}~\bibnamefont {Takahashi}},
  \bibinfo {author} {\bibfnamefont {M.}~\bibnamefont {Mogi}}, \bibinfo {author}
  {\bibfnamefont {R.}~\bibnamefont {Yoshimi}}, \bibinfo {author} {\bibfnamefont
  {A.}~\bibnamefont {Tsukazaki}}, \bibinfo {author} {\bibfnamefont {K.~S.}\
  \bibnamefont {Takahashi}}, \bibinfo {author} {\bibfnamefont {N.}~\bibnamefont
  {Ogawa}}, \bibinfo {author} {\bibfnamefont {M.}~\bibnamefont {Kawasaki}},\
  and\ \bibinfo {author} {\bibfnamefont {Y.}~\bibnamefont {Tokura}},\ }\href
  {https://doi.org/10.1038/ncomms12245} {\bibfield  {journal} {\bibinfo
  {journal} {Nature Communications}\ }\textbf {\bibinfo {volume} {7}},\
  \bibinfo {pages} {12245} (\bibinfo {year} {2016})}\BibitemShut {NoStop}%
\bibitem [{\citenamefont {Mogi}\ \emph
  {et~al.}(2017{\natexlab{a}})\citenamefont {Mogi}, \citenamefont {Kawamura},
  \citenamefont {Tsukazaki}, \citenamefont {Yoshimi}, \citenamefont
  {Takahashi}, \citenamefont {Kawasaki},\ and\ \citenamefont
  {Tokura}}]{Mogi1669}%
  \BibitemOpen
  \bibfield  {author} {\bibinfo {author} {\bibfnamefont {M.}~\bibnamefont
  {Mogi}}, \bibinfo {author} {\bibfnamefont {M.}~\bibnamefont {Kawamura}},
  \bibinfo {author} {\bibfnamefont {A.}~\bibnamefont {Tsukazaki}}, \bibinfo
  {author} {\bibfnamefont {R.}~\bibnamefont {Yoshimi}}, \bibinfo {author}
  {\bibfnamefont {K.~S.}\ \bibnamefont {Takahashi}}, \bibinfo {author}
  {\bibfnamefont {M.}~\bibnamefont {Kawasaki}},\ and\ \bibinfo {author}
  {\bibfnamefont {Y.}~\bibnamefont {Tokura}},\ }\bibfield  {journal} {\bibinfo
  {journal} {Science Advances}\ }\textbf {\bibinfo {volume} {3}},\ \href
  {https://doi.org/10.1126/sciadv.aao1669} {10.1126/sciadv.aao1669} (\bibinfo
  {year} {2017}{\natexlab{a}})\BibitemShut {NoStop}%
\bibitem [{\citenamefont {Mogi}\ \emph
  {et~al.}(2017{\natexlab{b}})\citenamefont {Mogi}, \citenamefont {Kawamura},
  \citenamefont {Yoshimi}, \citenamefont {Tsukazaki}, \citenamefont {Kozuka},
  \citenamefont {Shirakawa}, \citenamefont {Takahashi}, \citenamefont
  {Kawasaki},\ and\ \citenamefont {Tokura}}]{Mogi2017}%
  \BibitemOpen
  \bibfield  {author} {\bibinfo {author} {\bibfnamefont {M.}~\bibnamefont
  {Mogi}}, \bibinfo {author} {\bibfnamefont {M.}~\bibnamefont {Kawamura}},
  \bibinfo {author} {\bibfnamefont {R.}~\bibnamefont {Yoshimi}}, \bibinfo
  {author} {\bibfnamefont {A.}~\bibnamefont {Tsukazaki}}, \bibinfo {author}
  {\bibfnamefont {Y.}~\bibnamefont {Kozuka}}, \bibinfo {author} {\bibfnamefont
  {N.}~\bibnamefont {Shirakawa}}, \bibinfo {author} {\bibfnamefont {K.~S.}\
  \bibnamefont {Takahashi}}, \bibinfo {author} {\bibfnamefont {M.}~\bibnamefont
  {Kawasaki}},\ and\ \bibinfo {author} {\bibfnamefont {Y.}~\bibnamefont
  {Tokura}},\ }\href {https://doi.org/10.1038/nmat4855} {\bibfield  {journal}
  {\bibinfo  {journal} {Nature Materials}\ }\textbf {\bibinfo {volume} {16}},\
  \bibinfo {pages} {516} (\bibinfo {year} {2017}{\natexlab{b}})}\BibitemShut
  {NoStop}%
\bibitem [{\citenamefont {Xiao}\ \emph {et~al.}(2018)\citenamefont {Xiao},
  \citenamefont {Jiang}, \citenamefont {Shin}, \citenamefont {Wang},
  \citenamefont {Wang}, \citenamefont {Zhao}, \citenamefont {Liu},
  \citenamefont {Wu}, \citenamefont {Chan}, \citenamefont {Samarth},\ and\
  \citenamefont {Chang}}]{Xiao_2018}%
  \BibitemOpen
  \bibfield  {author} {\bibinfo {author} {\bibfnamefont {D.}~\bibnamefont
  {Xiao}}, \bibinfo {author} {\bibfnamefont {J.}~\bibnamefont {Jiang}},
  \bibinfo {author} {\bibfnamefont {J.-H.}\ \bibnamefont {Shin}}, \bibinfo
  {author} {\bibfnamefont {W.}~\bibnamefont {Wang}}, \bibinfo {author}
  {\bibfnamefont {F.}~\bibnamefont {Wang}}, \bibinfo {author} {\bibfnamefont
  {Y.-F.}\ \bibnamefont {Zhao}}, \bibinfo {author} {\bibfnamefont
  {C.}~\bibnamefont {Liu}}, \bibinfo {author} {\bibfnamefont {W.}~\bibnamefont
  {Wu}}, \bibinfo {author} {\bibfnamefont {M.~H.~W.}\ \bibnamefont {Chan}},
  \bibinfo {author} {\bibfnamefont {N.}~\bibnamefont {Samarth}},\ and\ \bibinfo
  {author} {\bibfnamefont {C.-Z.}\ \bibnamefont {Chang}},\ }\href
  {https://doi.org/10.1103/PhysRevLett.120.056801} {\bibfield  {journal}
  {\bibinfo  {journal} {Phys. Rev. Lett.}\ }\textbf {\bibinfo {volume} {120}},\
  \bibinfo {pages} {056801} (\bibinfo {year} {2018})}\BibitemShut {NoStop}%
\bibitem [{\citenamefont {Watanabe}\ \emph {et~al.}(2019)\citenamefont
  {Watanabe}, \citenamefont {Yoshimi}, \citenamefont {Kawamura}, \citenamefont
  {Mogi}, \citenamefont {Tsukazaki}, \citenamefont {Yu}, \citenamefont
  {Nakajima}, \citenamefont {Takahashi}, \citenamefont {Kawasaki},\ and\
  \citenamefont {Tokura}}]{Watanabe2019}%
  \BibitemOpen
  \bibfield  {author} {\bibinfo {author} {\bibfnamefont {R.}~\bibnamefont
  {Watanabe}}, \bibinfo {author} {\bibfnamefont {R.}~\bibnamefont {Yoshimi}},
  \bibinfo {author} {\bibfnamefont {M.}~\bibnamefont {Kawamura}}, \bibinfo
  {author} {\bibfnamefont {M.}~\bibnamefont {Mogi}}, \bibinfo {author}
  {\bibfnamefont {A.}~\bibnamefont {Tsukazaki}}, \bibinfo {author}
  {\bibfnamefont {X.~Z.}\ \bibnamefont {Yu}}, \bibinfo {author} {\bibfnamefont
  {K.}~\bibnamefont {Nakajima}}, \bibinfo {author} {\bibfnamefont {K.~S.}\
  \bibnamefont {Takahashi}}, \bibinfo {author} {\bibfnamefont {M.}~\bibnamefont
  {Kawasaki}},\ and\ \bibinfo {author} {\bibfnamefont {Y.}~\bibnamefont
  {Tokura}},\ }\href {https://doi.org/10.1063/1.5111891} {\bibfield  {journal}
  {\bibinfo  {journal} {Applied Physics Letters}\ }\textbf {\bibinfo {volume}
  {115}},\ \bibinfo {pages} {102403} (\bibinfo {year} {2019})}\BibitemShut
  {NoStop}%
\bibitem [{\citenamefont {Otrokov}\ \emph
  {et~al.}(2017{\natexlab{a}})\citenamefont {Otrokov}, \citenamefont
  {Menshchikova}, \citenamefont {Vergniory}, \citenamefont {Rusinov},
  \citenamefont {Vyazovskaya}, \citenamefont {Koroteev}, \citenamefont
  {Bihlmayer}, \citenamefont {Ernst}, \citenamefont {Echenique}, \citenamefont
  {Arnau},\ and\ \citenamefont {Chulkov}}]{Otrokov_HO2017}%
  \BibitemOpen
  \bibfield  {author} {\bibinfo {author} {\bibfnamefont {M.~M.}\ \bibnamefont
  {Otrokov}}, \bibinfo {author} {\bibfnamefont {T.~V.}\ \bibnamefont
  {Menshchikova}}, \bibinfo {author} {\bibfnamefont {M.~G.}\ \bibnamefont
  {Vergniory}}, \bibinfo {author} {\bibfnamefont {I.~P.}\ \bibnamefont
  {Rusinov}}, \bibinfo {author} {\bibfnamefont {A.~Y.}\ \bibnamefont
  {Vyazovskaya}}, \bibinfo {author} {\bibfnamefont {Y.~M.}\ \bibnamefont
  {Koroteev}}, \bibinfo {author} {\bibfnamefont {G.}~\bibnamefont {Bihlmayer}},
  \bibinfo {author} {\bibfnamefont {A.}~\bibnamefont {Ernst}}, \bibinfo
  {author} {\bibfnamefont {P.~M.}\ \bibnamefont {Echenique}}, \bibinfo {author}
  {\bibfnamefont {A.}~\bibnamefont {Arnau}},\ and\ \bibinfo {author}
  {\bibfnamefont {E.~V.}\ \bibnamefont {Chulkov}},\ }\href
  {https://doi.org/10.1088/2053-1583/aa6bec} {\bibfield  {journal} {\bibinfo
  {journal} {2D Materials}\ }\textbf {\bibinfo {volume} {4}},\ \bibinfo {pages}
  {025082} (\bibinfo {year} {2017}{\natexlab{a}})}\BibitemShut {NoStop}%
\bibitem [{\citenamefont {Otrokov}\ \emph
  {et~al.}(2017{\natexlab{b}})\citenamefont {Otrokov}, \citenamefont
  {Menshchikova}, \citenamefont {Rusinov}, \citenamefont {Vergniory},
  \citenamefont {Kuznetsov},\ and\ \citenamefont
  {Chulkov}}]{Otrokov.jetpl2017}%
  \BibitemOpen
  \bibfield  {author} {\bibinfo {author} {\bibfnamefont {M.~M.}\ \bibnamefont
  {Otrokov}}, \bibinfo {author} {\bibfnamefont {T.~V.}\ \bibnamefont
  {Menshchikova}}, \bibinfo {author} {\bibfnamefont {I.~P.}\ \bibnamefont
  {Rusinov}}, \bibinfo {author} {\bibfnamefont {M.~G.}\ \bibnamefont
  {Vergniory}}, \bibinfo {author} {\bibfnamefont {V.~M.}\ \bibnamefont
  {Kuznetsov}},\ and\ \bibinfo {author} {\bibfnamefont {E.~V.}\ \bibnamefont
  {Chulkov}},\ }\href@noop {} {\bibfield  {journal} {\bibinfo  {journal} {JETP
  Lett.}\ }\textbf {\bibinfo {volume} {105}},\ \bibinfo {pages} {297} (\bibinfo
  {year} {2017}{\natexlab{b}})}\BibitemShut {NoStop}%
\bibitem [{\citenamefont {Deng}\ \emph {et~al.}(2021)\citenamefont {Deng},
  \citenamefont {Chen}, \citenamefont {Wo{\l}o{\'{s}}}, \citenamefont
  {Konczykowski}, \citenamefont {Sobczak}, \citenamefont {Sitnicka},
  \citenamefont {Fedorchenko}, \citenamefont {Borysiuk}, \citenamefont
  {Heider}, \citenamefont {Pluci{\'{n}}ski}, \citenamefont {Park},
  \citenamefont {Georgescu}, \citenamefont {Cano},\ and\ \citenamefont
  {Krusin-Elbaum}}]{Deng2021}%
  \BibitemOpen
  \bibfield  {author} {\bibinfo {author} {\bibfnamefont {H.}~\bibnamefont
  {Deng}}, \bibinfo {author} {\bibfnamefont {Z.}~\bibnamefont {Chen}}, \bibinfo
  {author} {\bibfnamefont {A.}~\bibnamefont {Wo{\l}o{\'{s}}}}, \bibinfo
  {author} {\bibfnamefont {M.}~\bibnamefont {Konczykowski}}, \bibinfo {author}
  {\bibfnamefont {K.}~\bibnamefont {Sobczak}}, \bibinfo {author} {\bibfnamefont
  {J.}~\bibnamefont {Sitnicka}}, \bibinfo {author} {\bibfnamefont {I.~V.}\
  \bibnamefont {Fedorchenko}}, \bibinfo {author} {\bibfnamefont
  {J.}~\bibnamefont {Borysiuk}}, \bibinfo {author} {\bibfnamefont
  {T.}~\bibnamefont {Heider}}, \bibinfo {author} {\bibfnamefont
  {{\L}.}~\bibnamefont {Pluci{\'{n}}ski}}, \bibinfo {author} {\bibfnamefont
  {K.}~\bibnamefont {Park}}, \bibinfo {author} {\bibfnamefont {A.~B.}\
  \bibnamefont {Georgescu}}, \bibinfo {author} {\bibfnamefont {J.}~\bibnamefont
  {Cano}},\ and\ \bibinfo {author} {\bibfnamefont {L.}~\bibnamefont
  {Krusin-Elbaum}},\ }\href {https://doi.org/10.1038/s41567-020-0998-2}
  {\bibfield  {journal} {\bibinfo  {journal} {Nature Physics}\ }\textbf
  {\bibinfo {volume} {17}},\ \bibinfo {pages} {36} (\bibinfo {year}
  {2021})}\BibitemShut {NoStop}%
\bibitem [{\citenamefont {Deng}\ \emph {et~al.}(2020)\citenamefont {Deng},
  \citenamefont {Yu}, \citenamefont {Shi}, \citenamefont {Guo}, \citenamefont
  {Xu}, \citenamefont {Wang}, \citenamefont {Chen},\ and\ \citenamefont
  {Zhang}}]{Deng895}%
  \BibitemOpen
  \bibfield  {author} {\bibinfo {author} {\bibfnamefont {Y.}~\bibnamefont
  {Deng}}, \bibinfo {author} {\bibfnamefont {Y.}~\bibnamefont {Yu}}, \bibinfo
  {author} {\bibfnamefont {M.~Z.}\ \bibnamefont {Shi}}, \bibinfo {author}
  {\bibfnamefont {Z.}~\bibnamefont {Guo}}, \bibinfo {author} {\bibfnamefont
  {Z.}~\bibnamefont {Xu}}, \bibinfo {author} {\bibfnamefont {J.}~\bibnamefont
  {Wang}}, \bibinfo {author} {\bibfnamefont {X.~H.}\ \bibnamefont {Chen}},\
  and\ \bibinfo {author} {\bibfnamefont {Y.}~\bibnamefont {Zhang}},\ }\href
  {https://doi.org/10.1126/science.aax8156} {\bibfield  {journal} {\bibinfo
  {journal} {Science}\ }\textbf {\bibinfo {volume} {367}},\ \bibinfo {pages}
  {895} (\bibinfo {year} {2020})}\BibitemShut {NoStop}%
\bibitem [{\citenamefont {Otrokov}\ \emph
  {et~al.}(2019{\natexlab{a}})\citenamefont {Otrokov}, \citenamefont {Rusinov},
  \citenamefont {Blanco-Rey}, \citenamefont {Hoffmann}, \citenamefont
  {Vyazovskaya}, \citenamefont {Eremeev}, \citenamefont {Ernst}, \citenamefont
  {Echenique}, \citenamefont {Arnau},\ and\ \citenamefont
  {Chulkov}}]{Otrokov.prl2019}%
  \BibitemOpen
  \bibfield  {author} {\bibinfo {author} {\bibfnamefont {M.~M.}\ \bibnamefont
  {Otrokov}}, \bibinfo {author} {\bibfnamefont {I.~P.}\ \bibnamefont
  {Rusinov}}, \bibinfo {author} {\bibfnamefont {M.}~\bibnamefont {Blanco-Rey}},
  \bibinfo {author} {\bibfnamefont {M.}~\bibnamefont {Hoffmann}}, \bibinfo
  {author} {\bibfnamefont {A.~Y.}\ \bibnamefont {Vyazovskaya}}, \bibinfo
  {author} {\bibfnamefont {S.~V.}\ \bibnamefont {Eremeev}}, \bibinfo {author}
  {\bibfnamefont {A.}~\bibnamefont {Ernst}}, \bibinfo {author} {\bibfnamefont
  {P.~M.}\ \bibnamefont {Echenique}}, \bibinfo {author} {\bibfnamefont
  {A.}~\bibnamefont {Arnau}},\ and\ \bibinfo {author} {\bibfnamefont {E.~V.}\
  \bibnamefont {Chulkov}},\ }\href
  {https://doi.org/10.1103/PhysRevLett.122.107202} {\bibfield  {journal}
  {\bibinfo  {journal} {Phys. Rev. Lett.}\ }\textbf {\bibinfo {volume} {122}},\
  \bibinfo {pages} {107202} (\bibinfo {year} {2019}{\natexlab{a}})}\BibitemShut
  {NoStop}%
\bibitem [{\citenamefont {Liu}\ \emph {et~al.}(2020)\citenamefont {Liu},
  \citenamefont {Wang}, \citenamefont {Li}, \citenamefont {Wu}, \citenamefont
  {Li}, \citenamefont {Li}, \citenamefont {He}, \citenamefont {Xu},
  \citenamefont {Zhang},\ and\ \citenamefont {Wang}}]{Liu2020}%
  \BibitemOpen
  \bibfield  {author} {\bibinfo {author} {\bibfnamefont {C.}~\bibnamefont
  {Liu}}, \bibinfo {author} {\bibfnamefont {Y.}~\bibnamefont {Wang}}, \bibinfo
  {author} {\bibfnamefont {H.}~\bibnamefont {Li}}, \bibinfo {author}
  {\bibfnamefont {Y.}~\bibnamefont {Wu}}, \bibinfo {author} {\bibfnamefont
  {Y.}~\bibnamefont {Li}}, \bibinfo {author} {\bibfnamefont {J.}~\bibnamefont
  {Li}}, \bibinfo {author} {\bibfnamefont {K.}~\bibnamefont {He}}, \bibinfo
  {author} {\bibfnamefont {Y.}~\bibnamefont {Xu}}, \bibinfo {author}
  {\bibfnamefont {J.}~\bibnamefont {Zhang}},\ and\ \bibinfo {author}
  {\bibfnamefont {Y.}~\bibnamefont {Wang}},\ }\href
  {https://doi.org/10.1038/s41563-019-0573-3} {\bibfield  {journal} {\bibinfo
  {journal} {Nature Materials}\ }\textbf {\bibinfo {volume} {19}},\ \bibinfo
  {pages} {522} (\bibinfo {year} {2020})}\BibitemShut {NoStop}%
\bibitem [{\citenamefont {Shikin}\ \emph {et~al.}(2018)\citenamefont {Shikin},
  \citenamefont {Rybkina}, \citenamefont {Estyunin}, \citenamefont {Sostina},
  \citenamefont {Voroshnin}, \citenamefont {Klimovskikh}, \citenamefont
  {Rybkin}, \citenamefont {Surnin}, \citenamefont {Kokh}, \citenamefont
  {Tereshchenko}, \citenamefont {Petaccia}, \citenamefont {Di~Santo},
  \citenamefont {Skirdkov}, \citenamefont {Zvezdin}, \citenamefont {Zvezdin},
  \citenamefont {Kimura}, \citenamefont {Chulkov},\ and\ \citenamefont
  {Krasovskii}}]{Shikin2018}%
  \BibitemOpen
  \bibfield  {author} {\bibinfo {author} {\bibfnamefont {A.~M.}\ \bibnamefont
  {Shikin}}, \bibinfo {author} {\bibfnamefont {A.~A.}\ \bibnamefont {Rybkina}},
  \bibinfo {author} {\bibfnamefont {D.~A.}\ \bibnamefont {Estyunin}}, \bibinfo
  {author} {\bibfnamefont {D.~M.}\ \bibnamefont {Sostina}}, \bibinfo {author}
  {\bibfnamefont {V.~Y.}\ \bibnamefont {Voroshnin}}, \bibinfo {author}
  {\bibfnamefont {I.~I.}\ \bibnamefont {Klimovskikh}}, \bibinfo {author}
  {\bibfnamefont {A.~G.}\ \bibnamefont {Rybkin}}, \bibinfo {author}
  {\bibfnamefont {Y.~A.}\ \bibnamefont {Surnin}}, \bibinfo {author}
  {\bibfnamefont {K.~A.}\ \bibnamefont {Kokh}}, \bibinfo {author}
  {\bibfnamefont {O.~E.}\ \bibnamefont {Tereshchenko}}, \bibinfo {author}
  {\bibfnamefont {L.}~\bibnamefont {Petaccia}}, \bibinfo {author}
  {\bibfnamefont {G.}~\bibnamefont {Di~Santo}}, \bibinfo {author}
  {\bibfnamefont {P.~N.}\ \bibnamefont {Skirdkov}}, \bibinfo {author}
  {\bibfnamefont {K.~A.}\ \bibnamefont {Zvezdin}}, \bibinfo {author}
  {\bibfnamefont {A.~K.}\ \bibnamefont {Zvezdin}}, \bibinfo {author}
  {\bibfnamefont {A.}~\bibnamefont {Kimura}}, \bibinfo {author} {\bibfnamefont
  {E.~V.}\ \bibnamefont {Chulkov}},\ and\ \bibinfo {author} {\bibfnamefont
  {E.~E.}\ \bibnamefont {Krasovskii}},\ }\href
  {https://doi.org/10.1103/PhysRevB.97.245407} {\bibfield  {journal} {\bibinfo
  {journal} {Phys. Rev. B}\ }\textbf {\bibinfo {volume} {97}},\ \bibinfo
  {pages} {245407} (\bibinfo {year} {2018})}\BibitemShut {NoStop}%
\bibitem [{\citenamefont {Rakhmilevich}\ \emph {et~al.}(2018)\citenamefont
  {Rakhmilevich}, \citenamefont {Wang}, \citenamefont {Zhao}, \citenamefont
  {Chan}, \citenamefont {Moodera}, \citenamefont {Liu},\ and\ \citenamefont
  {Chang}}]{Rakhmilevich_2018}%
  \BibitemOpen
  \bibfield  {author} {\bibinfo {author} {\bibfnamefont {D.}~\bibnamefont
  {Rakhmilevich}}, \bibinfo {author} {\bibfnamefont {F.}~\bibnamefont {Wang}},
  \bibinfo {author} {\bibfnamefont {W.}~\bibnamefont {Zhao}}, \bibinfo {author}
  {\bibfnamefont {M.~H.~W.}\ \bibnamefont {Chan}}, \bibinfo {author}
  {\bibfnamefont {J.~S.}\ \bibnamefont {Moodera}}, \bibinfo {author}
  {\bibfnamefont {C.}~\bibnamefont {Liu}},\ and\ \bibinfo {author}
  {\bibfnamefont {C.-Z.}\ \bibnamefont {Chang}},\ }\href
  {https://doi.org/10.1103/PhysRevB.98.094404} {\bibfield  {journal} {\bibinfo
  {journal} {Phys. Rev. B}\ }\textbf {\bibinfo {volume} {98}},\ \bibinfo
  {pages} {094404} (\bibinfo {year} {2018})}\BibitemShut {NoStop}%
\bibitem [{\citenamefont {Chen}\ \emph {et~al.}(2015)\citenamefont {Chen},
  \citenamefont {Liu}, \citenamefont {Zheng}, \citenamefont {Gao},
  \citenamefont {Pan}, \citenamefont {van~der Laan}, \citenamefont {Wang},
  \citenamefont {Zhang}, \citenamefont {Song}, \citenamefont {Wang},
  \citenamefont {Wang}, \citenamefont {Xu}, \citenamefont {Wang},\ and\
  \citenamefont {Zhang}}]{Chen_2015_adv}%
  \BibitemOpen
  \bibfield  {author} {\bibinfo {author} {\bibfnamefont {T.}~\bibnamefont
  {Chen}}, \bibinfo {author} {\bibfnamefont {W.}~\bibnamefont {Liu}}, \bibinfo
  {author} {\bibfnamefont {F.}~\bibnamefont {Zheng}}, \bibinfo {author}
  {\bibfnamefont {M.}~\bibnamefont {Gao}}, \bibinfo {author} {\bibfnamefont
  {X.}~\bibnamefont {Pan}}, \bibinfo {author} {\bibfnamefont {G.}~\bibnamefont
  {van~der Laan}}, \bibinfo {author} {\bibfnamefont {X.}~\bibnamefont {Wang}},
  \bibinfo {author} {\bibfnamefont {Q.}~\bibnamefont {Zhang}}, \bibinfo
  {author} {\bibfnamefont {F.}~\bibnamefont {Song}}, \bibinfo {author}
  {\bibfnamefont {B.}~\bibnamefont {Wang}}, \bibinfo {author} {\bibfnamefont
  {B.}~\bibnamefont {Wang}}, \bibinfo {author} {\bibfnamefont {Y.}~\bibnamefont
  {Xu}}, \bibinfo {author} {\bibfnamefont {G.}~\bibnamefont {Wang}},\ and\
  \bibinfo {author} {\bibfnamefont {R.}~\bibnamefont {Zhang}},\ }\href
  {https://doi.org/https://doi.org/10.1002/adma.201501254} {\bibfield
  {journal} {\bibinfo  {journal} {Advanced Materials}\ }\textbf {\bibinfo
  {volume} {27}},\ \bibinfo {pages} {4823} (\bibinfo {year}
  {2015})}\BibitemShut {NoStop}%
\bibitem [{\citenamefont {Assaf}\ \emph {et~al.}(2015)\citenamefont {Assaf},
  \citenamefont {Katmis}, \citenamefont {Wei}, \citenamefont {Chang},
  \citenamefont {Satpati}, \citenamefont {Moodera},\ and\ \citenamefont
  {Heiman}}]{Assaf_2015}%
  \BibitemOpen
  \bibfield  {author} {\bibinfo {author} {\bibfnamefont {B.~A.}\ \bibnamefont
  {Assaf}}, \bibinfo {author} {\bibfnamefont {F.}~\bibnamefont {Katmis}},
  \bibinfo {author} {\bibfnamefont {P.}~\bibnamefont {Wei}}, \bibinfo {author}
  {\bibfnamefont {C.-Z.}\ \bibnamefont {Chang}}, \bibinfo {author}
  {\bibfnamefont {B.}~\bibnamefont {Satpati}}, \bibinfo {author} {\bibfnamefont
  {J.~S.}\ \bibnamefont {Moodera}},\ and\ \bibinfo {author} {\bibfnamefont
  {D.}~\bibnamefont {Heiman}},\ }\href
  {https://doi.org/10.1103/PhysRevB.91.195310} {\bibfield  {journal} {\bibinfo
  {journal} {Phys. Rev. B}\ }\textbf {\bibinfo {volume} {91}},\ \bibinfo
  {pages} {195310} (\bibinfo {year} {2015})}\BibitemShut {NoStop}%
\bibitem [{\citenamefont {Lee}\ \emph {et~al.}(2019)\citenamefont {Lee},
  \citenamefont {Zhu}, \citenamefont {Wang}, \citenamefont {Miao},
  \citenamefont {Pillsbury}, \citenamefont {Yi}, \citenamefont {Kempinger},
  \citenamefont {Hu}, \citenamefont {Heikes}, \citenamefont {Quarterman},
  \citenamefont {Ratcliff}, \citenamefont {Borchers}, \citenamefont {Zhang},
  \citenamefont {Ke}, \citenamefont {Graf}, \citenamefont {Alem}, \citenamefont
  {Chang}, \citenamefont {Samarth},\ and\ \citenamefont {Mao}}]{LeeZhu_2019}%
  \BibitemOpen
  \bibfield  {author} {\bibinfo {author} {\bibfnamefont {S.~H.}\ \bibnamefont
  {Lee}}, \bibinfo {author} {\bibfnamefont {Y.}~\bibnamefont {Zhu}}, \bibinfo
  {author} {\bibfnamefont {Y.}~\bibnamefont {Wang}}, \bibinfo {author}
  {\bibfnamefont {L.}~\bibnamefont {Miao}}, \bibinfo {author} {\bibfnamefont
  {T.}~\bibnamefont {Pillsbury}}, \bibinfo {author} {\bibfnamefont
  {H.}~\bibnamefont {Yi}}, \bibinfo {author} {\bibfnamefont {S.}~\bibnamefont
  {Kempinger}}, \bibinfo {author} {\bibfnamefont {J.}~\bibnamefont {Hu}},
  \bibinfo {author} {\bibfnamefont {C.~A.}\ \bibnamefont {Heikes}}, \bibinfo
  {author} {\bibfnamefont {P.}~\bibnamefont {Quarterman}}, \bibinfo {author}
  {\bibfnamefont {W.}~\bibnamefont {Ratcliff}}, \bibinfo {author}
  {\bibfnamefont {J.~A.}\ \bibnamefont {Borchers}}, \bibinfo {author}
  {\bibfnamefont {H.}~\bibnamefont {Zhang}}, \bibinfo {author} {\bibfnamefont
  {X.}~\bibnamefont {Ke}}, \bibinfo {author} {\bibfnamefont {D.}~\bibnamefont
  {Graf}}, \bibinfo {author} {\bibfnamefont {N.}~\bibnamefont {Alem}}, \bibinfo
  {author} {\bibfnamefont {C.-Z.}\ \bibnamefont {Chang}}, \bibinfo {author}
  {\bibfnamefont {N.}~\bibnamefont {Samarth}},\ and\ \bibinfo {author}
  {\bibfnamefont {Z.}~\bibnamefont {Mao}},\ }\href
  {https://doi.org/10.1103/PhysRevResearch.1.012011} {\bibfield  {journal}
  {\bibinfo  {journal} {Phys. Rev. Research}\ }\textbf {\bibinfo {volume}
  {1}},\ \bibinfo {pages} {012011(R)} (\bibinfo {year} {2019})}\BibitemShut
  {NoStop}%
\bibitem [{\citenamefont {Petrov}\ \emph {et~al.}(2021)\citenamefont {Petrov},
  \citenamefont {Men'shov}, \citenamefont {Rusinov}, \citenamefont {Hoffmann},
  \citenamefont {Ernst}, \citenamefont {Otrokov}, \citenamefont {Dugaev},
  \citenamefont {Menshchikova},\ and\ \citenamefont {Chulkov}}]{Petrov_2021}%
  \BibitemOpen
  \bibfield  {author} {\bibinfo {author} {\bibfnamefont {E.~K.}\ \bibnamefont
  {Petrov}}, \bibinfo {author} {\bibfnamefont {V.~M.}\ \bibnamefont
  {Men'shov}}, \bibinfo {author} {\bibfnamefont {I.~P.}\ \bibnamefont
  {Rusinov}}, \bibinfo {author} {\bibfnamefont {M.}~\bibnamefont {Hoffmann}},
  \bibinfo {author} {\bibfnamefont {A.}~\bibnamefont {Ernst}}, \bibinfo
  {author} {\bibfnamefont {M.~M.}\ \bibnamefont {Otrokov}}, \bibinfo {author}
  {\bibfnamefont {V.~K.}\ \bibnamefont {Dugaev}}, \bibinfo {author}
  {\bibfnamefont {T.~V.}\ \bibnamefont {Menshchikova}},\ and\ \bibinfo {author}
  {\bibfnamefont {E.~V.}\ \bibnamefont {Chulkov}},\ }\href@noop {} {\bibfield
  {journal} {\bibinfo  {journal} {arXiv:2001,06433}\ } (\bibinfo {year}
  {2021})}\BibitemShut {NoStop}%
\bibitem [{\citenamefont {Weng}\ \emph
  {et~al.}(2015{\natexlab{b}})\citenamefont {Weng}, \citenamefont {Yu},
  \citenamefont {Hu}, \citenamefont {Dai},\ and\ \citenamefont
  {Fang}}]{Weng_Rui_2015}%
  \BibitemOpen
  \bibfield  {author} {\bibinfo {author} {\bibfnamefont {H.}~\bibnamefont
  {Weng}}, \bibinfo {author} {\bibfnamefont {R.}~\bibnamefont {Yu}}, \bibinfo
  {author} {\bibfnamefont {X.}~\bibnamefont {Hu}}, \bibinfo {author}
  {\bibfnamefont {X.}~\bibnamefont {Dai}},\ and\ \bibinfo {author}
  {\bibfnamefont {Z.}~\bibnamefont {Fang}},\ }\href
  {https://doi.org/10.1080/00018732.2015.1068524} {\bibfield  {journal}
  {\bibinfo  {journal} {Advances in Physics}\ }\textbf {\bibinfo {volume}
  {64}},\ \bibinfo {pages} {227} (\bibinfo {year}
  {2015}{\natexlab{b}})}\BibitemShut {NoStop}%
\bibitem [{\citenamefont {Getzlaff}(2008)}]{Getzlaff_2008}%
  \BibitemOpen
  \bibfield  {author} {\bibinfo {author} {\bibfnamefont {M.}~\bibnamefont
  {Getzlaff}},\ }\href {https://doi.org/10.1007/978-3-540-31152-2} {\emph
  {\bibinfo {title} {Fundamentals of Magnetism}}}\ (\bibinfo  {publisher}
  {Springer-Verlag Berlin Heidelberg},\ \bibinfo {address} {Berlin},\ \bibinfo
  {year} {2008})\BibitemShut {NoStop}%
\bibitem [{\citenamefont {Bertotti}(1998)}]{Bertotti_1998}%
  \BibitemOpen
  \bibfield  {author} {\bibinfo {author} {\bibfnamefont {G.}~\bibnamefont
  {Bertotti}},\ }\href@noop {} {\emph {\bibinfo {title} {Hysteresis in
  Magnetism}}}\ (\bibinfo  {publisher} {Academic Press},\ \bibinfo {address}
  {San Diego},\ \bibinfo {year} {1998})\BibitemShut {NoStop}%
\bibitem [{\citenamefont {Gong}\ \emph {et~al.}(2019)\citenamefont {Gong},
  \citenamefont {Guo}, \citenamefont {Li}, \citenamefont {Zhu}, \citenamefont
  {Liao}, \citenamefont {Liu}, \citenamefont {Zhang}, \citenamefont {Gu},
  \citenamefont {Tang}, \citenamefont {Feng}, \citenamefont {Zhang},
  \citenamefont {Li}, \citenamefont {Song}, \citenamefont {Wang}, \citenamefont
  {Yu}, \citenamefont {Chen}, \citenamefont {Wang}, \citenamefont {Yao},
  \citenamefont {Duan}, \citenamefont {Xu}, \citenamefont {Zhang},
  \citenamefont {Ma}, \citenamefont {Xue},\ and\ \citenamefont
  {He}}]{Gong_2019}%
  \BibitemOpen
  \bibfield  {author} {\bibinfo {author} {\bibfnamefont {Y.}~\bibnamefont
  {Gong}}, \bibinfo {author} {\bibfnamefont {J.}~\bibnamefont {Guo}}, \bibinfo
  {author} {\bibfnamefont {J.}~\bibnamefont {Li}}, \bibinfo {author}
  {\bibfnamefont {K.}~\bibnamefont {Zhu}}, \bibinfo {author} {\bibfnamefont
  {M.}~\bibnamefont {Liao}}, \bibinfo {author} {\bibfnamefont {X.}~\bibnamefont
  {Liu}}, \bibinfo {author} {\bibfnamefont {Q.}~\bibnamefont {Zhang}}, \bibinfo
  {author} {\bibfnamefont {L.}~\bibnamefont {Gu}}, \bibinfo {author}
  {\bibfnamefont {L.}~\bibnamefont {Tang}}, \bibinfo {author} {\bibfnamefont
  {X.}~\bibnamefont {Feng}}, \bibinfo {author} {\bibfnamefont {D.}~\bibnamefont
  {Zhang}}, \bibinfo {author} {\bibfnamefont {W.}~\bibnamefont {Li}}, \bibinfo
  {author} {\bibfnamefont {C.}~\bibnamefont {Song}}, \bibinfo {author}
  {\bibfnamefont {L.}~\bibnamefont {Wang}}, \bibinfo {author} {\bibfnamefont
  {P.}~\bibnamefont {Yu}}, \bibinfo {author} {\bibfnamefont {X.}~\bibnamefont
  {Chen}}, \bibinfo {author} {\bibfnamefont {Y.}~\bibnamefont {Wang}}, \bibinfo
  {author} {\bibfnamefont {H.}~\bibnamefont {Yao}}, \bibinfo {author}
  {\bibfnamefont {W.}~\bibnamefont {Duan}}, \bibinfo {author} {\bibfnamefont
  {Y.}~\bibnamefont {Xu}}, \bibinfo {author} {\bibfnamefont {S.-C.}\
  \bibnamefont {Zhang}}, \bibinfo {author} {\bibfnamefont {X.}~\bibnamefont
  {Ma}}, \bibinfo {author} {\bibfnamefont {Q.-K.}\ \bibnamefont {Xue}},\ and\
  \bibinfo {author} {\bibfnamefont {K.}~\bibnamefont {He}},\ }\href
  {https://doi.org/10.1088/0256-307x/36/7/076801} {\bibfield  {journal}
  {\bibinfo  {journal} {Chinese Physics Letters}\ }\textbf {\bibinfo {volume}
  {36}},\ \bibinfo {pages} {076801} (\bibinfo {year} {2019})}\BibitemShut
  {NoStop}%
\bibitem [{\citenamefont {Shikin}\ \emph {et~al.}(2020)\citenamefont {Shikin},
  \citenamefont {Estyunin}, \citenamefont {Klimovskikh}, \citenamefont
  {Filnov}, \citenamefont {Schwier}, \citenamefont {Kumar}, \citenamefont
  {Miyamoto}, \citenamefont {Okuda}, \citenamefont {Kimura}, \citenamefont
  {Kuroda}, \citenamefont {Yaji}, \citenamefont {Shin}, \citenamefont {Takeda},
  \citenamefont {Saitoh}, \citenamefont {Aliev}, \citenamefont {Mamedov},
  \citenamefont {Amiraslanov}, \citenamefont {Babanly}, \citenamefont
  {Otrokov}, \citenamefont {Eremeev},\ and\ \citenamefont
  {Chulkov}}]{Shikin2020}%
  \BibitemOpen
  \bibfield  {author} {\bibinfo {author} {\bibfnamefont {A.~M.}\ \bibnamefont
  {Shikin}}, \bibinfo {author} {\bibfnamefont {D.~A.}\ \bibnamefont
  {Estyunin}}, \bibinfo {author} {\bibfnamefont {I.~I.}\ \bibnamefont
  {Klimovskikh}}, \bibinfo {author} {\bibfnamefont {S.~O.}\ \bibnamefont
  {Filnov}}, \bibinfo {author} {\bibfnamefont {E.~F.}\ \bibnamefont {Schwier}},
  \bibinfo {author} {\bibfnamefont {S.}~\bibnamefont {Kumar}}, \bibinfo
  {author} {\bibfnamefont {K.}~\bibnamefont {Miyamoto}}, \bibinfo {author}
  {\bibfnamefont {T.}~\bibnamefont {Okuda}}, \bibinfo {author} {\bibfnamefont
  {A.}~\bibnamefont {Kimura}}, \bibinfo {author} {\bibfnamefont
  {K.}~\bibnamefont {Kuroda}}, \bibinfo {author} {\bibfnamefont
  {K.}~\bibnamefont {Yaji}}, \bibinfo {author} {\bibfnamefont {S.}~\bibnamefont
  {Shin}}, \bibinfo {author} {\bibfnamefont {Y.}~\bibnamefont {Takeda}},
  \bibinfo {author} {\bibfnamefont {Y.}~\bibnamefont {Saitoh}}, \bibinfo
  {author} {\bibfnamefont {Z.~S.}\ \bibnamefont {Aliev}}, \bibinfo {author}
  {\bibfnamefont {N.~T.}\ \bibnamefont {Mamedov}}, \bibinfo {author}
  {\bibfnamefont {I.~R.}\ \bibnamefont {Amiraslanov}}, \bibinfo {author}
  {\bibfnamefont {M.~B.}\ \bibnamefont {Babanly}}, \bibinfo {author}
  {\bibfnamefont {M.~M.}\ \bibnamefont {Otrokov}}, \bibinfo {author}
  {\bibfnamefont {S.~V.}\ \bibnamefont {Eremeev}},\ and\ \bibinfo {author}
  {\bibfnamefont {E.~V.}\ \bibnamefont {Chulkov}},\ }\href
  {https://doi.org/10.1038/s41598-020-70089-9} {\bibfield  {journal} {\bibinfo
  {journal} {Scientific Reports}\ }\textbf {\bibinfo {volume} {10}},\ \bibinfo
  {pages} {13226} (\bibinfo {year} {2020})}\BibitemShut {NoStop}%
\bibitem [{\citenamefont {Sass}\ \emph {et~al.}(2020)\citenamefont {Sass},
  \citenamefont {Kim}, \citenamefont {Vanderbilt}, \citenamefont {Yan},\ and\
  \citenamefont {Wu}}]{Saas_2020}%
  \BibitemOpen
  \bibfield  {author} {\bibinfo {author} {\bibfnamefont {P.~M.}\ \bibnamefont
  {Sass}}, \bibinfo {author} {\bibfnamefont {J.}~\bibnamefont {Kim}}, \bibinfo
  {author} {\bibfnamefont {D.}~\bibnamefont {Vanderbilt}}, \bibinfo {author}
  {\bibfnamefont {J.}~\bibnamefont {Yan}},\ and\ \bibinfo {author}
  {\bibfnamefont {W.}~\bibnamefont {Wu}},\ }\href
  {https://doi.org/10.1103/PhysRevLett.125.037201} {\bibfield  {journal}
  {\bibinfo  {journal} {Phys. Rev. Lett.}\ }\textbf {\bibinfo {volume} {125}},\
  \bibinfo {pages} {037201} (\bibinfo {year} {2020})}\BibitemShut {NoStop}%
\bibitem [{\citenamefont {Hou}\ \emph {et~al.}(2020)\citenamefont {Hou},
  \citenamefont {Yao}, \citenamefont {Zhou}, \citenamefont {Ma}, \citenamefont
  {Han}, \citenamefont {Hao}, \citenamefont {Wu}, \citenamefont {Zhang},
  \citenamefont {Sun}, \citenamefont {Liu}, \citenamefont {Zhao}, \citenamefont
  {Liu},\ and\ \citenamefont {Lin}}]{Hou2020}%
  \BibitemOpen
  \bibfield  {author} {\bibinfo {author} {\bibfnamefont {F.}~\bibnamefont
  {Hou}}, \bibinfo {author} {\bibfnamefont {Q.}~\bibnamefont {Yao}}, \bibinfo
  {author} {\bibfnamefont {C.-S.}\ \bibnamefont {Zhou}}, \bibinfo {author}
  {\bibfnamefont {X.-M.}\ \bibnamefont {Ma}}, \bibinfo {author} {\bibfnamefont
  {M.}~\bibnamefont {Han}}, \bibinfo {author} {\bibfnamefont {Y.-J.}\
  \bibnamefont {Hao}}, \bibinfo {author} {\bibfnamefont {X.}~\bibnamefont
  {Wu}}, \bibinfo {author} {\bibfnamefont {Y.}~\bibnamefont {Zhang}}, \bibinfo
  {author} {\bibfnamefont {H.}~\bibnamefont {Sun}}, \bibinfo {author}
  {\bibfnamefont {C.}~\bibnamefont {Liu}}, \bibinfo {author} {\bibfnamefont
  {Y.}~\bibnamefont {Zhao}}, \bibinfo {author} {\bibfnamefont {Q.}~\bibnamefont
  {Liu}},\ and\ \bibinfo {author} {\bibfnamefont {J.}~\bibnamefont {Lin}},\
  }\href {https://doi.org/10.1021/acsnano.0c03149} {\bibfield  {journal}
  {\bibinfo  {journal} {ACS Nano}\ }\textbf {\bibinfo {volume} {14}},\ \bibinfo
  {pages} {11262} (\bibinfo {year} {2020})}\BibitemShut {NoStop}%
\bibitem [{\citenamefont {Liu}\ \emph {et~al.}(2016{\natexlab{b}})\citenamefont
  {Liu}, \citenamefont {Wang}, \citenamefont {Richardella}, \citenamefont
  {Kandala}, \citenamefont {Li}, \citenamefont {Yazdani}, \citenamefont
  {Samarth},\ and\ \citenamefont {Ong}}]{LiuMinhao2016}%
  \BibitemOpen
  \bibfield  {author} {\bibinfo {author} {\bibfnamefont {M.}~\bibnamefont
  {Liu}}, \bibinfo {author} {\bibfnamefont {W.}~\bibnamefont {Wang}}, \bibinfo
  {author} {\bibfnamefont {A.~R.}\ \bibnamefont {Richardella}}, \bibinfo
  {author} {\bibfnamefont {A.}~\bibnamefont {Kandala}}, \bibinfo {author}
  {\bibfnamefont {J.}~\bibnamefont {Li}}, \bibinfo {author} {\bibfnamefont
  {A.}~\bibnamefont {Yazdani}}, \bibinfo {author} {\bibfnamefont
  {N.}~\bibnamefont {Samarth}},\ and\ \bibinfo {author} {\bibfnamefont {N.~P.}\
  \bibnamefont {Ong}},\ }\bibfield  {journal} {\bibinfo  {journal} {Science
  Advances}\ }\textbf {\bibinfo {volume} {2}},\ \href
  {https://doi.org/10.1126/sciadv.1600167} {10.1126/sciadv.1600167} (\bibinfo
  {year} {2016}{\natexlab{b}})\BibitemShut {NoStop}%
\bibitem [{\citenamefont {Lachman}\ \emph {et~al.}(2015)\citenamefont
  {Lachman}, \citenamefont {Young}, \citenamefont {Richardella}, \citenamefont
  {Cuppens}, \citenamefont {Naren}, \citenamefont {Anahory}, \citenamefont
  {Meltzer}, \citenamefont {Kandala}, \citenamefont {Kempinger}, \citenamefont
  {Myasoedov}, \citenamefont {Huber}, \citenamefont {Samarth},\ and\
  \citenamefont {Zeldov}}]{Lachmane1500740}%
  \BibitemOpen
  \bibfield  {author} {\bibinfo {author} {\bibfnamefont {E.~O.}\ \bibnamefont
  {Lachman}}, \bibinfo {author} {\bibfnamefont {A.~F.}\ \bibnamefont {Young}},
  \bibinfo {author} {\bibfnamefont {A.}~\bibnamefont {Richardella}}, \bibinfo
  {author} {\bibfnamefont {J.}~\bibnamefont {Cuppens}}, \bibinfo {author}
  {\bibfnamefont {H.~R.}\ \bibnamefont {Naren}}, \bibinfo {author}
  {\bibfnamefont {Y.}~\bibnamefont {Anahory}}, \bibinfo {author} {\bibfnamefont
  {A.~Y.}\ \bibnamefont {Meltzer}}, \bibinfo {author} {\bibfnamefont
  {A.}~\bibnamefont {Kandala}}, \bibinfo {author} {\bibfnamefont
  {S.}~\bibnamefont {Kempinger}}, \bibinfo {author} {\bibfnamefont
  {Y.}~\bibnamefont {Myasoedov}}, \bibinfo {author} {\bibfnamefont {M.~E.}\
  \bibnamefont {Huber}}, \bibinfo {author} {\bibfnamefont {N.}~\bibnamefont
  {Samarth}},\ and\ \bibinfo {author} {\bibfnamefont {E.}~\bibnamefont
  {Zeldov}},\ }\href@noop {} {\bibfield  {journal} {\bibinfo  {journal}
  {Science Advances}\ }\textbf {\bibinfo {volume} {1}} (\bibinfo {year}
  {2015})}\BibitemShut {NoStop}%
\bibitem [{\citenamefont {Wang}\ \emph {et~al.}(2014)\citenamefont {Wang},
  \citenamefont {Lian},\ and\ \citenamefont {Zhang}}]{WangLian2014}%
  \BibitemOpen
  \bibfield  {author} {\bibinfo {author} {\bibfnamefont {J.}~\bibnamefont
  {Wang}}, \bibinfo {author} {\bibfnamefont {B.}~\bibnamefont {Lian}},\ and\
  \bibinfo {author} {\bibfnamefont {S.-C.}\ \bibnamefont {Zhang}},\ }\href
  {https://doi.org/10.1103/PhysRevB.89.085106} {\bibfield  {journal} {\bibinfo
  {journal} {Phys. Rev. B}\ }\textbf {\bibinfo {volume} {89}},\ \bibinfo
  {pages} {085106} (\bibinfo {year} {2014})}\BibitemShut {NoStop}%
\bibitem [{\citenamefont {Allen}\ \emph {et~al.}(2019)\citenamefont {Allen},
  \citenamefont {Cui}, \citenamefont {Yue~Ma}, \citenamefont {Mogi},
  \citenamefont {Kawamura}, \citenamefont {Fulga}, \citenamefont
  {Goldhaber-Gordon}, \citenamefont {Tokura},\ and\ \citenamefont
  {Shen}}]{Allen14511}%
  \BibitemOpen
  \bibfield  {author} {\bibinfo {author} {\bibfnamefont {M.}~\bibnamefont
  {Allen}}, \bibinfo {author} {\bibfnamefont {Y.}~\bibnamefont {Cui}}, \bibinfo
  {author} {\bibfnamefont {E.}~\bibnamefont {Yue~Ma}}, \bibinfo {author}
  {\bibfnamefont {M.}~\bibnamefont {Mogi}}, \bibinfo {author} {\bibfnamefont
  {M.}~\bibnamefont {Kawamura}}, \bibinfo {author} {\bibfnamefont {I.~C.}\
  \bibnamefont {Fulga}}, \bibinfo {author} {\bibfnamefont {D.}~\bibnamefont
  {Goldhaber-Gordon}}, \bibinfo {author} {\bibfnamefont {Y.}~\bibnamefont
  {Tokura}},\ and\ \bibinfo {author} {\bibfnamefont {Z.-X.}\ \bibnamefont
  {Shen}},\ }\href {https://doi.org/10.1073/pnas.1818255116} {\bibfield
  {journal} {\bibinfo  {journal} {Proceedings of the National Academy of
  Sciences}\ }\textbf {\bibinfo {volume} {116}},\ \bibinfo {pages} {14511}
  (\bibinfo {year} {2019})}\BibitemShut {NoStop}%
\bibitem [{\citenamefont {Wu}\ \emph {et~al.}(2020)\citenamefont {Wu},
  \citenamefont {Xiao}, \citenamefont {Chen}, \citenamefont {Sun},
  \citenamefont {Zhang}, \citenamefont {Chan}, \citenamefont {Samarth},
  \citenamefont {Xie}, \citenamefont {Lin},\ and\ \citenamefont
  {Chang}}]{WuXiao2020}%
  \BibitemOpen
  \bibfield  {author} {\bibinfo {author} {\bibfnamefont {X.}~\bibnamefont
  {Wu}}, \bibinfo {author} {\bibfnamefont {D.}~\bibnamefont {Xiao}}, \bibinfo
  {author} {\bibfnamefont {C.-Z.}\ \bibnamefont {Chen}}, \bibinfo {author}
  {\bibfnamefont {J.}~\bibnamefont {Sun}}, \bibinfo {author} {\bibfnamefont
  {L.}~\bibnamefont {Zhang}}, \bibinfo {author} {\bibfnamefont {M.~H.~W.}\
  \bibnamefont {Chan}}, \bibinfo {author} {\bibfnamefont {N.}~\bibnamefont
  {Samarth}}, \bibinfo {author} {\bibfnamefont {X.~C.}\ \bibnamefont {Xie}},
  \bibinfo {author} {\bibfnamefont {X.}~\bibnamefont {Lin}},\ and\ \bibinfo
  {author} {\bibfnamefont {C.-Z.}\ \bibnamefont {Chang}},\ }\href
  {https://doi.org/10.1038/s41467-020-18312-z} {\bibfield  {journal} {\bibinfo
  {journal} {Nature Communications}\ }\textbf {\bibinfo {volume} {11}},\
  \bibinfo {pages} {4532} (\bibinfo {year} {2020})}\BibitemShut {NoStop}%
\bibitem [{\citenamefont {Li}\ \emph {et~al.}(2019)\citenamefont {Li},
  \citenamefont {Gao}, \citenamefont {Duan}, \citenamefont {Xu}, \citenamefont
  {Zhu}, \citenamefont {Tian}, \citenamefont {Gao}, \citenamefont {Fan},
  \citenamefont {Rao}, \citenamefont {Huang}, \citenamefont {Li}, \citenamefont
  {Yan}, \citenamefont {Liu}, \citenamefont {Liu}, \citenamefont {Huang},
  \citenamefont {Li}, \citenamefont {Liu}, \citenamefont {Zhang}, \citenamefont
  {Zhang}, \citenamefont {Kondo}, \citenamefont {Shin}, \citenamefont {Lei},
  \citenamefont {Shi}, \citenamefont {Zhang}, \citenamefont {Weng},
  \citenamefont {Qian},\ and\ \citenamefont {Ding}}]{Stability1}%
  \BibitemOpen
  \bibfield  {author} {\bibinfo {author} {\bibfnamefont {H.}~\bibnamefont
  {Li}}, \bibinfo {author} {\bibfnamefont {S.-Y.}\ \bibnamefont {Gao}},
  \bibinfo {author} {\bibfnamefont {S.-F.}\ \bibnamefont {Duan}}, \bibinfo
  {author} {\bibfnamefont {Y.-F.}\ \bibnamefont {Xu}}, \bibinfo {author}
  {\bibfnamefont {K.-J.}\ \bibnamefont {Zhu}}, \bibinfo {author} {\bibfnamefont
  {S.-J.}\ \bibnamefont {Tian}}, \bibinfo {author} {\bibfnamefont {J.-C.}\
  \bibnamefont {Gao}}, \bibinfo {author} {\bibfnamefont {W.-H.}\ \bibnamefont
  {Fan}}, \bibinfo {author} {\bibfnamefont {Z.-C.}\ \bibnamefont {Rao}},
  \bibinfo {author} {\bibfnamefont {J.-R.}\ \bibnamefont {Huang}}, \bibinfo
  {author} {\bibfnamefont {J.-J.}\ \bibnamefont {Li}}, \bibinfo {author}
  {\bibfnamefont {D.-Y.}\ \bibnamefont {Yan}}, \bibinfo {author} {\bibfnamefont
  {Z.-T.}\ \bibnamefont {Liu}}, \bibinfo {author} {\bibfnamefont {W.-L.}\
  \bibnamefont {Liu}}, \bibinfo {author} {\bibfnamefont {Y.-B.}\ \bibnamefont
  {Huang}}, \bibinfo {author} {\bibfnamefont {Y.-L.}\ \bibnamefont {Li}},
  \bibinfo {author} {\bibfnamefont {Y.}~\bibnamefont {Liu}}, \bibinfo {author}
  {\bibfnamefont {G.-B.}\ \bibnamefont {Zhang}}, \bibinfo {author}
  {\bibfnamefont {P.}~\bibnamefont {Zhang}}, \bibinfo {author} {\bibfnamefont
  {T.}~\bibnamefont {Kondo}}, \bibinfo {author} {\bibfnamefont
  {S.}~\bibnamefont {Shin}}, \bibinfo {author} {\bibfnamefont {H.-C.}\
  \bibnamefont {Lei}}, \bibinfo {author} {\bibfnamefont {Y.-G.}\ \bibnamefont
  {Shi}}, \bibinfo {author} {\bibfnamefont {W.-T.}\ \bibnamefont {Zhang}},
  \bibinfo {author} {\bibfnamefont {H.-M.}\ \bibnamefont {Weng}}, \bibinfo
  {author} {\bibfnamefont {T.}~\bibnamefont {Qian}},\ and\ \bibinfo {author}
  {\bibfnamefont {H.}~\bibnamefont {Ding}},\ }\href
  {https://doi.org/10.1103/PhysRevX.9.041039} {\bibfield  {journal} {\bibinfo
  {journal} {Phys. Rev. X}\ }\textbf {\bibinfo {volume} {9}},\ \bibinfo {pages}
  {041039} (\bibinfo {year} {2019})}\BibitemShut {NoStop}%
\bibitem [{\citenamefont {Hao}\ \emph {et~al.}(2019)\citenamefont {Hao},
  \citenamefont {Liu}, \citenamefont {Feng}, \citenamefont {Ma}, \citenamefont
  {Schwier}, \citenamefont {Arita}, \citenamefont {Kumar}, \citenamefont {Hu},
  \citenamefont {Lu}, \citenamefont {Zeng}, \citenamefont {Wang}, \citenamefont
  {Hao}, \citenamefont {Sun}, \citenamefont {Zhang}, \citenamefont {Mei},
  \citenamefont {Ni}, \citenamefont {Wu}, \citenamefont {Shimada},
  \citenamefont {Chen}, \citenamefont {Liu},\ and\ \citenamefont
  {Liu}}]{Stability2}%
  \BibitemOpen
  \bibfield  {author} {\bibinfo {author} {\bibfnamefont {Y.-J.}\ \bibnamefont
  {Hao}}, \bibinfo {author} {\bibfnamefont {P.}~\bibnamefont {Liu}}, \bibinfo
  {author} {\bibfnamefont {Y.}~\bibnamefont {Feng}}, \bibinfo {author}
  {\bibfnamefont {X.-M.}\ \bibnamefont {Ma}}, \bibinfo {author} {\bibfnamefont
  {E.~F.}\ \bibnamefont {Schwier}}, \bibinfo {author} {\bibfnamefont
  {M.}~\bibnamefont {Arita}}, \bibinfo {author} {\bibfnamefont
  {S.}~\bibnamefont {Kumar}}, \bibinfo {author} {\bibfnamefont
  {C.}~\bibnamefont {Hu}}, \bibinfo {author} {\bibfnamefont {R.}~\bibnamefont
  {Lu}}, \bibinfo {author} {\bibfnamefont {M.}~\bibnamefont {Zeng}}, \bibinfo
  {author} {\bibfnamefont {Y.}~\bibnamefont {Wang}}, \bibinfo {author}
  {\bibfnamefont {Z.}~\bibnamefont {Hao}}, \bibinfo {author} {\bibfnamefont
  {H.-Y.}\ \bibnamefont {Sun}}, \bibinfo {author} {\bibfnamefont
  {K.}~\bibnamefont {Zhang}}, \bibinfo {author} {\bibfnamefont
  {J.}~\bibnamefont {Mei}}, \bibinfo {author} {\bibfnamefont {N.}~\bibnamefont
  {Ni}}, \bibinfo {author} {\bibfnamefont {L.}~\bibnamefont {Wu}}, \bibinfo
  {author} {\bibfnamefont {K.}~\bibnamefont {Shimada}}, \bibinfo {author}
  {\bibfnamefont {C.}~\bibnamefont {Chen}}, \bibinfo {author} {\bibfnamefont
  {Q.}~\bibnamefont {Liu}},\ and\ \bibinfo {author} {\bibfnamefont
  {C.}~\bibnamefont {Liu}},\ }\href {https://doi.org/10.1103/PhysRevX.9.041038}
  {\bibfield  {journal} {\bibinfo  {journal} {Phys. Rev. X}\ }\textbf {\bibinfo
  {volume} {9}},\ \bibinfo {pages} {041038} (\bibinfo {year}
  {2019})}\BibitemShut {NoStop}%
\bibitem [{\citenamefont {Chen}\ \emph {et~al.}(2019)\citenamefont {Chen},
  \citenamefont {Xu}, \citenamefont {Li}, \citenamefont {Li}, \citenamefont
  {Wang}, \citenamefont {Zhang}, \citenamefont {Li}, \citenamefont {Wu},
  \citenamefont {Liang}, \citenamefont {Chen}, \citenamefont {Jung},
  \citenamefont {Cacho}, \citenamefont {Mao}, \citenamefont {Liu},
  \citenamefont {Wang}, \citenamefont {Guo}, \citenamefont {Xu}, \citenamefont
  {Liu}, \citenamefont {Yang},\ and\ \citenamefont {Chen}}]{Stability3}%
  \BibitemOpen
  \bibfield  {author} {\bibinfo {author} {\bibfnamefont {Y.~J.}\ \bibnamefont
  {Chen}}, \bibinfo {author} {\bibfnamefont {L.~X.}\ \bibnamefont {Xu}},
  \bibinfo {author} {\bibfnamefont {J.~H.}\ \bibnamefont {Li}}, \bibinfo
  {author} {\bibfnamefont {Y.~W.}\ \bibnamefont {Li}}, \bibinfo {author}
  {\bibfnamefont {H.~Y.}\ \bibnamefont {Wang}}, \bibinfo {author}
  {\bibfnamefont {C.~F.}\ \bibnamefont {Zhang}}, \bibinfo {author}
  {\bibfnamefont {H.}~\bibnamefont {Li}}, \bibinfo {author} {\bibfnamefont
  {Y.}~\bibnamefont {Wu}}, \bibinfo {author} {\bibfnamefont {A.~J.}\
  \bibnamefont {Liang}}, \bibinfo {author} {\bibfnamefont {C.}~\bibnamefont
  {Chen}}, \bibinfo {author} {\bibfnamefont {S.~W.}\ \bibnamefont {Jung}},
  \bibinfo {author} {\bibfnamefont {C.}~\bibnamefont {Cacho}}, \bibinfo
  {author} {\bibfnamefont {Y.~H.}\ \bibnamefont {Mao}}, \bibinfo {author}
  {\bibfnamefont {S.}~\bibnamefont {Liu}}, \bibinfo {author} {\bibfnamefont
  {M.~X.}\ \bibnamefont {Wang}}, \bibinfo {author} {\bibfnamefont {Y.~F.}\
  \bibnamefont {Guo}}, \bibinfo {author} {\bibfnamefont {Y.}~\bibnamefont
  {Xu}}, \bibinfo {author} {\bibfnamefont {Z.~K.}\ \bibnamefont {Liu}},
  \bibinfo {author} {\bibfnamefont {L.~X.}\ \bibnamefont {Yang}},\ and\
  \bibinfo {author} {\bibfnamefont {Y.~L.}\ \bibnamefont {Chen}},\ }\href
  {https://doi.org/10.1103/PhysRevX.9.041040} {\bibfield  {journal} {\bibinfo
  {journal} {Phys. Rev. X}\ }\textbf {\bibinfo {volume} {9}},\ \bibinfo {pages}
  {041040} (\bibinfo {year} {2019})}\BibitemShut {NoStop}%
\bibitem [{\citenamefont {Swatek}\ \emph {et~al.}(2020)\citenamefont {Swatek},
  \citenamefont {Wu}, \citenamefont {Wang}, \citenamefont {Lee}, \citenamefont
  {Schrunk}, \citenamefont {Yan},\ and\ \citenamefont {Kaminski}}]{Stability4}%
  \BibitemOpen
  \bibfield  {author} {\bibinfo {author} {\bibfnamefont {P.}~\bibnamefont
  {Swatek}}, \bibinfo {author} {\bibfnamefont {Y.}~\bibnamefont {Wu}}, \bibinfo
  {author} {\bibfnamefont {L.-L.}\ \bibnamefont {Wang}}, \bibinfo {author}
  {\bibfnamefont {K.}~\bibnamefont {Lee}}, \bibinfo {author} {\bibfnamefont
  {B.}~\bibnamefont {Schrunk}}, \bibinfo {author} {\bibfnamefont
  {J.}~\bibnamefont {Yan}},\ and\ \bibinfo {author} {\bibfnamefont
  {A.}~\bibnamefont {Kaminski}},\ }\href
  {https://doi.org/10.1103/PhysRevB.101.161109} {\bibfield  {journal} {\bibinfo
   {journal} {Phys. Rev. B}\ }\textbf {\bibinfo {volume} {101}},\ \bibinfo
  {pages} {161109(R)} (\bibinfo {year} {2020})}\BibitemShut {NoStop}%
\bibitem [{\citenamefont {Otrokov}\ \emph
  {et~al.}(2019{\natexlab{b}})\citenamefont {Otrokov}, \citenamefont
  {Klimovskikh}, \citenamefont {Bentmann}, \citenamefont {Estyunin},
  \citenamefont {Zeugner}, \citenamefont {Aliev}, \citenamefont {Ga{\ss}},
  \citenamefont {Wolter}, \citenamefont {Koroleva}, \citenamefont {Shikin},
  \citenamefont {Blanco-Rey}, \citenamefont {Hoffmann}, \citenamefont
  {Rusinov}, \citenamefont {Vyazovskaya}, \citenamefont {Eremeev},
  \citenamefont {Koroteev}, \citenamefont {Kuznetsov}, \citenamefont {Freyse},
  \citenamefont {S{\'a}nchez-Barriga}, \citenamefont {Amiraslanov},
  \citenamefont {Babanly}, \citenamefont {Mamedov}, \citenamefont {Abdullayev},
  \citenamefont {Zverev}, \citenamefont {Alfonsov}, \citenamefont {Kataev},
  \citenamefont {B{\"u}chner}, \citenamefont {Schwier}, \citenamefont {Kumar},
  \citenamefont {Kimura}, \citenamefont {Petaccia}, \citenamefont {Di~Santo},
  \citenamefont {Vidal}, \citenamefont {Schatz}, \citenamefont {Ki{\ss}ner},
  \citenamefont {{\"U}nzelmann}, \citenamefont {Min}, \citenamefont {Moser},
  \citenamefont {Peixoto}, \citenamefont {Reinert}, \citenamefont {Ernst},
  \citenamefont {Echenique}, \citenamefont {Isaeva},\ and\ \citenamefont
  {Chulkov}}]{OtrokovN2019}%
  \BibitemOpen
  \bibfield  {author} {\bibinfo {author} {\bibfnamefont {M.~M.}\ \bibnamefont
  {Otrokov}}, \bibinfo {author} {\bibfnamefont {I.~I.}\ \bibnamefont
  {Klimovskikh}}, \bibinfo {author} {\bibfnamefont {H.}~\bibnamefont
  {Bentmann}}, \bibinfo {author} {\bibfnamefont {D.}~\bibnamefont {Estyunin}},
  \bibinfo {author} {\bibfnamefont {A.}~\bibnamefont {Zeugner}}, \bibinfo
  {author} {\bibfnamefont {Z.~S.}\ \bibnamefont {Aliev}}, \bibinfo {author}
  {\bibfnamefont {S.}~\bibnamefont {Ga{\ss}}}, \bibinfo {author} {\bibfnamefont
  {A.~U.~B.}\ \bibnamefont {Wolter}}, \bibinfo {author} {\bibfnamefont {A.~V.}\
  \bibnamefont {Koroleva}}, \bibinfo {author} {\bibfnamefont {A.~M.}\
  \bibnamefont {Shikin}}, \bibinfo {author} {\bibfnamefont {M.}~\bibnamefont
  {Blanco-Rey}}, \bibinfo {author} {\bibfnamefont {M.}~\bibnamefont
  {Hoffmann}}, \bibinfo {author} {\bibfnamefont {I.~P.}\ \bibnamefont
  {Rusinov}}, \bibinfo {author} {\bibfnamefont {A.~Y.}\ \bibnamefont
  {Vyazovskaya}}, \bibinfo {author} {\bibfnamefont {S.~V.}\ \bibnamefont
  {Eremeev}}, \bibinfo {author} {\bibfnamefont {Y.~M.}\ \bibnamefont
  {Koroteev}}, \bibinfo {author} {\bibfnamefont {V.~M.}\ \bibnamefont
  {Kuznetsov}}, \bibinfo {author} {\bibfnamefont {F.}~\bibnamefont {Freyse}},
  \bibinfo {author} {\bibfnamefont {J.}~\bibnamefont {S{\'a}nchez-Barriga}},
  \bibinfo {author} {\bibfnamefont {I.~R.}\ \bibnamefont {Amiraslanov}},
  \bibinfo {author} {\bibfnamefont {M.~B.}\ \bibnamefont {Babanly}}, \bibinfo
  {author} {\bibfnamefont {N.~T.}\ \bibnamefont {Mamedov}}, \bibinfo {author}
  {\bibfnamefont {N.~A.}\ \bibnamefont {Abdullayev}}, \bibinfo {author}
  {\bibfnamefont {V.~N.}\ \bibnamefont {Zverev}}, \bibinfo {author}
  {\bibfnamefont {A.}~\bibnamefont {Alfonsov}}, \bibinfo {author}
  {\bibfnamefont {V.}~\bibnamefont {Kataev}}, \bibinfo {author} {\bibfnamefont
  {B.}~\bibnamefont {B{\"u}chner}}, \bibinfo {author} {\bibfnamefont {E.~F.}\
  \bibnamefont {Schwier}}, \bibinfo {author} {\bibfnamefont {S.}~\bibnamefont
  {Kumar}}, \bibinfo {author} {\bibfnamefont {A.}~\bibnamefont {Kimura}},
  \bibinfo {author} {\bibfnamefont {L.}~\bibnamefont {Petaccia}}, \bibinfo
  {author} {\bibfnamefont {G.}~\bibnamefont {Di~Santo}}, \bibinfo {author}
  {\bibfnamefont {R.~C.}\ \bibnamefont {Vidal}}, \bibinfo {author}
  {\bibfnamefont {S.}~\bibnamefont {Schatz}}, \bibinfo {author} {\bibfnamefont
  {K.}~\bibnamefont {Ki{\ss}ner}}, \bibinfo {author} {\bibfnamefont
  {M.}~\bibnamefont {{\"U}nzelmann}}, \bibinfo {author} {\bibfnamefont {C.~H.}\
  \bibnamefont {Min}}, \bibinfo {author} {\bibfnamefont {S.}~\bibnamefont
  {Moser}}, \bibinfo {author} {\bibfnamefont {T.~R.~F.}\ \bibnamefont
  {Peixoto}}, \bibinfo {author} {\bibfnamefont {F.}~\bibnamefont {Reinert}},
  \bibinfo {author} {\bibfnamefont {A.}~\bibnamefont {Ernst}}, \bibinfo
  {author} {\bibfnamefont {P.~M.}\ \bibnamefont {Echenique}}, \bibinfo {author}
  {\bibfnamefont {A.}~\bibnamefont {Isaeva}},\ and\ \bibinfo {author}
  {\bibfnamefont {E.~V.}\ \bibnamefont {Chulkov}},\ }\href
  {https://doi.org/10.1038/s41586-019-1840-9} {\bibfield  {journal} {\bibinfo
  {journal} {Nature}\ }\textbf {\bibinfo {volume} {576}},\ \bibinfo {pages}
  {416} (\bibinfo {year} {2019}{\natexlab{b}})}\BibitemShut {NoStop}%
\bibitem [{\citenamefont {Zeugner}\ \emph {et~al.}(2019)\citenamefont
  {Zeugner}, \citenamefont {Nietschke}, \citenamefont {Wolter}, \citenamefont
  {Gaß}, \citenamefont {Vidal}, \citenamefont {Peixoto}, \citenamefont {Pohl},
  \citenamefont {Damm}, \citenamefont {Lubk}, \citenamefont {Hentrich},
  \citenamefont {Moser}, \citenamefont {Fornari}, \citenamefont {Min},
  \citenamefont {Schatz}, \citenamefont {Kißner}, \citenamefont {Ünzelmann},
  \citenamefont {Kaiser}, \citenamefont {Scaravaggi}, \citenamefont
  {Rellinghaus}, \citenamefont {Nielsch}, \citenamefont {Hess}, \citenamefont
  {Büchner}, \citenamefont {Reinert}, \citenamefont {Bentmann}, \citenamefont
  {Oeckler}, \citenamefont {Doert}, \citenamefont {Ruck},\ and\ \citenamefont
  {Isaeva}}]{Zeugner_2019_AFM}%
  \BibitemOpen
  \bibfield  {author} {\bibinfo {author} {\bibfnamefont {A.}~\bibnamefont
  {Zeugner}}, \bibinfo {author} {\bibfnamefont {F.}~\bibnamefont {Nietschke}},
  \bibinfo {author} {\bibfnamefont {A.~U.~B.}\ \bibnamefont {Wolter}}, \bibinfo
  {author} {\bibfnamefont {S.}~\bibnamefont {Gaß}}, \bibinfo {author}
  {\bibfnamefont {R.~C.}\ \bibnamefont {Vidal}}, \bibinfo {author}
  {\bibfnamefont {T.~R.~F.}\ \bibnamefont {Peixoto}}, \bibinfo {author}
  {\bibfnamefont {D.}~\bibnamefont {Pohl}}, \bibinfo {author} {\bibfnamefont
  {C.}~\bibnamefont {Damm}}, \bibinfo {author} {\bibfnamefont {A.}~\bibnamefont
  {Lubk}}, \bibinfo {author} {\bibfnamefont {R.}~\bibnamefont {Hentrich}},
  \bibinfo {author} {\bibfnamefont {S.~K.}\ \bibnamefont {Moser}}, \bibinfo
  {author} {\bibfnamefont {C.}~\bibnamefont {Fornari}}, \bibinfo {author}
  {\bibfnamefont {C.~H.}\ \bibnamefont {Min}}, \bibinfo {author} {\bibfnamefont
  {S.}~\bibnamefont {Schatz}}, \bibinfo {author} {\bibfnamefont
  {K.}~\bibnamefont {Kißner}}, \bibinfo {author} {\bibfnamefont
  {M.}~\bibnamefont {Ünzelmann}}, \bibinfo {author} {\bibfnamefont
  {M.}~\bibnamefont {Kaiser}}, \bibinfo {author} {\bibfnamefont
  {F.}~\bibnamefont {Scaravaggi}}, \bibinfo {author} {\bibfnamefont
  {B.}~\bibnamefont {Rellinghaus}}, \bibinfo {author} {\bibfnamefont
  {K.}~\bibnamefont {Nielsch}}, \bibinfo {author} {\bibfnamefont
  {C.}~\bibnamefont {Hess}}, \bibinfo {author} {\bibfnamefont {B.}~\bibnamefont
  {Büchner}}, \bibinfo {author} {\bibfnamefont {F.}~\bibnamefont {Reinert}},
  \bibinfo {author} {\bibfnamefont {H.}~\bibnamefont {Bentmann}}, \bibinfo
  {author} {\bibfnamefont {O.}~\bibnamefont {Oeckler}}, \bibinfo {author}
  {\bibfnamefont {T.}~\bibnamefont {Doert}}, \bibinfo {author} {\bibfnamefont
  {M.}~\bibnamefont {Ruck}},\ and\ \bibinfo {author} {\bibfnamefont
  {A.}~\bibnamefont {Isaeva}},\ }\href
  {https://doi.org/10.1021/acs.chemmater.8b05017} {\bibfield  {journal}
  {\bibinfo  {journal} {Chemistry of Materials}\ }\textbf {\bibinfo {volume}
  {31}},\ \bibinfo {pages} {2795} (\bibinfo {year} {2019})}\BibitemShut
  {NoStop}%
\bibitem [{\citenamefont {Vidal}\ \emph {et~al.}(2019)\citenamefont {Vidal},
  \citenamefont {Bentmann}, \citenamefont {Peixoto}, \citenamefont {Zeugner},
  \citenamefont {Moser}, \citenamefont {Min}, \citenamefont {Schatz},
  \citenamefont {Ki\ss{}ner}, \citenamefont {\"Unzelmann}, \citenamefont
  {Fornari}, \citenamefont {Vasili}, \citenamefont {Valvidares}, \citenamefont
  {Sakamoto}, \citenamefont {Mondal}, \citenamefont {Fujii}, \citenamefont
  {Vobornik}, \citenamefont {Jung}, \citenamefont {Cacho}, \citenamefont {Kim},
  \citenamefont {Koch}, \citenamefont {Jozwiak}, \citenamefont {Bostwick},
  \citenamefont {Denlinger}, \citenamefont {Rotenberg}, \citenamefont {Buck},
  \citenamefont {Hoesch}, \citenamefont {Diekmann}, \citenamefont {Rohlf},
  \citenamefont {Kall\"ane}, \citenamefont {Rossnagel}, \citenamefont
  {Otrokov}, \citenamefont {Chulkov}, \citenamefont {Ruck}, \citenamefont
  {Isaeva},\ and\ \citenamefont {Reinert}}]{Vidal_2019_MBT}%
  \BibitemOpen
  \bibfield  {author} {\bibinfo {author} {\bibfnamefont {R.~C.}\ \bibnamefont
  {Vidal}}, \bibinfo {author} {\bibfnamefont {H.}~\bibnamefont {Bentmann}},
  \bibinfo {author} {\bibfnamefont {T.~R.~F.}\ \bibnamefont {Peixoto}},
  \bibinfo {author} {\bibfnamefont {A.}~\bibnamefont {Zeugner}}, \bibinfo
  {author} {\bibfnamefont {S.}~\bibnamefont {Moser}}, \bibinfo {author}
  {\bibfnamefont {C.-H.}\ \bibnamefont {Min}}, \bibinfo {author} {\bibfnamefont
  {S.}~\bibnamefont {Schatz}}, \bibinfo {author} {\bibfnamefont
  {K.}~\bibnamefont {Ki\ss{}ner}}, \bibinfo {author} {\bibfnamefont
  {M.}~\bibnamefont {\"Unzelmann}}, \bibinfo {author} {\bibfnamefont {C.~I.}\
  \bibnamefont {Fornari}}, \bibinfo {author} {\bibfnamefont {H.~B.}\
  \bibnamefont {Vasili}}, \bibinfo {author} {\bibfnamefont {M.}~\bibnamefont
  {Valvidares}}, \bibinfo {author} {\bibfnamefont {K.}~\bibnamefont
  {Sakamoto}}, \bibinfo {author} {\bibfnamefont {D.}~\bibnamefont {Mondal}},
  \bibinfo {author} {\bibfnamefont {J.}~\bibnamefont {Fujii}}, \bibinfo
  {author} {\bibfnamefont {I.}~\bibnamefont {Vobornik}}, \bibinfo {author}
  {\bibfnamefont {S.}~\bibnamefont {Jung}}, \bibinfo {author} {\bibfnamefont
  {C.}~\bibnamefont {Cacho}}, \bibinfo {author} {\bibfnamefont {T.~K.}\
  \bibnamefont {Kim}}, \bibinfo {author} {\bibfnamefont {R.~J.}\ \bibnamefont
  {Koch}}, \bibinfo {author} {\bibfnamefont {C.}~\bibnamefont {Jozwiak}},
  \bibinfo {author} {\bibfnamefont {A.}~\bibnamefont {Bostwick}}, \bibinfo
  {author} {\bibfnamefont {J.~D.}\ \bibnamefont {Denlinger}}, \bibinfo {author}
  {\bibfnamefont {E.}~\bibnamefont {Rotenberg}}, \bibinfo {author}
  {\bibfnamefont {J.}~\bibnamefont {Buck}}, \bibinfo {author} {\bibfnamefont
  {M.}~\bibnamefont {Hoesch}}, \bibinfo {author} {\bibfnamefont
  {F.}~\bibnamefont {Diekmann}}, \bibinfo {author} {\bibfnamefont
  {S.}~\bibnamefont {Rohlf}}, \bibinfo {author} {\bibfnamefont
  {M.}~\bibnamefont {Kall\"ane}}, \bibinfo {author} {\bibfnamefont
  {K.}~\bibnamefont {Rossnagel}}, \bibinfo {author} {\bibfnamefont {M.~M.}\
  \bibnamefont {Otrokov}}, \bibinfo {author} {\bibfnamefont {E.~V.}\
  \bibnamefont {Chulkov}}, \bibinfo {author} {\bibfnamefont {M.}~\bibnamefont
  {Ruck}}, \bibinfo {author} {\bibfnamefont {A.}~\bibnamefont {Isaeva}},\ and\
  \bibinfo {author} {\bibfnamefont {F.}~\bibnamefont {Reinert}},\ }\href
  {https://doi.org/10.1103/PhysRevB.100.121104} {\bibfield  {journal} {\bibinfo
   {journal} {Phys. Rev. B}\ }\textbf {\bibinfo {volume} {100}},\ \bibinfo
  {pages} {121104(R)} (\bibinfo {year} {2019})}\BibitemShut {NoStop}%
\bibitem [{\citenamefont {Yasuda}\ \emph {et~al.}(2017)\citenamefont {Yasuda},
  \citenamefont {Mogi}, \citenamefont {Yoshimi}, \citenamefont {Tsukazaki},
  \citenamefont {Takahashi}, \citenamefont {Kawasaki}, \citenamefont {Kagawa},\
  and\ \citenamefont {Tokura}}]{Yasuda2017_MTI}%
  \BibitemOpen
  \bibfield  {author} {\bibinfo {author} {\bibfnamefont {K.}~\bibnamefont
  {Yasuda}}, \bibinfo {author} {\bibfnamefont {M.}~\bibnamefont {Mogi}},
  \bibinfo {author} {\bibfnamefont {R.}~\bibnamefont {Yoshimi}}, \bibinfo
  {author} {\bibfnamefont {A.}~\bibnamefont {Tsukazaki}}, \bibinfo {author}
  {\bibfnamefont {K.~S.}\ \bibnamefont {Takahashi}}, \bibinfo {author}
  {\bibfnamefont {M.}~\bibnamefont {Kawasaki}}, \bibinfo {author}
  {\bibfnamefont {F.}~\bibnamefont {Kagawa}},\ and\ \bibinfo {author}
  {\bibfnamefont {Y.}~\bibnamefont {Tokura}},\ }\href
  {https://doi.org/10.1126/science.aan5991} {\bibfield  {journal} {\bibinfo
  {journal} {Science}\ }\textbf {\bibinfo {volume} {358}},\ \bibinfo {pages}
  {1311} (\bibinfo {year} {2017})}\BibitemShut {NoStop}%
\bibitem [{\citenamefont {Rosen}\ \emph {et~al.}(2017)\citenamefont {Rosen},
  \citenamefont {Fox}, \citenamefont {Kou}, \citenamefont {Pan}, \citenamefont
  {Wang},\ and\ \citenamefont {Goldhaber-Gordon}}]{Rosen2017}%
  \BibitemOpen
  \bibfield  {author} {\bibinfo {author} {\bibfnamefont {I.~T.}\ \bibnamefont
  {Rosen}}, \bibinfo {author} {\bibfnamefont {E.~J.}\ \bibnamefont {Fox}},
  \bibinfo {author} {\bibfnamefont {X.}~\bibnamefont {Kou}}, \bibinfo {author}
  {\bibfnamefont {L.}~\bibnamefont {Pan}}, \bibinfo {author} {\bibfnamefont
  {K.~L.}\ \bibnamefont {Wang}},\ and\ \bibinfo {author} {\bibfnamefont
  {D.}~\bibnamefont {Goldhaber-Gordon}},\ }\href
  {https://doi.org/10.1038/s41535-017-0073-0} {\bibfield  {journal} {\bibinfo
  {journal} {npj Quantum Materials}\ }\textbf {\bibinfo {volume} {2}},\
  \bibinfo {pages} {69} (\bibinfo {year} {2017})}\BibitemShut {NoStop}%
\bibitem [{\citenamefont {Araki}\ \emph {et~al.}(2016)\citenamefont {Araki},
  \citenamefont {Yoshida},\ and\ \citenamefont {Nomura}}]{Araki_2016}%
  \BibitemOpen
  \bibfield  {author} {\bibinfo {author} {\bibfnamefont {Y.}~\bibnamefont
  {Araki}}, \bibinfo {author} {\bibfnamefont {A.}~\bibnamefont {Yoshida}},\
  and\ \bibinfo {author} {\bibfnamefont {K.}~\bibnamefont {Nomura}},\ }\href
  {https://doi.org/10.1103/PhysRevB.94.115312} {\bibfield  {journal} {\bibinfo
  {journal} {Phys. Rev. B}\ }\textbf {\bibinfo {volume} {94}},\ \bibinfo
  {pages} {115312} (\bibinfo {year} {2016})}\BibitemShut {NoStop}%
\bibitem [{\citenamefont {Zhang}\ \emph {et~al.}(2019)\citenamefont {Zhang},
  \citenamefont {Liu},\ and\ \citenamefont {Wang}}]{Zhang_MDW_2019}%
  \BibitemOpen
  \bibfield  {author} {\bibinfo {author} {\bibfnamefont {J.}~\bibnamefont
  {Zhang}}, \bibinfo {author} {\bibfnamefont {Z.}~\bibnamefont {Liu}},\ and\
  \bibinfo {author} {\bibfnamefont {J.}~\bibnamefont {Wang}},\ }\href
  {https://doi.org/10.1103/PhysRevB.100.165117} {\bibfield  {journal} {\bibinfo
   {journal} {Phys. Rev. B}\ }\textbf {\bibinfo {volume} {100}},\ \bibinfo
  {pages} {165117} (\bibinfo {year} {2019})}\BibitemShut {NoStop}%
\bibitem [{\citenamefont {Varnava}\ \emph {et~al.}(2020)\citenamefont
  {Varnava}, \citenamefont {Wilson}, \citenamefont {Pixley},\ and\
  \citenamefont {Vanderbilt}}]{VarnavaVander_MDW}%
  \BibitemOpen
  \bibfield  {author} {\bibinfo {author} {\bibfnamefont {N.}~\bibnamefont
  {Varnava}}, \bibinfo {author} {\bibfnamefont {J.~H.}\ \bibnamefont {Wilson}},
  \bibinfo {author} {\bibfnamefont {J.~H.}\ \bibnamefont {Pixley}},\ and\
  \bibinfo {author} {\bibfnamefont {D.}~\bibnamefont {Vanderbilt}},\
  }\href@noop {} {\bibfield  {journal} {\bibinfo  {journal} {arXiv preprint
  arXiv:2008.03316}\ } (\bibinfo {year} {2020})}\BibitemShut {NoStop}%
\bibitem [{\citenamefont {Sedlmayr}\ \emph {et~al.}(2020)\citenamefont
  {Sedlmayr}, \citenamefont {Sedlmayr}, \citenamefont
  {Barna\ifmmode~\acute{s}\else \'{s}\fi{}},\ and\ \citenamefont
  {Dugaev}}]{Dugaev_2020}%
  \BibitemOpen
  \bibfield  {author} {\bibinfo {author} {\bibfnamefont {M.}~\bibnamefont
  {Sedlmayr}}, \bibinfo {author} {\bibfnamefont {N.}~\bibnamefont {Sedlmayr}},
  \bibinfo {author} {\bibfnamefont {J.}~\bibnamefont
  {Barna\ifmmode~\acute{s}\else \'{s}\fi{}}},\ and\ \bibinfo {author}
  {\bibfnamefont {V.~K.}\ \bibnamefont {Dugaev}},\ }\href
  {https://doi.org/10.1103/PhysRevB.101.155420} {\bibfield  {journal} {\bibinfo
   {journal} {Phys. Rev. B}\ }\textbf {\bibinfo {volume} {101}},\ \bibinfo
  {pages} {155420} (\bibinfo {year} {2020})}\BibitemShut {NoStop}%
\bibitem [{\citenamefont {Garrity}\ \emph {et~al.}(2021)\citenamefont
  {Garrity}, \citenamefont {Chowdhury},\ and\ \citenamefont
  {Tavazza}}]{Tavazza_2021}%
  \BibitemOpen
  \bibfield  {author} {\bibinfo {author} {\bibfnamefont {K.~F.}\ \bibnamefont
  {Garrity}}, \bibinfo {author} {\bibfnamefont {S.}~\bibnamefont {Chowdhury}},\
  and\ \bibinfo {author} {\bibfnamefont {F.~M.}\ \bibnamefont {Tavazza}},\
  }\href {https://doi.org/10.1103/PhysRevMaterials.5.024207} {\bibfield
  {journal} {\bibinfo  {journal} {Phys. Rev. Materials}\ }\textbf {\bibinfo
  {volume} {5}},\ \bibinfo {pages} {024207} (\bibinfo {year}
  {2021})}\BibitemShut {NoStop}%
\bibitem [{\citenamefont {Fu}(2009)}]{LFU_hex_warping_2009}%
  \BibitemOpen
  \bibfield  {author} {\bibinfo {author} {\bibfnamefont {L.}~\bibnamefont
  {Fu}},\ }\href {https://doi.org/10.1103/PhysRevLett.103.266801} {\bibfield
  {journal} {\bibinfo  {journal} {Phys. Rev. Lett.}\ }\textbf {\bibinfo
  {volume} {103}},\ \bibinfo {pages} {266801} (\bibinfo {year}
  {2009})}\BibitemShut {NoStop}%
\bibitem [{\citenamefont {Rauch}\ \emph {et~al.}(2014)\citenamefont {Rauch},
  \citenamefont {Flieger}, \citenamefont {Henk}, \citenamefont {Mertig},\ and\
  \citenamefont {Ernst}}]{TRauch_DualTop_2014}%
  \BibitemOpen
  \bibfield  {author} {\bibinfo {author} {\bibfnamefont {T.}~\bibnamefont
  {Rauch}}, \bibinfo {author} {\bibfnamefont {M.}~\bibnamefont {Flieger}},
  \bibinfo {author} {\bibfnamefont {J.}~\bibnamefont {Henk}}, \bibinfo {author}
  {\bibfnamefont {I.}~\bibnamefont {Mertig}},\ and\ \bibinfo {author}
  {\bibfnamefont {A.}~\bibnamefont {Ernst}},\ }\href
  {https://doi.org/10.1103/PhysRevLett.112.016802} {\bibfield  {journal}
  {\bibinfo  {journal} {Phys. Rev. Lett.}\ }\textbf {\bibinfo {volume} {112}},\
  \bibinfo {pages} {016802} (\bibinfo {year} {2014})}\BibitemShut {NoStop}%
\bibitem [{\citenamefont {Messias~de Resende}\ \emph
  {et~al.}(2017)\citenamefont {Messias~de Resende}, \citenamefont {de~Lima},
  \citenamefont {Miwa}, \citenamefont {Vernek},\ and\ \citenamefont
  {Ferreira}}]{Resende_2017}%
  \BibitemOpen
  \bibfield  {author} {\bibinfo {author} {\bibfnamefont {B.}~\bibnamefont
  {Messias~de Resende}}, \bibinfo {author} {\bibfnamefont {F.~C.}\ \bibnamefont
  {de~Lima}}, \bibinfo {author} {\bibfnamefont {R.~H.}\ \bibnamefont {Miwa}},
  \bibinfo {author} {\bibfnamefont {E.}~\bibnamefont {Vernek}},\ and\ \bibinfo
  {author} {\bibfnamefont {G.~J.}\ \bibnamefont {Ferreira}},\ }\href
  {https://doi.org/10.1103/PhysRevB.96.161113} {\bibfield  {journal} {\bibinfo
  {journal} {Phys. Rev. B}\ }\textbf {\bibinfo {volume} {96}},\ \bibinfo
  {pages} {161113(R)} (\bibinfo {year} {2017})}\BibitemShut {NoStop}%
\bibitem [{\citenamefont {Sancho}\ \emph {et~al.}(1985)\citenamefont {Sancho},
  \citenamefont {Sancho}, \citenamefont {Sancho},\ and\ \citenamefont
  {Rubio}}]{Sancho:85}%
  \BibitemOpen
  \bibfield  {author} {\bibinfo {author} {\bibfnamefont {M.~P.~L.}\
  \bibnamefont {Sancho}}, \bibinfo {author} {\bibfnamefont {J.~M.~L.}\
  \bibnamefont {Sancho}}, \bibinfo {author} {\bibfnamefont {J.~M.~L.}\
  \bibnamefont {Sancho}},\ and\ \bibinfo {author} {\bibfnamefont
  {J.}~\bibnamefont {Rubio}},\ }\href
  {https://doi.org/10.1088/0305-4608/15/4/009} {\bibfield  {journal} {\bibinfo
  {journal} {Journal of Physics F: Metal Physics}\ }\textbf {\bibinfo {volume}
  {15}},\ \bibinfo {pages} {851} (\bibinfo {year} {1985})}\BibitemShut
  {NoStop}%
\bibitem [{\citenamefont {Henk}\ and\ \citenamefont
  {Schattke}(1993)}]{Henk:93}%
  \BibitemOpen
  \bibfield  {author} {\bibinfo {author} {\bibfnamefont {J.}~\bibnamefont
  {Henk}}\ and\ \bibinfo {author} {\bibfnamefont {W.}~\bibnamefont
  {Schattke}},\ }\href
  {https://doi.org/https://doi.org/10.1016/0010-4655(93)90038-E} {\bibfield
  {journal} {\bibinfo  {journal} {Computer Physics Communications}\ }\textbf
  {\bibinfo {volume} {77}},\ \bibinfo {pages} {69 } (\bibinfo {year}
  {1993})}\BibitemShut {NoStop}%
\bibitem [{\citenamefont {Nomura}\ and\ \citenamefont
  {Nagaosa}(2011)}]{Namura2011}%
  \BibitemOpen
  \bibfield  {author} {\bibinfo {author} {\bibfnamefont {K.}~\bibnamefont
  {Nomura}}\ and\ \bibinfo {author} {\bibfnamefont {N.}~\bibnamefont
  {Nagaosa}},\ }\href {https://doi.org/10.1103/PhysRevLett.106.166802}
  {\bibfield  {journal} {\bibinfo  {journal} {Phys. Rev. Lett.}\ }\textbf
  {\bibinfo {volume} {106}},\ \bibinfo {pages} {166802} (\bibinfo {year}
  {2011})}\BibitemShut {NoStop}%
\bibitem [{\citenamefont {Upadhyaya}\ and\ \citenamefont
  {Tserkovnyak}(2016)}]{Upadhyaya_2016}%
  \BibitemOpen
  \bibfield  {author} {\bibinfo {author} {\bibfnamefont {P.}~\bibnamefont
  {Upadhyaya}}\ and\ \bibinfo {author} {\bibfnamefont {Y.}~\bibnamefont
  {Tserkovnyak}},\ }\href {https://doi.org/10.1103/PhysRevB.94.020411}
  {\bibfield  {journal} {\bibinfo  {journal} {Phys. Rev. B}\ }\textbf {\bibinfo
  {volume} {94}},\ \bibinfo {pages} {020411(R)} (\bibinfo {year}
  {2016})}\BibitemShut {NoStop}%
\bibitem [{\citenamefont {Volkov}\ \emph {et~al.}(1995)\citenamefont {Volkov},
  \citenamefont {Idlis},\ and\ \citenamefont {Usmanov}}]{VIU_1995}%
  \BibitemOpen
  \bibfield  {author} {\bibinfo {author} {\bibfnamefont {B.~A.}\ \bibnamefont
  {Volkov}}, \bibinfo {author} {\bibfnamefont {B.~G.}\ \bibnamefont {Idlis}},\
  and\ \bibinfo {author} {\bibfnamefont {M.~S.}\ \bibnamefont {Usmanov}},\
  }\href@noop {} {\bibfield  {journal} {\bibinfo  {journal} {Phys. Usp.}\
  }\textbf {\bibinfo {volume} {38}},\ \bibinfo {pages} {761} (\bibinfo {year}
  {1995})}\BibitemShut {NoStop}%
\bibitem [{\citenamefont {Volkov}\ and\ \citenamefont
  {Pankratov}(1986)}]{VP_1986}%
  \BibitemOpen
  \bibfield  {author} {\bibinfo {author} {\bibfnamefont {B.~A.}\ \bibnamefont
  {Volkov}}\ and\ \bibinfo {author} {\bibfnamefont {O.~A.}\ \bibnamefont
  {Pankratov}},\ }\href@noop {} {\bibfield  {journal} {\bibinfo  {journal}
  {JETP Letters}\ }\textbf {\bibinfo {volume} {43}},\ \bibinfo {pages} {130}
  (\bibinfo {year} {1986})}\BibitemShut {NoStop}%
\bibitem [{\citenamefont {Hubert}\ and\ \citenamefont
  {Schafer}(1998)}]{Hubert_book}%
  \BibitemOpen
  \bibfield  {author} {\bibinfo {author} {\bibfnamefont {A.}~\bibnamefont
  {Hubert}}\ and\ \bibinfo {author} {\bibfnamefont {R.}~\bibnamefont
  {Schafer}},\ }\href {https://doi.org/10.1007/978-3-540-85054-0} {\emph
  {\bibinfo {title} {Magnetic domains}}}\ (\bibinfo  {publisher}
  {Springer-Verlag Berlin Heidelberg},\ \bibinfo {address} {Berlin},\ \bibinfo
  {year} {1998})\BibitemShut {NoStop}%
\bibitem [{\citenamefont {Nolting}\ and\ \citenamefont
  {Ramakanth}(2009)}]{Nolting_book}%
  \BibitemOpen
  \bibfield  {author} {\bibinfo {author} {\bibfnamefont {W.}~\bibnamefont
  {Nolting}}\ and\ \bibinfo {author} {\bibfnamefont {A.}~\bibnamefont
  {Ramakanth}},\ }\href@noop {} {\emph {\bibinfo {title} {Quantum theory of
  magnetism}}}\ (\bibinfo  {publisher} {Springer-Verlag Berlin Heidelberg},\
  \bibinfo {address} {Berlin},\ \bibinfo {year} {2009})\BibitemShut {NoStop}%
\bibitem [{\citenamefont {Kim}\ \emph {et~al.}(2020)\citenamefont {Kim},
  \citenamefont {Denlinger}, \citenamefont {Kundu}, \citenamefont {Gu},\ and\
  \citenamefont {Valla}}]{KimDenliner2020}%
  \BibitemOpen
  \bibfield  {author} {\bibinfo {author} {\bibfnamefont {C.}~\bibnamefont
  {Kim}}, \bibinfo {author} {\bibfnamefont {J.}~\bibnamefont {Denlinger}},
  \bibinfo {author} {\bibfnamefont {A.}~\bibnamefont {Kundu}}, \bibinfo
  {author} {\bibfnamefont {G.}~\bibnamefont {Gu}},\ and\ \bibinfo {author}
  {\bibfnamefont {T.}~\bibnamefont {Valla}},\ }\href@noop {} {\bibfield
  {journal} {\bibinfo  {journal} {arXiv preprint arXiv:2012.05884}\ } (\bibinfo
  {year} {2020})}\BibitemShut {NoStop}%
\bibitem [{\citenamefont {Men'shov}\ \emph
  {et~al.}(2019{\natexlab{b}})\citenamefont {Men'shov}, \citenamefont
  {Shvets},\ and\ \citenamefont {Chulkov}}]{MSC_2019}%
  \BibitemOpen
  \bibfield  {author} {\bibinfo {author} {\bibfnamefont {V.~N.}\ \bibnamefont
  {Men'shov}}, \bibinfo {author} {\bibfnamefont {I.~A.}\ \bibnamefont
  {Shvets}},\ and\ \bibinfo {author} {\bibfnamefont {E.~V.}\ \bibnamefont
  {Chulkov}},\ }\href {https://doi.org/10.1103/PhysRevB.99.115301} {\bibfield
  {journal} {\bibinfo  {journal} {Phys. Rev. B}\ }\textbf {\bibinfo {volume}
  {99}},\ \bibinfo {pages} {115301} (\bibinfo {year}
  {2019}{\natexlab{b}})}\BibitemShut {NoStop}%
\bibitem [{\citenamefont {Kawamura}\ \emph {et~al.}(2018)\citenamefont
  {Kawamura}, \citenamefont {Mogi}, \citenamefont {Yoshimi}, \citenamefont
  {Tsukazaki}, \citenamefont {Kozuka}, \citenamefont {Takahashi}, \citenamefont
  {Kawasaki},\ and\ \citenamefont {Tokura}}]{Kawamura_2018}%
  \BibitemOpen
  \bibfield  {author} {\bibinfo {author} {\bibfnamefont {M.}~\bibnamefont
  {Kawamura}}, \bibinfo {author} {\bibfnamefont {M.}~\bibnamefont {Mogi}},
  \bibinfo {author} {\bibfnamefont {R.}~\bibnamefont {Yoshimi}}, \bibinfo
  {author} {\bibfnamefont {A.}~\bibnamefont {Tsukazaki}}, \bibinfo {author}
  {\bibfnamefont {Y.}~\bibnamefont {Kozuka}}, \bibinfo {author} {\bibfnamefont
  {K.~S.}\ \bibnamefont {Takahashi}}, \bibinfo {author} {\bibfnamefont
  {M.}~\bibnamefont {Kawasaki}},\ and\ \bibinfo {author} {\bibfnamefont
  {Y.}~\bibnamefont {Tokura}},\ }\href
  {https://doi.org/10.1103/PhysRevB.98.140404} {\bibfield  {journal} {\bibinfo
  {journal} {Phys. Rev. B}\ }\textbf {\bibinfo {volume} {98}},\ \bibinfo
  {pages} {140404(R)} (\bibinfo {year} {2018})}\BibitemShut {NoStop}%
\bibitem [{\citenamefont {Mackenzie}\ and\ \citenamefont
  {Maeno}(2003)}]{Mackanzie_2003}%
  \BibitemOpen
  \bibfield  {author} {\bibinfo {author} {\bibfnamefont {A.~P.}\ \bibnamefont
  {Mackenzie}}\ and\ \bibinfo {author} {\bibfnamefont {Y.}~\bibnamefont
  {Maeno}},\ }\href {https://doi.org/10.1103/RevModPhys.75.657} {\bibfield
  {journal} {\bibinfo  {journal} {Rev. Mod. Phys.}\ }\textbf {\bibinfo {volume}
  {75}},\ \bibinfo {pages} {657} (\bibinfo {year} {2003})}\BibitemShut
  {NoStop}%
\bibitem [{\citenamefont {Sethi}\ \emph {et~al.}(2021)\citenamefont {Sethi},
  \citenamefont {Zhou}, \citenamefont {Zhu}, \citenamefont {Yang},\ and\
  \citenamefont {Liu}}]{Sethi_2021}%
  \BibitemOpen
  \bibfield  {author} {\bibinfo {author} {\bibfnamefont {G.}~\bibnamefont
  {Sethi}}, \bibinfo {author} {\bibfnamefont {Y.}~\bibnamefont {Zhou}},
  \bibinfo {author} {\bibfnamefont {L.}~\bibnamefont {Zhu}}, \bibinfo {author}
  {\bibfnamefont {L.}~\bibnamefont {Yang}},\ and\ \bibinfo {author}
  {\bibfnamefont {F.}~\bibnamefont {Liu}},\ }\href@noop {} {\bibfield
  {journal} {\bibinfo  {journal} {arXiv preprint arXiv:2102.08593}\ } (\bibinfo
  {year} {2021})}\BibitemShut {NoStop}%
\end{thebibliography}%
\end{document}